\documentclass[10pt]{article}
\usepackage[sort&compress,numbers,comma]{natbib}
\usepackage[margin=0.75in, paperwidth=6in, paperheight=9in]{geometry}

\makeatletter
\newcommand\appendix@section[1]{%
  \refstepcounter{section}%
  \orig@section*{Appendix \@Alph\c@section: #1}%
  \addcontentsline{toc}{section}{Appendix \@Alph\c@section: #1}%
}
\let\orig@section\section
\g@addto@macro\appendix{\let\section\appendix@section}
\makeatother

\usepackage{amsmath}

\usepackage{amsfonts}
\usepackage{mathrsfs}

\usepackage[T1]{fontenc}
\usepackage[protrusion=true,expansion,final]{microtype}

\usepackage[hidelinks]{hyperref}

\usepackage{lineno}
\modulolinenumbers[5]

\usepackage{xargs}
\usepackage{ifthen}
\usepackage{mathtools}
\usepackage{xfrac}

\usepackage{graphicx}

\usepackage{pgf}
\usepackage{tikz}
\input{pgfsettings}

\definecolor{background}{HTML}{272822}
\definecolor{caret}{HTML}{F8F8F0}
\definecolor{foreground}{HTML}{F8F8F2}
\definecolor{invisibles}{HTML}{3B3A32}
\definecolor{lineHighlight}{HTML}{3E3D32}
\definecolor{selection}{HTML}{49483E}
\definecolor{findHighlight}{HTML}{FFE792}
\definecolor{findHighlightForeground}{HTML}{000000}
\definecolor{selectionBorder}{HTML}{222218}
\definecolor{activeGuide}{HTML}{9D550F}
\definecolor{bracketsForeground}{HTML}{F8F8F2}
\definecolor{bracketContentsForeground}{HTML}{F8F8F2}

\definecolor{comment.foreground}{HTML}{75715E}
\definecolor{string.foreground}{HTML}{E6DB74}
\definecolor{number.foreground}{HTML}{AE81FF}

\definecolor{builtInConstant.foreground}{HTML}{AE81FF}
\definecolor{userDefinedConstant.foreground}{HTML}{AE81FF}
\definecolor{keyword.foreground}{HTML}{F92672}
\definecolor{storage.foreground}{HTML}{F92672}
\definecolor{storageType.foreground}{HTML}{66D9EF}
\definecolor{className.foreground}{HTML}{A6E22E}
\definecolor{inheritedClass.foreground}{HTML}{A6E22E}
\definecolor{functionName.foreground}{HTML}{A6E22E}
\definecolor{functionArgument.foreground}{HTML}{FD971F}
\definecolor{tagName.foreground}{HTML}{F92672}
\definecolor{tagAttribute.foreground}{HTML}{A6E22E}
\definecolor{libraryFunction.foreground}{HTML}{66D9EF}
\definecolor{libraryConstant.foreground}{HTML}{66D9EF}
\definecolor{libraryClass.foreground}{HTML}{66D9EF}
\definecolor{invalid.background}{HTML}{F92672}
\definecolor{invalid.foreground}{HTML}{F8F8F0}
\definecolor{invalid.deprecated.background}{HTML}{AE81FF}
\definecolor{invalid.deprecated.foreground}{HTML}{F8F8F0}
\definecolor{JSONString.foreground}{HTML}{CFCFC2}
\definecolor{diffHeader.foreground}{HTML}{75715E}
\definecolor{diffDeleted.foreground}{HTML}{F92672}
\definecolor{diffInserted.foreground}{HTML}{A6E22E}
\definecolor{diffChanged.foreground}{HTML}{E6DB74}

\usepackage[inline]{enumitem}

\usepackage{xspace}
\newcommand*{\ie}{i.e.\@\xspace}
\newcommand*{\eg}{e.g.\@\xspace}
\newcommand*{\cf}{cf.\@\xspace}
\newcommand*{\cp}{cp.\@\xspace}

\newcommand*{\etal}{et al.\@\xspace}

\newcommand*{\vs}{vs.\@\xspace}
\makeatletter
\newcommand*{\etc}{%
  \@ifnextchar{.}%
      {etc}%
      {etc.\@\xspace}%
}
\newcommand*{\hc}{%
  \@ifnextchar{.}%
    {\ifmmode\text{h.c}\else{h.c}\fi}%
    {\ifmmode\text{h.c.}\else{h.c.\@\xspace}\fi}%
}
\makeatother

\let\originalleft\left
\let\originalmiddle\middle
\let\originalright\right
\renewcommand{\left}{\mathopen{}\mathclose\bgroup\originalleft}
\renewcommand{\middle}{\originalmiddle}
\renewcommand{\right}{\aftergroup\egroup\originalright}

\newcommand{\ii}{\ensuremath{\mathrm{i}}}
\newcommand{\ee}{\ensuremath{\mathrm{e}}}

\DeclareMathOperator{\Tr}{Tr}
\DeclarePairedDelimiter\abs{\lvert}{\rvert}
\DeclarePairedDelimiter\norm{\lVert}{\rVert}
\DeclareMathOperator{\sgn}{sgn}

\DeclareMathOperator{\arccsc}{arccsc}

\renewcommand{\Re}{\operatorname{Re}}
\renewcommand{\Im}{\operatorname{Im}}
\DeclareMathOperator{\Arg}{Arg}

\DeclarePairedDelimiterX\sproduct[2]{(}{)}{#1,#2}

\DeclarePairedDelimiterX\expval[1]{\langle}{\rangle}{#1}
\DeclarePairedDelimiterX\bra[1]{\langle}{\vert}{#1}
\DeclarePairedDelimiterX\ket[1]{\vert}{\rangle}{#1}
\DeclarePairedDelimiterX\overlap[2]{\langle}{\rangle}{#1\delimsize\vert#2}
\DeclarePairedDelimiterX\bracket[3]{\langle}{\rangle}{#1\delimsize\vert#2\delimsize\vert#3}
\DeclarePairedDelimiterX\clebschgordan[3]{\langle}{\rangle}{#1\,#2\delimsize\vert#3}

\makeatletter
\DeclareFontFamily{OMX}{MnSymbolE}{}
\DeclareSymbolFont{MnLargeSymbols}{OMX}{MnSymbolE}{m}{n}
\SetSymbolFont{MnLargeSymbols}{bold}{OMX}{MnSymbolE}{b}{n}
\DeclareFontShape{OMX}{MnSymbolE}{m}{n}{
    <-6>  MnSymbolE5
   <6-7>  MnSymbolE6
   <7-8>  MnSymbolE7
   <8-9>  MnSymbolE8
   <9-10> MnSymbolE9
  <10-12> MnSymbolE10
  <12->   MnSymbolE12
}{}
\DeclareFontShape{OMX}{MnSymbolE}{b}{n}{
    <-6>  MnSymbolE-Bold5
   <6-7>  MnSymbolE-Bold6
   <7-8>  MnSymbolE-Bold7
   <8-9>  MnSymbolE-Bold8
   <9-10> MnSymbolE-Bold9
  <10-12> MnSymbolE-Bold10
  <12->   MnSymbolE-Bold12
}{}
\let\llangle\@undefined
\DeclareMathDelimiter{\llangle}{\mathopen}%
                     {MnLargeSymbols}{'164}{MnLargeSymbols}{'164}
\let\rrangle\@undefined
\DeclareMathDelimiter{\rrangle}{\mathclose}%
                     {MnLargeSymbols}{'171}{MnLargeSymbols}{'171}
\makeatother

\DeclareMathOperator*{\sumint}{%
\mathchoice%
  {\ooalign{$\displaystyle\sum$\cr\hidewidth$\displaystyle\int$\hidewidth\cr}}
  {\ooalign{\raisebox{.14\height}{\scalebox{.7}{$\textstyle\sum$}}\cr\hidewidth$\textstyle\int$\hidewidth\cr}}
  {\ooalign{\raisebox{.2\height}{\scalebox{.6}{$\scriptstyle\sum$}}\cr$\scriptstyle\int$\cr}}
  {\ooalign{\raisebox{.2\height}{\scalebox{.6}{$\scriptstyle\sum$}}\cr$\scriptstyle\int$\cr}}
}

\DeclarePairedDelimiterX\lbra[1]{\llangle}{\vert}{#1}
\DeclarePairedDelimiterX\lket[1]{\vert}{\rrangle}{#1}
\DeclarePairedDelimiterX\loverlap[2]{\llangle}{\rrangle}{#1\delimsize\vert#2}
\DeclarePairedDelimiterX\lbracket[3]{\llangle}{\rrangle}{#1\delimsize\vert#2\delimsize\vert#3}

\usepackage{dsfont}
\newcommand{\operatorone}{\ensuremath{\mathds{1}}}
\renewcommand{\vec}[1]{\ensuremath{\boldsymbol{#1}}}
\newcommand{\uvec}[1]{\ensuremath{\hat{\vec{#1}} \protect\vphantom{\vec{#1}}}}

\usepackage[affil-it]{authblk}

\newif\iftikz
\usetikzlibrary{arrows,automata,calc,positioning}
\usepackage{pgf}
\usepackage{tikz}
\tikzfalse

\begin{document}
\bibliographystyle{apsrev4-1}

\title{An Approach to ``Quantumness'' in Coherent Control}

\author{%
  Torsten Scholak%
  \thanks{%
    Electronic address: \texttt{tscholak@chem.utoronto.ca}%;
    %Corresponding author
  }
}%
\author{%
  Paul Brumer%
  \thanks{Electronic address: \texttt{pbrumer@chem.utoronto.ca}}
}%
\affil{
  Chemical Physics Theory Group, Department of Chemistry, and
  Center for Quantum Information and Quantum Control,
  University of Toronto, Toronto, Ontario, Canada, M5S~3H6
}

\date{Dated: \today}

\maketitle

\begin{abstract}
  Developments in the foundations of quantum mechanics have
  identified several attributes and tests associated with the
  ``quantumness" of systems,
  including entanglement,
  nonlocality,
  quantum erasure,
  Bell test, \etc{}.
  Here we introduce and utilize these tools
  to examine the role of quantum coherence and nonclassical effects
  in $1$ \vs{} $N$ photon coherent phase control,
  a paradigm for an all-optical method for manipulating molecular dynamics.
  In addition,
  truly quantum control scenarios are introduced and examined.
  The approach adopted here serves as a template
  for studies of the role
  of quantum mechanics in other coherent control
  and optimal control scenarios.
\end{abstract}

\tableofcontents

\section{Introduction}
\label{sec:introduction}

The goal of coherent control
is to manipulate the dynamics of matter at
the molecular, atomic, and even subatomic level.
The methodology,
which has been extensively reviewed \cite{Shapiro:2012ly,Rice}
relies on an external driving field,
the characteristics of the material target system,
and the coupling between them.
For example,
in two-color phase control \cite{Shapiro:1988cr},
control over the material system is achieved by
irradiating the system with a field of the form:
\begin{align}
  E(t) & = \varepsilon_1 \cos\left(\omega t + \phi_1\right) + \varepsilon_{2} \cos\left(N \omega t + \phi_2\right)~,
\end{align}
and then varying the relative
phase $\phi = N \phi_1 - \phi_2$
between the two incident weak laser fields
of frequencies $\omega$ and $N \omega$, 
with $N = 2, 3, \ldots$~.
This particular control scenario has proven of interest
in applications ranging from
control of electric currents
\cite{Kurizki:1989aa,Dupont:1995ss,Franco:2006rq,Franco:2007}
to the control of laser phase envelopes in precision measurements
\cite{Apolonski:2000ct,Jones:2000ce},
and is the primary scenario upon which we focus in this Advances.

Two-color phase control has repeatedly
\cite{Jackson:1982bf,Brumer:1986uq,Baskin:1988kk,Shapiro:1988cr,Baranova:1990xy,Chen:1990ve,Chen:1990kx,Chan:1991sf,Dupont:1995ss,Kleiman:1995vl,Zhu:1995fu,Shapiro:1997fk,Hache:1997aa,Wang:2001pi,Ehlotzky:2001ij,Bhat:2005sf,Wahlstrand:2006yg,Bolovinos:2008nr,Pasenow:2008rm,Kuroda:2009aa,Rao:2011lh,Antypas:2014ph,Muniz:2015aa}
been explained as a quantum analog of Young's double slit experiment
\cite{Feynman:1963yu},
with the slits replaced by two independent and mutually exclusive
dynamical pathways, $\text{A}$ and $\text{B}$,
both leading from an initial material
state $\ket{\psi\left(t = -\infty\right)}$
to a final state $\ket{\psi\left(t = \infty\right)}$.
In this picture, path $\text{A}$ is given by a
$1$-photon transition induced by the
laser field of frequency $N \omega$,
and path $\text{B}$ by an $N$-photon
transition caused by the second field of
frequency $\omega$.
In accord with the standard quantum description,
the probability of excitation at long times $P(\infty)$ is then
obtained by adding the
transition amplitudes $a_{\text{A}}(\infty)$
and $a_{\text{B}}(\infty)$ for the pathways
and taking the square of the absolute value to give:
\begin{align*}
  P\left(\infty\right) & = \abs*{a_{\text{A}}^{\vphantom{*}}\left(\infty\right)}^{2} +
    \abs*{a_{\text{B}}^{\vphantom{*}}\left(\infty\right)}^{2} +
    a_{\text{A}}^{\vphantom{*}}\left(\infty\right) a_{\text{B}}^{*}\left(\infty\right) +
    a_{\text{B}}^{\vphantom{*}}\left(\infty\right) a_{\text{A}}^{*}\left(\infty\right) \\
  & = P_{\text{A}}\left(\infty\right) + P_{\text{B}}\left(\infty\right) +
    2 \sqrt{P_{\text{A}}\left(\infty\right) P_{\text{B}}\left(\infty\right)} \cos\left(\phi - \theta\right),
\end{align*}
where $\theta$ denotes a material phase.
Phase control, a subset of coherent control effects,
is manifest only in the latter term,
which depends on the external control phase $\phi$.
This interference term can contribute
either constructively or destructively,
resulting in nonadditive probabilities, so that
$P\left(\infty\right) \neq P_{\text{A}}\left(\infty\right) + P_{\text{B}}\left(\infty\right)$.

Arguments for the validity and applicability of coherent phase control
have primarily relied upon formulations of this kind,
which emphasize the importance of quantum interference in control.
In particular, they rely on the functional form of this equation,
which is reminiscent of double slit interference,
the phase sensitivity of the control, and
the display of nonadditivity of the contributions.
Early two-color photoionization experiments with alkali atoms,
\cite{Chen:1990kx,Chen:1990ve,Yin:1992fu,Yin:1993fk},
serve as seminal examples of this scenario,
as do experiments
on current directionality in quantum wells \cite{Dupont:1995ss}
and semiconductors \cite{Hache:1997aa}.
However, phase sensitivity and nonadditivity can also be of classical origin
arising, for example, in transport processes\cite{Flach:2000ys}.
Hence, one can ask \cite{Franco:2010kx,Spanner:2009fk} whether
$1$ \vs{} $N$ phase control can also be viewed
as a classical interference phenomenon
or, for example, as a system's collective response
to shaped incident laser fields \cite{Bucksbaum1990,Han2005}.
Related concerns are raised by a number of additional studies
\cite{Constantoudis:2005qo,Lima:2008vn,Ivanov:2012aa}
that consider $1$ \vs{} $2$ photon phase control
as an intrinsically classical phenomenon.

In this article, we justify this concern and then
provide a formal approach to addressing this issue.
Specifically, we examine quantum control from a
\emph{foundational perspective},
relying upon modern concepts in
quantum mechanics,
such as entanglement,
quantum erasure,
nonlocality,
and Bell tests
(for an introduction, see \cite{Greenstein})
to see how a truly quantum phase control scenario can
be tested and built.
Although this article focuses explicitly on the $1$ \vs{} $N$
photon weak-field phase control, our goal is to provide a detailed approach to
examining the role of quantum mechanics in control scenarios that is
applicable to other coherent control, and to optimal control scenarios.
In the case of coherent control, such a degree of foundational rigor
has so far not been considered necessary.
Rather, one has been satisfied with
``bona fide nonclassicality'', \ie{}, the expectation that
a classical theory can not be written down
that also conforms to what one believes to be
the physics of the observed phenomenon.
From a foundational perspective,
this approach is problematic
\cite{Leggett:2002fk,Leggett:2008wd},
as discussed below.

Hence, the central focus of this Advances is the use of modern
foundational ideas in quantum mechanics to explore ``quantumness''
in coherent control.
In our examination of the $1$ \vs{} $N$ scenario,
we begin by noting that no features of interference are displayed that go beyond
the nonadditivity and the phase sensitivity
of nonlinear response to the incident laser field.
Indeed, if nonclassical effects play a role
in this coherent control scenario,
then it is by virtue of the fact that the material upon which
the lasers are incident display quantum features.
This is, of course, distinct from saying
that the \emph{control scenario} itself reflects principles
that are non-classical.
This distinction requires clarification through two significant comments.
\begin{enumerate}[label=(\roman{enumi})]
  \item
    There is absolutely no question
    that quantum mechanics is necessary to
    \emph{quantitatively} describe the outcome of
    coherent control scenarios.
    For example, a classical
    computation gives products of $\text{HD}^+$ dissociation
    in a $1$ \vs{} $2$ photon
    experiment that are antiparallel to the correct
    quantum result \cite{Sheehy:1995ff}.
    Indeed, even the absorption of light by matter requires quantum
    mechanics to obtain quantitatively accurate results, although,
    a classical description is often qualitatively satisfactory
    \cite{kauzmann,Chance:1975co,Duque:2015ca}. 
    Here, however, our focus is on the \emph{conceptual} relationship
    between the quantum and classical perspectives.
  \item
    From a \emph{formal foundational perspective},
    in both dynamics and control,
    quantitative agreement with quantum theory does not provide a
    \emph{proof} of nonclassicality.
    Rather, formally,
    proving the role of quantum mechanics requires that
    all classical descriptions and explanations
    be falsified, \ie{},
    all classical models must be unable to
    reproduce the phenomenon.
    As a consequence,
    the rigorous certification of nonclassicality
    is a challenging task which has, for example,
    been the subject of intense efforts
    in quantum optics
    \cite{Aspect:1982rr,Ou:1988aa,Kaiser:2012fk},
    quantum information
    \cite{Ekert:1991aa,Bennett:1993aa,Steane:1998aa,Duan:2001aa,Gisin:2002aa,Guhne:2009vn,Ekert:2014aa},
    and, most notably, quantum foundations
    \cite{Zurek:1991aa,Salart:2008aa,Winter:2010aa,Hensen:2015lf,Giustina:2015sl,Shalm:2015sl}.
    Studies of this type take the form of carefully crafted
    experimental protocols that close all loopholes
    that would prevent rigorous assertions about the degree to
    which quantum mechanics is necessary to describe the phenomenon---%
    called here, and below, ``quantumness''.
\end{enumerate}
If conventional $1$ \vs{} $N$ phase control,
at its core, conceptually admits a classical description,
and thus does not show
nontrivial aspects of quantum mechanics,
what new, extended verifiably quantum coherent phase control scenarios
can be designed?
Here, we introduce tools to answer these questions
in any coherent control scenario
and apply them specifically to the $1$ \vs{} $N$ photon case.
Finally, we note that understanding the nonclassicality of coherent control
has practical applications that
arise from a recognition that decoherence effects \cite{Schlosshauer}
often bring a system to the classical limit.
Hence, if control is indeed (at least partially) classical,
then it may well survive in the often unavoidable
decohering environments
associated with realistic molecular processes.

\subsection{A General Overview of This Article}
\label{sec:introduction:general_overview}

Sections \ref{sec:introduction:review_quantumclassical_correspondence}
to \ref{sec:introduction:interference_and_complementarity}
review arguments that
standard two-color weak-field phase control
is, conceptually, an analog of
a purely classical control phenomenon
in which quantum interference
and the double-slit analogy play no role.
As noted below, quantum physics only influences
the material component of the system,
expressed by response functions to the external field.
These quantities contain the complete information that is
necessary to calculate
the perturbation of a system property
in response to classical driving radiation,
particularly weak two-color fields.

In the main body of this article,
we address the question of how and when
nontrivial aspects of \emph{quantum interference}
affect phase control.
Note that by ``quantum interference''
we do not just mean that interference is observed,
but rather that the interference is verifiably quantum,
as described below.
Indeed, in this context quantum interference
and quantum mechanics are essentially synonymous.
Since these aspects will be seen to be absent in conventional phase control,
we extend our search to unconventional scenarios.
To do so,
we introduce, in Sec.~\ref{sec:CCI},
a novel type of two-path interferometer,
the ``coherent control interferometer'' (the CCI).
As in a Mach-Zehnder interferometer (MZI),
the CCI will be seen to have two input ports, two output ports,
and two path alternatives.
Each port is identified with one
of two possible outcomes in the measurement
of a matter observable
that is diagonal in the material eigenbasis,
\eg{}, spin, parity, \etc.
For example, the path alternatives
are given by the two excitation processes
in a $1$ \vs{} $N$ coherent control scenario.
The resultant CCI framework is general and
applicable to many systems, from atoms
\cite{Chen:1990kx,Zhuang:2013aa} to molecules
\cite{Sheehy:1995ff,Ohmura:2004aa,Ohmura:2004ab,Ohmura:2006qf,Kuroda:2009aa}
to bulk solids
\cite{Kurizki:1989aa,Dupont:1995ss,Hache:1997aa,Pronin:2004aa,Sun:2010aa,Mele:2000aa,Najmaie:2003aa,Rioux:2011aa,Muniz:2014aa,Muniz:2015aa}.
Further, the control field can be in a semiclassical coherent
or in a quantum state of light, such as a multimode Fock state
or a squeezed state.

The CCI will prove central to the study of
nontrivial quantum interference effects in phase control,
insofar as it formalizes the relationship between
coherent phase control and quantum interference phenomena.
With it, we are able to study
the fundamental complementarity
of wave- and particle-like behavior
in the context of coherent control
(Sec.~\ref{sec:path_distinguishability}).
Further, the trade-off between these two behaviors
is quantified via duality relations of the form
$\mathfrak{V}^{2} + \mathfrak{K}^{2} \le 1$
that provide upper bounds on the
interference fringe contrast $\mathfrak{V}$,
a wave property,
and the simultaneously acquirable
amount of path knowledge $\mathfrak{K}$,
a particle property.
Relations of this type
are not only conducive to an
experimental demonstration of complementarity
and thus one of the hallmarks of quantum interference,
but they are of immediate
practical significance in coherent phase control.
This is because
these relations show the maximum amount of control
achievable through ``quantum erasure'',
a measurement strategy by which
path knowledge
can be ``erased'' and interference restored
(Sec.~\ref{sec:quantum_erasure})
and that there is a strict equivalence between
controllability and interference contrast
in the CCI that can be indicative of
nonclassical correlations and entanglement.

Another hallmark of quantum interference
is ``delayed choice'' \cite{Zeilinger2016}.
In the context of quantum erasure,
delayed choice refers to the fact
that the time at which the experimenter decides
to restore wave-like interference
(naturally, at the expense of any particle-like path knowledge)
is of no consequence to the measurement results.
The choice can be made
even well after detecting the output from the CCI.
Below, we exploit the CCI to
design a novel coherent control implementation
of ``quantum delayed choice''
\cite{Ionicioiu:2011fk}
that has thus far only been observed
in experiments with photons
\cite{Tang:2012kx,Kaiser:2012fk,Peruzzo:2012mw}
and nuclear spins
\cite{Roy:2012uq,Auccaise:2012kx}
(Sec.~\ref{sec:delayed_choice}).
Note that quantum delayed choice
will be seen to differ from delayed choice
in that the system is in a quantum superposition of both
particle- and wavelike behaviors
that, in the CCI, are controlled
by the incident light field.
As such, prior to the measurement it can be physically unknowable
whether control is achieved or not.

Rigorous experimental proofs for
the nonclassicality of
our proposed quantum erasure and
quantum delayed choice coherent phase control scenarios
are discussed in
Sec.~\ref{sec:certification_of_nonclassicality}.
These verifications are made possible
with the Bell-CHSH inequality tests introduced below.
If these inequalities are unambiguously violated in an experiment,
it would guarantee that the observed measurement statistics
cannot be described by \emph{any} classical theory.
But even without these tests,
these new scenarios come significantly
closer to demonstrating quantum coherence and
nonclassical interference effects
in a $1$ \vs{} $N$ coherent phase control experiment.

Section \ref{sec:application}
applies the CCI formalism to
an example of control over the
spin polarization of the ejected photoelectron
in the photoionization of an alkali atom
by a $1$ \vs{} $2$ control field.
We derive explicit settings
that facilitate the implementation of
the new control scenarios and,
in particular, we design control fields
for wave- and particle-like CCI output statistics.

Loopholes are discussed in Sec.~\ref{sec:application:discussion},
and conclusions are provided
in Sec.~\ref{sec:conclusions}.

\subsection{Quantum-Classical Correspondence in Coherent Phase Control}
\label{sec:introduction:review_quantumclassical_correspondence}

The issue of classical \vs{} nonclassical effects
in coherent control
has been recently addressed for a number of control scenarios.
For example \cite{Franco:2006rq,Franco:2008fc,Franco:2010kx},
a classical, two-color cw driving field with frequency components
$\omega$ and $2 \omega$ of the form
$E(t) = E_1(t) + E_2(t) = \varepsilon_1 \cos\left(\omega t + \phi_1\right) + \varepsilon_2 \cos\left(2\omega t + \phi_2\right)$
can induce an asymmetry in the
position or momentum distribution of a single charged particle
trapped in a symmetric potential with quartic anharmonicity.
The asymmetry is controllable by
changing the relative phase
$\phi = 2 \phi_1 - \phi_2$
of the two field components.
In the quantum case,
the time-averaged position $\overline{x}_q$
is given by \cite{Franco:2006rq}
\begin{align}
  \overline{x}_q & = \chi^{(3)}_q \tfrac{3}{4} \varepsilon_1^{2} \varepsilon_2^{\vphantom{2}} \cos \phi,
  \label{eq:overlinex_quantum}
\end{align}
where the average position $\overline{x}_q$
is seen to be proportional to $\cos \phi$
and to the field amplitudes
$\varepsilon_1$, $\varepsilon_2$,
with a coefficient given by the quantum susceptibility $\chi^{(3)}_q$.
The quantum-mechanical description of control
was found \cite{Franco:2006rq} to
cross over smoothly to the classical limit,
\begin{align}
  \lim_{\hbar \to 0} \overline{x}_c & = \chi_{\mathrm{c}}^{(3)} \tfrac{3}{4} \varepsilon_1^{2} \varepsilon_2^{\vphantom{2}} \cos \phi.
  \label{eq:overlinex_classical}
\end{align}
Here the quantum and classical results are seen
to be of exactly the same form,
albeit with a \emph{quantitatively} different
susceptibilities $\displaystyle \chi_{\mathrm{c}}^{(3)}$ and $\chi^{(3)}_q$.
In particular, the quantum result adds a series of correction terms,
$\displaystyle \Delta\chi = ( \chi^{(3)}_q - \chi_{\mathrm{c}}^{(3)})$,
that describe the $\hbar$-dependent resonance structure
of the system.
Further, in the specific example discussed above,
both quantum and classical control can be explained by
solely considering classical arguments
based on reflection symmetry,
in contrast to earlier explanations
\cite{Shapiro:2012ly}
that attributed control to the breaking of parity,
an inherently quantum effect.
In addition, consider uncoupled, charged, massive particles
in static, symmetric, one-dimensional potentials
subject to external driving fields in the dipole approximation.
In these general cases, the necessary conditions for
 the creation of net dipoles or currents is
identical in the quantum and the classical regime
if the driving fields are time-periodic
and have vanishing temporal average
\cite{Franco:2008fc}---%
a standard case in scenarios of coherent phase control.
That is, both quantum and classical control
rely on the same temporal asymmetries in the driving field
and on the presence of anharmonicities in the static potential.
All of these results suggest that, conceptually,
the phase control is consistent with classical mechanics.

The quantative quantum-classical correspondence breaks down
when the initial state becomes semi-classically inadmissible
or when the driving frequencies approach the quantum resonances.
In such cases, classical and quantum control can be wildly different
and may not necessarily correspond to the same physical phenomenon.
Such a conclusion has been reached in a
particular case of two-color strong-field photodissociation
of a diatomic molecule modeled as a Morse oscillator
\cite{Constantoudis:2005qo,Lima:2008vn}.
Away from the resonances, however,
the classical and quantum calculations have been found to agree
and further to reproduce the control achieved in
a related photoionization experiment
\cite{Sirko:2002ye,Ehlotzky:2001ij}.

Notwithstanding,
the case of the quartic oscillator discussed above
illustrates that quantum-classical correspondence is
not restricted to weak-field coherent control
(\ie{}, the perturbative limit),
but can also occur for intense driving.
The results make clear that a general approach
to the role of quantum mechanics in control is necessary.

\subsection{The Nonlinear Response Perspective}
\label{sec:introduction:nonlinear_response}

Such an approach to analyzing quantum control scenarios
and their possible classical analog,
formulates the issue in terms of nonlinear perturbative response
\cite{Franco:2010kx,Ivanov:2012aa}.
Non-linear response
provides an infinite series expansion
for the time-dependent ensemble average of
the outcome of a measurement performed on a material system
that is subject to an external
(and here for simplicity, scalar-valued)
driving signal $E(t)$.
If $O(t)$ is the response signal,
then \cite{Schetzen:2006ab}
\begin{multline}
  O(t) = R^{(0)} \\
  + \sum_{n = 1}^{\infty} \int_{\mathrlap{-\infty}}^{\mathrlap{t}} \mathrm{d}\tau_{n} \int_{\mathrlap{-\infty}}^{\mathrlap{\tau_{n}}} \mathrm{d}\tau_{n-1} \cdots \int_{\mathrlap{-\infty}}^{\mathrlap{\tau_{2}}} \mathrm{d}\tau_{1} \, R^{(n)}(\tau_{2} - \tau_{1}, \tau_{3} - \tau_{2}, \ldots, t - \tau_{n}) \\ \times E(\tau_{n}) E(\tau_{n-1}) \cdots E(\tau_{1}).
  \label{eq:volterra}
\end{multline}
If all response functions $R^{(n)}$ are known,
$O(t)$ can be obtained to arbitrary precision
for virtually any $E(t)$.
Eq.~\eqref{eq:volterra} is completely general
and applicable in both quantum and classical mechanics,
where the response functions $R^{(n)}$ are obtainable
from time-dependent perturbation theory
in the coupling between $E(t)$ and the material system.

Perturbation theory also suggests criteria for truncating
the series [Eq.~\eqref{eq:volterra}].
For weak two-color phase control,
the input signal $E(t)$ is a small parameter
and, typically, only low orders $n$ are relevant.
Further, in this case, the signal separates into two components,
$E = E_{1} + E_{2}$, according to their frequencies.
Hence, by suppressing one component,
the contribution
of each input component to the response $O$ can be assessed.
If $O_{1}$ and $O_{2}$ are the individual responses
to $E_{1}$ and $E_{2}$,
then, in general, if the response is
nonlinear in the field amplitude, the collective response $O$
will differ from the sum $O_{1} + O_{2}$.
With $\hat{R}^{(n)}$, $\hat{E}_{1}$ and $\hat{E}_{2}$
as the Fourier transforms of the response functions
and the field components,
the difference
\begin{multline}
  O\left(t\right) - O_{1}\left(t\right) - O_{2}\left(t\right) = \int \mathrm{d}\omega_{1} \int \mathrm{d}\omega_{2} \; \ee^{\ii \left(\omega_{1} + \omega_{2}\right) t} \, \hat{R}^{(2)}\left(\omega_{1}, \omega_{1} + \omega_{2}\right) \\
  \times \left[\hat{E}_{2}\left(\omega_{1}\right) \hat{E}_{1}\left(\omega_{2}\right) + \hat{E}_{1}\left(\omega_{1}\right) \hat{E}_{2}\left(\omega_{2}\right)\right] \\
  + \int \mathrm{d}\omega_{1} \int \mathrm{d}\omega_{2} \int \mathrm{d}\omega_{3} \; \ee^{\ii \left(\omega_{1} + \omega_{2} + \omega_{3}\right) t} \, \hat{R}^{(3)}\left(\omega_{1}, \omega_{1} + \omega_{2}, \omega_{1} + \omega_{2} + \omega_{3}\right) \\ \vphantom{\int} \times \left[\hat{E}_{2}\left(\omega_{1}\right) \hat{E}_{1}\left(\omega_{2}\right) \hat{E}_{1}\left(\omega_{3}\right) + \text{5 permutations}\right] + \text{higher orders (neglected)}
  \label{eq:volterra:crossed_contribution}
\end{multline}
quantifies the contribution to the response
due to the simultaneous presence of both fields.

The primary control parameter
in two-color control scenarios is the relative phase between two frequencies.
The direct terms $O_{1}$ and $O_{2}$
are typically not phase controllable
\cite{Spanner:2010ad} and the only way a phase shift $\phi_{1}$
of, \eg{}, the component $E_{1}$ can affect the response
is by means of the mixed cross terms given by
Eq.~\eqref{eq:volterra:crossed_contribution}.
Consider, specifically, the effect of the phase change:
\begin{align}
  \hat{E}_{i}\left(\omega\right) & \mapsto \hat{E}_{i}\left(\omega\right) \ee^{\ii \sgn\left(\omega\right) \phi_{i}}
  \label{eq:Ei_substitution}
\end{align}
The right-hand side of Eq.~\eqref{eq:Ei_substitution}
is the Fourier transform of
$a_{i}\left(t\right) \cos\left[\varphi_{i}\left(t\right) + \phi_{i}\right]$,
if $\hat{E}_{i}\left(\omega\right)$
is the Fourier transform of
$E_{i}\left(t\right) = a_{i}\left(t\right) \cos\left[\varphi_{i}\left(t\right)\right]$.
The functions $a_{i}$ and $\varphi_{i}$ are, respectively,
the instantaneous amplitude
and phase of the field $E_{i}$
before shifting its phase by $\phi_{i}$.
Equation \eqref{eq:volterra:crossed_contribution}
shows explicitly that
the phases $\phi_{i}$ affect the
field terms, but not the response functions,
which depend on the field frequencies, but not
on the field phases $\phi_{i}$.
This feature is common to both
the classical and the quantum descriptions
of conventional two-color weak-field control,
and generalizes the case of
the quartic oscillator \cite{Franco:2006rq} reviewed in
Sec.~\ref{sec:introduction:review_quantumclassical_correspondence}.
This approach shows
that the difference between
the classical and quantum $1$ \vs{} $N$ response
can only result from quantitative differences in
the response functions obtained in either regime.
Phenomena like
quantum resonances, tunneling, broken parity, \etc{}
only contribute to
the response functions,
but not to the fundamental structure of phase control, \ie{},
the dependence on the phase of the incident field.

The discussion above would apply to any coherent control scheme
for which the nonlinear response is valid;
those that do not fit into this category
merit additional consideration.

\subsection[Is There Quantum Mechanics in $1$ \vs{} $N$ Coherent Phase Control?]{Is There Quantum Mechanics in $\boldsymbol{1}$ \vs{} $\boldsymbol{N}$ Coherent Phase Control?}
\label{sec:introduction:interference_and_complementarity}

Section \ref{sec:introduction:nonlinear_response},
makes no reference to either ``interference'' or to ``quantum interference''.
That is, neither of these concepts
is necessary to understand the coherent phase control scenario.
Hence, the idea that coherent phase control results from
the quantum-coherent interference of
mutually exclusive excitation pathways
is challenged by the above treatment.
Furthermore, even if quantum interference were involved,
none of its hallmark features, discussed below, would be displayed;
rather one only finds
the nonadditivity and phase sensitivity of
the nonlinear response.
At this trivial level,
quantum interference and the double-slit analogy
are of questionable relevance.

Moreover,
the double-slit analogy can, at times,
even be misleading.
This is demonstrated in the example of
the bichromatically driven quartic oscillator
(see Refs.~\citealp{Franco:2006rq},
\citealp{Franco:2010kx}, and
Sec.~\ref{sec:introduction:review_quantumclassical_correspondence}
above),
a system that admits a description in terms of
quantum-coherently interfering dynamic pathways
that is formally analogous to single-particle interference
in the Young double-slit (YDS).
However, and significantly,
single-particle YDS interference does not survive in the classical limit;
rather, one obtains an average over
ballistic particle trajectories
that alternate probabilistically between the two slits.
This is not the case for
the $1$ \vs{} $2$ control process,
which retains its wavelike interference behavior in the classical limit.
Indeed, as we have explained elsewhere \cite{Franco:2010kx},
the primary difference between the YDS and coherent phase control is
that the latter occurs due to \emph{external driving},
\ie{}, $E(t)$, whereas the former requires no such external influence.
This result is in full agreement with the
nonlinear response description in
Sec.~\ref{sec:introduction:nonlinear_response},
but it is at odds with the double-slit analogy.

We have to conclude that,
for standard scenarios of $1$ \vs{} $2$ coherent phase control,
the double-slit analogy should be avoided,
and $1$ \vs{} $2$ phase control should not be advertised conceptually 
as a quantum interference effect.
The result motivates our efforts below,
in which we seek to develop
and test a genuinely and verifiable quantum
mechanical phase control scenario.

\section[The Coherent Control Interferometer (CCI)]{The Coherent Control Interferometer}
\label{sec:CCI}

To address this issue,
we focus on the quantum mechanical principle of complementarity,
and its role in coherent control.
This principle addresses a fundamental qualitative property
of quantum mechanics:
the violation of realism.
Realism is central to classical physics,
stating that the properties and configurations of classical systems are
independent of their observation,
and exist prior to their measurement.
In quantum physics, however, this concept is no longer expected.
Rather,
quantum mechanics describes systems
and measurements probabilistically.
A globally consistent classical picture of
a quantum system
is impossible because
sequential measurements give evidence for properties that
classically would be regarded as contradictory,
but are in fact equally valid.
One therefore cannot regard a system as having
``intrinsic" properties, \ie{}, properties that
exist independent of measurement.
Rather, since Bohr, it is known that
the true nature of a quantum phenomenon
can only be understood by obtaining evidence
through complementary measurements.
In practical terms, this means that
if some measurement is absolutely predictable
and its \emph{a priori} outcome certain,
then there is a complementary measurement
whose outcome will be completely uncertain
\cite{Scully:1991kl}.

In the context of interference,
Bohr's principle refers to the complementarity
of waves and particles.
According to the laws of classical physics,
these are intrinsic properties:
particles never interfere in an interferometer,
and, for waves, passage through slits is not mutually exclusive.
In quantum mechanics, however,
the notions of wave and particle exist only
as concepts that are borrowed from classical physics.
The chances of
a system displaying wave- or particle-like behavior
are intrinsic to the mathematical framework
of quantum superpositions
that, formally, ascribes these properties
to a quantum system simultaneously.
It is the choice of the experiment that determines what property
the system displays.
These properties are thus not intrinsic:
it is not just the ignorance of the measurement that
prevents an observer from knowing if something
is a wave or a particle.

The focus of the remaining part of this chapter discusses ways
to demonstrate complementarity
in new, unconventional scenarios of
true quantum coherent phase control,
scenarios in which nonlinear response theory
cannot be applied.
In contrast to previous coherent phase control experiments,
the scenarios proposed below strongly support an analogy between
coherent phase control
and the single-particle interference
in the Young double-slit.
Further, the central hallmarks of quantum interference---%
complementarity and the violation of realism in particular---%
are displayed prominently and unambiguously.
Essential to this approach is the framework of
the CCI
that is introduced in this section.
This CCI is inspired by the
optical MZI,
a standard two-path interferometer
that, like the Young double-slit (YDS),
can demonstrate complementarity
and the probabilistic nature of
quantum mechanics. 
Introducing the CCI will, however,
necessitate the introduction of a general structure
for the phase control scenario under consideration.

\subsection{A General Theory of Two-Color Weak-Field Coherent Phase Control}
\label{sec:cci:coherent_control_theory}

Here we introduce a fully general
quantum-mechanical description
of the two-color weak-field
$1$ \vs{} $N$ coherent control scheme,
where $N = 2$, $3$, \etc.
The treatment allows for realistic pulse profiles
that are compact in space and
thus have a spread in frequency.
Furthermore,
the full quantum treatment
accounts for the radiation field as
an additional quantum degree of freedom $\mathcal{R}$
\cite{Cohen-Tannoudji:1998kl}
that can be in a nonclassical state
and that can exhibit correlations with
the material degrees of freedom $\mathcal{M}$.
Here, the field will be subject to feedback and
change due to its interaction with the matter,
an extension of the coherent control approach described above,
where the driving field is external and predetermined.

Consider now a material system $\mathcal{M}$
upon excitation by light.
The material Hamiltonian $H_{\mathcal{M}}$
is assumed to be time-independent with
a bound and continuum spectrum $E_{\mu}$
with eigenstates $\left\{\ket{\mu}\right\}$.
That is,
\begin{align}
  H_{\mathcal{M}} \ket{\mu} = E_{\mu} \ket{\mu}.
\end{align}
Initially, in the distant past,
the material system is its ground state,
\begin{align}
  \varrho_{\mathcal{M}}\left(-\infty\right) & = \smashoperator{\sum_{\nu \in \mathcal{D}_{-}}} p_{\nu} \ket*{\nu} \bra*{\nu}
  \label{eq:varrhoM}
\end{align}
where $\sum_{\nu \in \mathcal{D}_{-}} p_{\nu} = 1$ with
$p_{\nu} \ge 0$, and the quantum numbers
$\nu$ from the set $\mathcal{D}_{-}$
count over any ground state degeneracies.

The material system
then interacts with the quantized radiation field,
which is described
via continuous multimode Fock space,
where the vacuum $\ket{\mathrm{vac}}$
and all $n$-photon states of the form
$\ket*{\vec{k}_{1} \lambda_{1}, \ldots, \vec{k}_{n} \lambda_{n}}$
are the preferred basis states.
Here the indices $\lambda_{i} = 1, 2$ label
two unit polarization vectors
$\uvec{\varepsilon}\left(\vec{k}_{i}, \lambda_{i}\right)$
that are each perpendicular to the corresponding wave vector
$\vec{k}_{i} = \omega\left(k_{i}\right) \, \uvec{k}_{i} / c$ and to one another.
The field Hamiltonian is
\begin{align}
  H_{\mathcal{R}} & = \int \mathrm{d}\vec{k} \smashoperator{\sum_{\lambda}} \, \hbar \omega\left(k\right) \, a^{\dagger}\left(\vec{k}, \lambda\right) a\left(\vec{k}, \lambda\right) ,
\end{align}
where the vacuum radiation field energy is ignored,
and $a^{\dagger}\left(\vec{k}, \lambda\right)$
and $a\left(\vec{k}, \lambda\right)$
are the photon creation,
\begin{subequations}
  \begin{gather}
    a^{\dagger}\left(\vec{k}, \lambda\right) \ket*{\mathrm{vac}} = \ket*{\vec{k} \lambda}, \\
    a^{\dagger}\left(\vec{k}, \lambda\right) \ket*{\vec{k}_{1} \lambda_{1}, \ldots, \vec{k}_{n} \lambda_{n}} = \sqrt{n+1} \, \ket*{\vec{k} \lambda, \vec{k}_{1} \lambda_{1}, \ldots, \vec{k}_{n} \lambda_{n}},
  \end{gather}
  and annihilation operators,
  \begin{multline}
    a\left(\vec{k}, \lambda\right) \ket*{\vec{k}_{1} \lambda_{1}, \ldots, \vec{k}_{n} \lambda_{n}} = \frac{1}{\sqrt{n}} \smashoperator{\sum_{i=1}^{n}} \delta\left(\vec{k} - \vec{k}_{i}\right) \delta_{\lambda \lambda_{i}} \\
    \times \ket*{\ldots, \vec{k}_{i-1} \lambda_{i-1}, \vec{k}_{i+1} \lambda_{i+1}, \ldots},
  \end{multline}
  \begin{gather}
    a\left(\vec{k}, \lambda\right) \ket*{\mathrm{vac}} = 0,
  \end{gather}
\end{subequations}
respectively. In this continuous Fock space,
wave vectors can take any value and
we have
\begin{multline}
  \overlap*{\vec{k}'_{1} \lambda'_{1}, \ldots, \vec{k}'_{m} \lambda'_{m}}{\vec{k}^{\vphantom{\prime}}_{1} \lambda^{\vphantom{\prime}}_{1}, \ldots, \vec{k}^{\vphantom{\prime}}_{n} \lambda^{\vphantom{\prime}}_{n}} = \\
  \delta_{n m} \frac{1}{n!} \sum_{\sigma \in S_{n}} \delta\bigl(\vec{k}^{\vphantom{\prime}}_{\vphantom{\sigma()}1} - \vec{k}'_{\sigma(1)}\bigr) \, \delta_{\lambda^{\vphantom{\prime}}_{\vphantom{\sigma()}1} \lambda'_{\sigma(1)}} \cdots \delta\bigl(\vec{k}^{\vphantom{\prime}}_{\vphantom{\sigma()}n} - \vec{k}'_{\sigma(n)}\bigr) \, \delta_{\lambda^{\vphantom{\prime}}_{\vphantom{\sigma()}n} \lambda'_{\sigma(n)}},
\end{multline}
where the sum runs over all $n!$ permutations $\sigma$
in the symmetric group $S_{n}$.
The creation and annihilation operators obey the standard commutation relations
\cite{Cohen-Tannoudji:1998kl}, \ie{},
\begin{gather}
  \bigl[a\left(\vec{k}, \lambda\right), a\left(\vec{k}', \lambda'\right)\bigr] = \bigl[a^{\dagger}\left(\vec{k}, \lambda\right), a^{\dagger}\left(\vec{k}', \lambda'\right)\bigr] = 0,
  \shortintertext{and}
  \bigl[a\left(\vec{k}, \lambda\right), a^{\dagger}\left(\vec{k}', \lambda'\right)\bigr] = \delta\left(\vec{k} - \vec{k}'\right) \, \delta_{\lambda \lambda'}.
\end{gather}

The total Hamiltonian is
\begin{align}
  H & = H_{\mathcal{M}} + H_{\mathcal{R}} + V,
\end{align}
with the light-matter interaction $V$,
treated here within the electric dipole approximation.
In the length gauge, and
when the material part is modeled
as a point scatterer, the interaction is given by
\begin{align}
  V & = - \vec{d} \cdot \int \mathrm{d}\vec{k} \smashoperator{\sum_{\lambda}} \, \ii \hbar \mathcal{E}\left(k\right) \, \left[a\left(\vec{k}, \lambda\right) \uvec{\varepsilon}\left(\vec{k}, \lambda\right) - \text{h.c.}\right],
  \label{eq:quantum_light-matter_interaction}
\end{align}
where $\vec{d}$ is the material dipole operator,
$\mathcal{E}\left(k\right) = \sqrt{\hbar \omega\left(k\right) / \epsilon_{0} \left(2\pi\right){}^{3}} / \hbar$
is the vacuum field strength,
and $\hc$ denotes the Hermitian conjugate of the preceeding expression.

At time $t = -\infty$,
the total system with density matrix $\rho(-\infty)$ assumes an
uncorrelated, separable state,
\begin{align}
  \varrho\left(-\infty\right) & = \varrho_{\mathcal{M}}\left(-\infty\right) \otimes \varrho_{\mathcal{R}}\left(-\infty\right),
  \label{eq:varrhominusinfty}
\end{align}
where $\mathcal{M}$ and $\mathcal{R}$ subscripts
denote the material and radiative subsystems.
Subsequently, the light and the matter interact, and
finally ($t=\infty$) totally separate from one another.
The resultant final state $\varrho\left(\infty\right)$
of the total system
will in general be correlated and inseparable,
$\varrho\left(\infty\right) \neq \varrho_{\mathcal{M}}\left(\infty\right) \otimes \varrho_{\mathcal{R}}\left(\infty\right)$.
Interest is then in a small fraction of the final states,
denoted $\tilde{\varrho}\left(\infty\right)$
that culminate with matter in
specific states $\ket{\mu}$ at particular energies $E_{\mu}$,
with $\mu \in \mathcal{D}_{+}$:
\begin{align}
 \varrho\left(\infty\right) & \mapsto \tilde{\varrho}\left(\infty\right).
\end{align}
[See also Eq.~\eqref{eq:tilde_varrho_infty:initial_definition}
in App.~\ref{app:derivation_total_density_operator}.]
Here $\mathcal{D}_{+}$ denotes the set of quantum numbers
that characterizes all interesting scattering events.
For example, in the case of molecular photodissociation,
a scattering event is interesting
if enough energy is absorbed
from the field to drive the material to the continuum.
The set $\mathcal{D}_{+}$
and the state $\tilde{\varrho}\left(\infty\right)$
are then associated with that particular energy range.
We assume that $\mathcal{D}_{-}$ and $\mathcal{D}_{+}$ do not overlap.

Scattering into $\ket{\mu}$
requires fields with frequencies
whose integer multiples are equal to
$\Delta_{\mu \nu} / \hbar = (E_{\mu} - E_{\nu}) / \hbar$,
where $E_{\nu}$ with $\nu \in \mathcal{D}_{-}$
refers to the ground state energy.
We focus attention on the case where the energy range of
$\Delta_{\mu \mu'}$
is sufficiently narrow so that
the resonance conditions are approximately the same
for all states associated
with $\mathcal{D}_{+}$,
greatly reducing the number of
possible optical excitation routes
from $\mathcal{D}_{-}$
to $\mathcal{D}_{+}$.

In the $1$ \vs{} $N$ control scheme,
one irradiates the system with two coherent wavepackets
that are well monochromatized,
have nonoverlapping frequency spectra,
and are initially localized
far away from the material system.
To account for these features,
we use multimode creation operators $a^{\dagger}[f]$
to describe the initial state
$\varrho_{\mathcal{R}}\left(-\infty\right)$
of the radiation field.
In particular,
we focus on pure states of light
and define
$\varrho_{\mathcal{R}}\left(-\infty\right) = \ket*{\chi} \bra*{\chi}$
with
\begin{align}
  \ket*{\chi} & = g_{1}\bigl(a^{\dagger}\bigl[\hat{f}_{1}\bigr]\bigr) \, g_{2}\bigl(a^{\dagger}\bigl[\hat{f}_{2}\bigr]\bigr) \, \ket*{\mathrm{vac}} \, .
  \label{eq:chi}
\end{align}
Here $g_{1}$, $g_{2}$ are two smooth functions
that characterize the photon statistics of the wavepackets.
The operator $a^{\dagger}[f]$ differs from
the photon creation operator
$a^{\dagger}\left(\vec{k}, \lambda\right)$
insofar as the photon is not created exactly
with wavevector $\vec{k}$ and polarization $\lambda$,
but with a spread in frequency,
orientation and polarization defined by
a complex-valued photonic amplitude function $f$:
\begin{align}
  a^{\dagger}\left[f\right] & = \textstyle \int \mathrm{d}\vec{k} \sum_{\lambda} f\left(\vec{k}, \lambda\right) \, a^{\dagger}\left(\vec{k}, \lambda\right).
\end{align}
The operator $a^{\dagger}\left[f\right]$
and its hermitian adjoint, $a\left[f\right]$,
allow for an elegant description
of pulses with realistic spectral profiles.
The amplitude functions
$f_{1}$ and $f_{2}$ in Eq.~\eqref{eq:chi} in have disjoint supports,
and thus the multimode operators
$a\left[f_{1}\right]$ and
$a^{\dagger}\left[f_{2}\right]$ commute.
Specifically,
\begin{gather}
  \left[\textstyle a\left[f_{1}\right], a^{\dagger}\left[f_{2}\right]\right] = \sproduct*{f_{1}}{f_{2}} = 0,
\end{gather}
where $\sproduct*{f_{1}}{f_{2}}$ is a scalar product,
\begin{align}
  \sproduct*{f_{1}}{f_{2}} & = \textstyle \int \mathrm{d}\vec{k} \sum_{\lambda} f_{1}^{*}\left(\vec{k}, \lambda\right) \, f_{2}^{\phantom{*}}\left(\vec{k}, \lambda\right).
\end{align}
In Eq.~\eqref{eq:chi},
the $g_{i}$ are functions of $a^{\dagger}\bigl[\hat{f}_{i}\bigr]$,
where $\hat{f}_{i} = f_{i} / \norm*{f_{i}}$ with
$\norm*{f_{i}} = \sqrt{\sproduct*{f_{i}}{f_{i}}}$.
Normalization of $\ket*{\chi}$ requires
\begin{align}
  \sum_{n = 0}^{\infty} \frac{1}{n!} \abs[\big]{g_{i}^{\left(n\right)}\left(0\right)}^{2} = 1,
\end{align}
where $g_{i}^{\left(n\right)}\left(0\right)$
denotes the $n^{\text{th}}$ derivative of $g_{i}\left(x\right)$
evaluated at $x=0$.
The introduction of the functions $g_{i}$ allows for
a solution of the scattering problem
that is universal with respect to the photon statistics
of the initial wavepackets.
For example, to obtain the result for
an initially sharply defined number of photons $n_{i}$,
we set $g_{i}\left(x\right) = x^{n_{i}} / \sqrt{n_{i}!}$.
Alternatively,
for coherent states of light, we choose
$g_{i}\left(x\right) = \exp\bigl(\norm*{f_{i}} x - \norm*{f_{i}}{}^{2} / 2\bigr)$,
where $\norm*{f_{i}}$ is the absolute value of the amplitude
of the coherent state. Analogously, setting
$g_{i}\left(x\right) = \exp\big[-\tanh(\rho_{i}) x^2 / 2\bigr] / \sqrt{\cosh\rho_{i}}$
gives a squeezed vacuum state with
quadrature fluctuations $\exp(-2 \rho_{i})$
and $\exp(2 \rho_{i})$ for arbitrary $\rho_{i}$
\cite{Mandel:1995zt}.

As noted above, we assume that
the energies of the incoming wavepackets
are narrowly clustered
around integer multiples of
$\Delta_{\mu \nu}$
in energy between the material asymptotic in-
($\ket{\nu}$, $\nu \in \mathcal{D}_{-}$)
and the asymptotic out-states
($\ket{\mu}$, $\mu \in \mathcal{D}_{+}$).
In particular,
$f_{1}$ is concentrated around
$\Delta_{\mu \nu} / (N \hbar)$,
whereas $f_{2}$ is concentrated around
$\Delta_{\mu \nu} / \hbar$.
For this scenario,
we show in Appendix \ref{app:derivation_total_density_operator}
that within perturbation theory
the longtime limit of
the total density operator
projected on the sector $\mathcal{D}_{+}$
of interesting scattering outcomes is
\begin{multline}
  \tilde{\varrho}\left(\infty\right) = \frac{1}{\tilde{p}} \smashoperator{\sumint_{\mathcal{D}_{+}}} \mathrm{d}\mu \smashoperator{\sumint_{\mathcal{D}_{+}}} \mathrm{d}\mu' \, \ket*{\mu^{\vphantom{\prime}}} \bra*{\mu'} \, \otimes \, \smashoperator{\sum_{\nu \in \mathcal{D}_{-}}} \, p_{\nu} \\
  \begin{aligned}[b]
    \times \Bigr\{ \, T_{1}\bigl[\hat{f}_{2}\bigr]\left(\mu, \nu\right) \, T_{1}^{*}\bigl[\hat{f}_{2}\bigr]\left(\mu', \nu\right) & \, \ket[\big]{\tilde{\chi}^{1}} \bra[\big]{\tilde{\chi}^{1}} \\
    + T_{1}^{\vphantom{*}}\bigl[\hat{f}_{2}\bigr]\left(\mu, \nu\right) \, T_{N}^{*}\bigl[\hat{f}_{1}\bigr]\left(\mu', \nu\right) & \, \ket[\big]{\tilde{\chi}^{1}} \bra[\big]{\tilde{\chi}^{N}} \\
    + T_{N}^{\vphantom{*}}\bigl[\hat{f}_{1}\bigr]\left(\mu, \nu\right) \, T_{1}^{*}\bigl[\hat{f}_{2}\bigr]\left(\mu', \nu\right) & \, \ket[\big]{\tilde{\chi}^{N}} \bra[\big]{\tilde{\chi}^{1}} \\
    + T_{N}^{\vphantom{*}}\bigl[\hat{f}_{1}\bigr]\left(\mu, \nu\right) \, T_{N}^{*}\bigl[\hat{f}_{1}\bigr]\left(\mu', \nu\right) & \, \ket[\big]{\tilde{\chi}^{N}} \bra[\big]{\tilde{\chi}^{N}} \, \Bigr\},
  \end{aligned}
  \label{eq:tilde_varrho_infty}%
\end{multline}
where
\begin{subequations}%
  \begin{align}%
    \ket[\big]{\tilde{\chi}^{1}} & = g_{1}^{\vphantom{\left(N\right)}}\bigl(a^{\dagger}\bigl[\hat{f}_{1}\bigr]\bigr) \, g_{2}^{\left(1\right)}\bigl(a^{\dagger}\bigl[\hat{f}_{2}\bigr]\bigr) \ket*{\mathrm{vac}}, \\
    \ket[\big]{\tilde{\chi}^{N}} & = g_{1}^{\left(N\right)}\bigl(a^{\dagger}\bigl[\hat{f}_{1}\bigr]\bigr) \, g_{2}^{\vphantom{\left(1\right)}}\bigl(a^{\dagger}\bigl[\hat{f}_{2}\bigr]\bigr) \ket*{\mathrm{vac}}
  \end{align}%
  \label{eqs:tilde_chi}%
\end{subequations}%
are the final (nonnormalized) states of radiation and
\begin{align}
  \tilde{p} & = \textstyle \sumint_{\mathcal{D}_{+}} \mathrm{d}\mu \Tr\left[\left(\ket*{\mu}\bra{\mu} \otimes \operatorone_{\mathcal{R}}\right) \varrho\left(\infty\right)\right]
  \label{eq:tilde_p}
\end{align}
is the probability of any interesting scattering event to occur.
The identity operator $\operatorone_{\mathcal{R}}$ on
the continuous multimode Fock space is
\begin{multline}
  \operatorone_{\mathcal{R}} = \sum_{n = 0}^{\infty} \frac{1}{n!} \int \mathrm{d}\vec{k}_{1} \sum_{\lambda_{1}} \cdots \int \mathrm{d}\vec{k}_{2} \sum_{\lambda_{2}} a^{\dagger}\left(\vec{k}_{1}, \lambda_{1}\right) \cdots a^{\dagger}\left(\vec{k}_{n}, \lambda_{n}\right) \ket*{\mathrm{vac}} \\
  \times \bra*{\mathrm{vac}} a\left(\vec{k}_{1}, \lambda_{1}\right) \cdots a\left(\vec{k}_{n}, \lambda_{n}\right).
\end{multline}
The symbol $\sumint$ denotes summation over all discrete states
and integration over all continuum states
associated with $\mathcal{D}_{+}$.

The $1$- and the $N$-photon transition amplitudes
are shown in App.~\ref{app:derivation_total_density_operator}
to be given by
\begin{subequations}%
  \begin{multline}%
    T_{1}\bigl[\hat{f}_{2}\bigr]\left(\mu, \nu\right) = - 2 \pi \ii \int \mathrm{d}\vec{k} \sum_{\lambda} \delta\left[\textstyle \Delta_{\nu \mu} + \hbar \omega\left(k\right)\right] \, \hat{f}_{2}\left(\vec{k}, \lambda\right) \, \left(-\ii\right) \hbar \mathcal{E}\left(k\right) \\
    \times \bracket{\mu}{\vec{d}}{\nu} \cdot \uvec{\varepsilon}\left(\vec{k}, \lambda\right)
    \label{eq:T1}
  \end{multline}%
  and
  \begin{multline}%
    T_{N}\bigl[\hat{f}_{1}\bigr]\left(\mu, \nu\right) = - 2 \pi \ii \int \mathrm{d}\vec{k}_{1} \sum_{\lambda_{1}} \cdots \int \mathrm{d}\vec{k}_{N} \sum_{\lambda_{N}} \delta\bigl[\textstyle \Delta_{\nu \mu} + \sum_{i = 1}^{N} \hbar \omega\left(k_{i}\right)\bigr] \\ \vphantom{\int} \times \hat{f}_{1}\left(\vec{k}_{1}, \lambda_{1}\right) \, \left(-\ii\right) \hbar \mathcal{E}\left(k_{1}\right) \cdots \hat{f}_{1}\left(\vec{k}_{N}, \lambda_{N}\right) \, \left(-\ii\right) \hbar \mathcal{E}\left(k_{N}\right) \, \smashoperator{\sumint_{\mathcal{D}_{+}}} \mathrm{d}\xi_{1} \cdots \smashoperator{\sumint_{\mathcal{D}_{+}}} \mathrm{d}\xi_{N - 1} \\ \times \frac{\bracket{\mu}{\vec{d}}{\xi_{1}} \cdot \uvec{\varepsilon}\left(\vec{k}_{1}, \lambda_{1}\right) \; \bracket{\xi_{1}}{\vec{d}}{\xi_{2}} \cdot \uvec{\varepsilon}\left(\vec{k}_{2}, \lambda_{2}\right) \; \cdots \; \bracket{\xi_{N - 1}}{\vec{d}}{\nu} \cdot \uvec{\varepsilon}\left(\vec{k}_{N}, \lambda_{N}\right)}{\left(\Delta_{\xi_{2} \xi_{1} \vphantom{\xi_{N - 1}}} + \hbar \omega\left(k_{2}\right) + \ii 0\right) \cdots \left(\Delta_{\nu \xi_{N - 1}} + \hbar \omega\left(k_{N}\right) + \ii 0\right)},
    \label{eq:TN}
  \end{multline}%
\end{subequations}%
respectively, where
the transition amplitudes are
complex-valued functionals of the
photonic amplitude functions $\hat{f}_{i}$.
The quantum resonance condition
is contained within $T_{1}$ and $T_{N}$
through the $\delta$-kernels.
Specifically, for $T_{1}$,
the kernel extracts a single frequency component from
the amplitude function $\hat{f}_{2}$.
For $T_{N}$,
the kernel is nonzero for $\hat{f}_{1}$
sharply peaked on the wave number shell with radius
$\Delta_{\nu \mu} / \hbar c N$.

Note first that
the first and the last term
in the braces, $\{\cdots\}$,
in Eq.~\eqref{eq:tilde_varrho_infty}
describes excitation processes
caused by either of the two incoming wavepackets,
contributions corresponding to
the individual nonlinear responses $O_{1}$ and $O_{2}$
discussed in Sec.~\ref{sec:introduction:nonlinear_response}.
Second,
the terms in the second line of
Eq.~\eqref{eq:tilde_varrho_infty} are ``cross contributions'',
containing both
the single-photon transition amplitude $T_{1}$
and the $N$-photon transition amplitude $T_{N}$.
They are contributions
that contain coherences between the
two possible final states of the radiation field
$\ket{\tilde{\chi}^{1}}$ and $\ket{\tilde{\chi}^{N}}$,
which arise when
the material target state is reached
via either a $1$- or $N$-photon process.
For the focus on coherent control,
these cross terms are the more relevant,
having been interpreted as
occurring due to the \emph{interference} between
the processes described by the ladder contributions.
It is usually argued that these processes interfere
because they simultaneously couple
the same initial and final states
\cite{Shapiro:1988cr,Shapiro:2012ly},
a role analogous to the one played by the joint contributions term
$O - O_{1} - O_{2}$
[\cf{} Eq.~\eqref{eq:volterra:crossed_contribution},
Sec.~\ref{sec:introduction:nonlinear_response}].
Finally, note that Eq.~\eqref{eq:tilde_varrho_infty} can take
the form of an entangled state of the radiation field and the matter,
an issue discussed below.

\subsection{Dichotomous Material Observables}
\label{sec:cci:dichotomous_material_observables}

In general,
the objective is to control population transfer
to a final region $\mathcal{D}_{+}$
so that material eigenstates
with a desired property are populated in preference to
other undesired states.
Here we define a material observable, $M$,
that labels material eigenstates
as either desired (with label $+1$)
or undesired (with label $-1$).
The role of control is then to
maximize the expectation value
$\expval*{M} = \Tr\left[\left(M \otimes \operatorone_{\mathcal{R}}\right) \tilde{\varrho}\left(\infty\right)\right]$
of $M$.
We define $M$ in the following general but implicit way:
\begin{align}
  M & = \sum_{\eta = \pm 1} \eta \; \tfrac{1}{2} \left(\operatorone + \eta M\right) = \sum_{\eta = \pm 1} \eta \, \sumint \mathrm{d}\mu \, \ket*{\eta \mu} \bra*{\eta \mu}.
  \label{eq:M}
\end{align}
The states $\ket*{\eta \mu}$ are defined
in terms of both $M$ and the eigenstates $\ket*{\mu}$
of the material Hamiltonian $H_{\mathcal{M}}$,
\begin{align}
  \ket*{\eta \mu} & = \tfrac{1}{2} \left(\operatorone + \eta M\right) \ket*{\mu} = \tfrac{1}{2} \left(1 + \eta M_{\mu}\right) \ket*{\mu} = \delta_{\eta M_{\mu}} \ket*{\mu}
\end{align}
with $M_{\mu} = \pm 1$ fixed and $\overlap*{\eta \mu}{\eta' \mu'} = \delta_{\eta \eta'} \delta_{\mu \mu'} \delta_{\eta M_{\mu}}$ .
Since $M$ has only two eigenvalues $M_{\mu}= \pm 1$,
it is termed a ``dichotomous'' observable.
In particular, $M$ is sensitive only to
whether a state is desired or undesired;
it remains completely ignorant about
all other properties of the material system.

The above definition of $M$ conveniently assures that
\begin{enumerate*}[label=(\roman{enumi})]
  \item
    $M$ is dichotomous, \ie{}, measuring $M$
    (on a single realization of the system) yields one
    of two possible outcomes: either $+1$ or $-1$;
  \item
    $M$ is diagonal in the basis $\left\{\ket*{\mu}\right\}$
    and thus commutes with the Hamiltonian $H_{\mathcal{M}}$,
    $\bracket*{\mu}{M}{\mu'} = M_{\mu} \delta_{\mu \mu'}$
    with $M_{\mu} = \pm 1$; and
  \item
    $M$ is its own inverse, $M^{2} = \operatorone$.
\end{enumerate*}
Effectively,
in defining $M$,
we assign a pseudospin label,
$M_{\mu} = +1$ or $-1$,
to every material eigenstate $\ket*{\mu}$.
Hence, the pseudospin label
admits a decomposition of the
material Hilbert space $\mathcal{H}_{\mathcal{M}}$
into two subspaces:
the subspace of states that are superpositions
of material eigenstates with label $+1$
and those states that are linear combinations
of eigenstates labelled $-1$, and
$\mathcal{H}_{\mathcal{M}} = \mathcal{H}_{\mathcal{M}, +1} \oplus \mathcal{H}_{\mathcal{M}, -1}$.
In the case of control,
this will divide the product space into
states associated with desired properties,
and those which are not.

The definition of $M$ is more intuitive
when the controlled physical property
can only have one of two possible outcomes.
An example is the coherent control of
a single-electron spin,
whose polarization
against any chosen observation axis
can have only two values,
a case focussed upon in Sec.~\ref{sec:application}.
In the absence of spin-orbit interaction,
the spin observable commutes with
$H_{\mathcal{M}}$,
and the system is
described by a two-dimensional Hilbert space.
Therefore, the direct sum decomposition
into subspaces
$\mathcal{H}_{\mathcal{M}, +1}$,
$\mathcal{H}_{\mathcal{M}, -1}$
of positive and negative spin
is equivalent to a direct product
of the two-dimensional spin Hilbert space
and the Hilbert space of the remainder of
the material system.

Below, we use these dichotomous labels to define both initial
and final states of the scattering problem.

\subsection[The Coherent Control Interferometer for $1$ \vs{} $N$]{The Coherent Control Interferometer for $\boldsymbol{1}$ \vs{} $\boldsymbol{N}$}
\label{sec:cci:definition}

\begin{figure}
  \centering%
  \begingroup%
    \iftikz%
      \tikzexternalenable%
      \input{tikz/mzi_closed.tex}%
    \else%
      \includegraphics{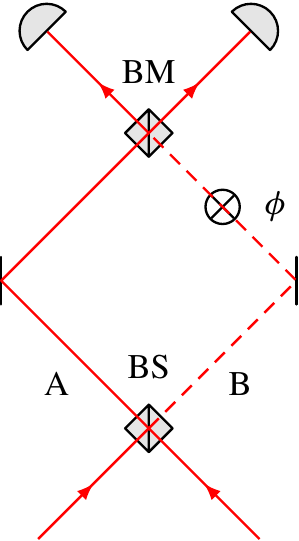}%
    \fi%
  \endgroup%
  \caption{%
    (color online)
    In an optical MZI,
    two incoming beams of light are
    first split at a half-transparent mirror,
    $\mathrm{BS}$,
    into two spatially well-separated daughter beams,
    $\text{A}$ and $\text{B}$,
    then allowed to accumulate a mutual phase shift $\phi$,
    and eventually recombined by another half-transparent mirror,
    $\mathrm{BM}$.
    At the two output ports of the MZI,
    detectors measure the intensity of the outgoing light.
    Observed is an oscillation of the intensity
    with the phase shift $\phi$.
    In units of the background intensity,
    this `fringe pattern' is, ideally,
    equal to $(1 \pm \cos \phi) / 2$.
  }%
  \label{fig:mzi:closed}%
\end{figure}

Given this formalism, we are ready to define
the coherent control interferometer (CCI).
Conceptually, it is borrowed from
the optical MZI, shown schematically in Fig.~\ref{fig:mzi:closed},
and the approach constitutes a significant reformulation of multichannel 
scattering theory \cite{Taylor:1972nx}.
Like an MZI,
the CCI has two input ports,
two output ports,
and two path alternatives.

\begin{figure}
  \centering%
  \begingroup%
    \iftikz%
      \tikzexternalenable%
      \input{tikz/cci_definition.tex}%
    \else%
      \includegraphics{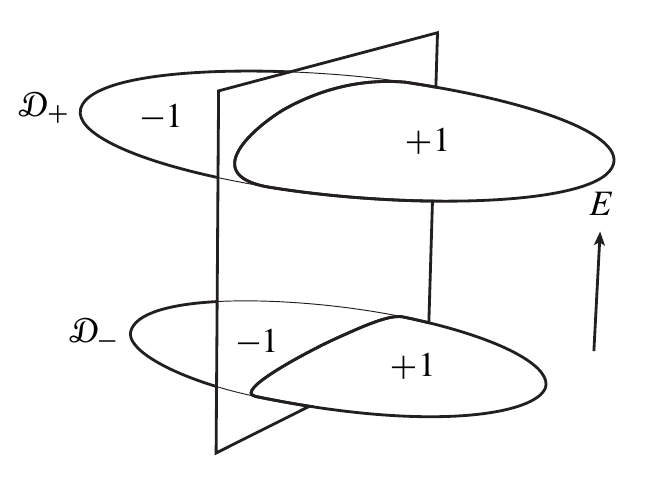}%
    \fi%
  \endgroup%
  \caption{%
    Schematic of the proposed
    coherent control interferometer (CCI).
    Shown are the sets $\mathcal{D}_{-}$
    and $\mathcal{D}_{+}$ as well as the
    CCI's input and output ports.
    The hyperplane separates the regions
    $\mathcal{H}_{\mathcal{M}, +1}$
    and $\mathcal{H}_{\mathcal{M}, -1}$.
  }%
  \label{fig:cci:definition}%
\end{figure}

The dichotomous observable $M$
allows us to directly define the \emph{input ports} of the CCI,
identified with
labels $\pm 1$, \cf{} Fig.~\ref{fig:cci:definition}.
The input port $\sigma$ is said to be occupied if
the material system is prepared
with population in the state $\ket*{\sigma \nu}$.
To formalize this,
we redefine the initial state
$\varrho_{\mathcal{M}}\left(-\infty\right)$
[originally introduced in Eq.~\eqref{eq:varrhoM} on page \pageref{eq:varrhoM}]
as
\begin{align}%
  \varrho_{\mathcal{M}}\left(-\infty\right) & = \sum_{\sigma = \pm 1} \frac{p_{\sigma}}{\sum_{\nu \in \mathcal{D}_{-}} \delta_{M_{\nu} \sigma}} \sum_{\nu \in \mathcal{D}_{-}} \ket*{\sigma \nu} \bra*{\sigma \nu}%
  \label{eq:varrhoM:sigma}%
\end{align}%
with $p_{+1} + p_{-1} = 1$.
The normalization factor in the denominator,
$\sum_{\nu \in \mathcal{D}_{-}} \delta_{M_{\nu} \sigma}$,
counts the number of states $\ket*{\nu}$
with quantum numbers
$\nu \in \mathcal{D}_{-}$
that are labeled with $\sigma$.
If, for instance,
the ground state is spin-$\tfrac{1}{2}$,
and $M$ measures ($2/\hbar$ times the) spin quantum-number,
then that count is $1$ for each of $\sigma = +1$ and $-1$.
Accordingly,
Eq.~\eqref{eq:varrhoM:sigma} reduces
to an incoherent sum of the two
ground state spin orientations.
This leaves us with an initial state $\varrho_{\mathcal{M}}\left(-\infty\right)$
far simpler than our original definition \eqref{eq:varrhoM}.
In cases with a larger count,
$\sum_{\nu \in \mathcal{D}_{-}} \delta_{M_{\nu} \sigma} > 1$,
our new initial state \eqref{eq:varrhoM:sigma}
is still simpler than the original \eqref{eq:varrhoM},
since the populations $p_{\sigma \nu}$ of all states
$\ket{\sigma \nu}$ with $\nu \in \mathcal{D}_{-}$
and fixed $\sigma$ will be the same, \ie{}
\begin{gather}
  p_{\sigma \nu} = \frac{p_{\sigma}}{\sum_{\nu' \in \mathcal{D}_{-}} \delta_{M_{\nu'} \sigma}}
\end{gather}
is independent of $\nu$.

Turning now to the \emph{output ports} of the CCI,
and incorporating the changes made to the initial state
$\varrho_{\mathcal{M}}\left(-\infty\right)$,
the state $\tilde{\varrho}\left(\infty\right)$
is given by
\begin{multline}
  \tilde{\varrho}\left(\infty\right) = \tilde{p}^{-1} \, \smashoperator[l]{\sum_{\eta, \eta' = \pm 1}} \sum_{\vphantom{\eta} \sigma = \pm 1} \frac{p_{\sigma}}{\sum_{\nu \in \mathcal{D}_{-}} \delta_{M_{\nu} \sigma}} \\
  \times \sum_{\nu \in \mathcal{D}_{-}} \left(\ket[\big]{\tilde{\psi}^{1}_{\eta, \sigma \nu}} \ket[\big]{\tilde{\chi}^{1}} + \ket[\big]{\tilde{\psi}^{N}_{\eta, \sigma \nu}} \ket[\big]{\tilde{\chi}^{N}}\right) \, \left(\bra[\big]{\tilde{\psi}^{1}_{\eta', \sigma \nu}} \bra[\big]{\tilde{\chi}^{1}} + \bra[\big]{\tilde{\psi}^{N}_{\eta', \sigma \nu}} \bra[\big]{\tilde{\chi}^{N}}\right),
\end{multline}
where
\begin{subequations}%
  \begin{align}%
    \ket[\big]{\tilde{\psi}^{1}_{\eta, \sigma \nu}} & = \sumint_{\mathcal{D}_{+}} \mathrm{d}\mu \; T_{1}\bigl[\hat{f}_{2}\bigr]\left(\eta \mu, \sigma \nu\right) \ket*{\eta \mu} \\
    & = \sumint_{\mathcal{D}_{+}} \mathrm{d}\mu \; \delta_{M_{\mu} \eta} \, \delta_{M_{\nu} \sigma} \, T_{1}\bigl[\hat{f}_{2}\bigr]\left(\mu, \nu\right) \ket*{\eta \mu}
    \shortintertext{and}
    \ket[\big]{\tilde{\psi}^{N}_{\eta, \sigma \nu}} & = \sumint_{\mathcal{D}_{+}} \mathrm{d}\mu \; T_{N}\bigl[\hat{f}_{1}\bigr]\left(\eta \mu, \sigma \nu\right) \ket*{\eta \mu} \\
    & = \sumint_{\mathcal{D}_{+}} \mathrm{d}\mu \; \delta_{M_{\mu} \eta} \, \delta_{M_{\nu} \sigma} \, T_{N}\bigl[\hat{f}_{1}\bigr]\left(\mu, \nu\right) \ket*{\eta \mu}
  \label{eqs:tilde_psi}%
  \end{align}%
  \end{subequations}%
where both $\ket[\big]{\tilde{\psi}^{1}_{\eta, \sigma \nu}}$
and $\ket[\big]{\tilde{\psi}^{N}_{\eta, \sigma \nu}}$ are not normalized.
Here the subscript $\eta = \pm 1$ denotes the value of $M$ of the final state,
and the $\sigma = \pm 1$ subscript denotes the $M$ value of the initial state $\nu$.
Below, we only refer to
the labeled transition amplitudes
$T_{i}\bigl[\hat{f}_{j}\bigr]\left(\eta \mu, \sigma \nu\right)$
defined in Eq.~\eqref{eqs:tilde_psi}.
The definition of the CCI is completed
by the following identifications.
First, the final labels $\eta = \pm 1$ are the \emph{output ports}
and, second, the population statistics in these product channels
are \emph{interference patterns}.
Specifically,
if we measure the observable $M$ and
sort the measurement events into two subensembles according to
the measurement outcome $\eta$,
then, following our convention,
the interference pattern of the output port $\eta$
is encoded in the expectation value
\begin{multline}
  \expval*{\tfrac{1}{2} \left(\operatorone + \eta M\right)} = \tilde{p}^{-1} \, \sum_{\vphantom{\eta} \sigma = \pm 1} \frac{p_{\sigma}}{\sum_{\nu \in \mathcal{D}_{-}} \delta_{M_{\nu} \sigma}} \sum_{\nu \in \mathcal{D}_{-}} \Bigl(\overlap[\big]{\tilde{\psi}^{1}_{\eta, \sigma \nu}}{\tilde{\psi}^{1}_{\eta, \sigma \nu}} \overlap[\big]{\tilde{\chi}^{1}}{\tilde{\chi}^{1}} \\ + \overlap[\big]{\tilde{\psi}^{1}_{\eta, \sigma \nu}}{\tilde{\psi}^{N}_{\eta, \sigma \nu}} \overlap[\big]{\tilde{\chi}^{1}}{\tilde{\chi}^{N}} + \overlap[\big]{\tilde{\psi}^{N}_{\eta, \sigma \nu}}{\tilde{\psi}^{1}_{\eta, \sigma \nu}} \overlap[\big]{\tilde{\chi}^{N}}{\tilde{\chi}^{1}} + \overlap[\big]{\tilde{\psi}^{N}_{\eta, \sigma \nu}}{\tilde{\psi}^{N}_{\eta, \sigma \nu}} \overlap[\big]{\tilde{\chi}^{N}}{\tilde{\chi}^{N}}\Bigr),
\end{multline}
where
\begin{multline}
  \tilde{p} = \sum_{\eta = \pm 1} \sum_{\vphantom{\eta} \sigma = \pm 1} \frac{p_{\sigma}}{\sum_{\nu \in \mathcal{D}_{-}} \delta_{M_{\nu} \sigma}} \sum_{\nu \in \mathcal{D}_{-}} \Bigl(\overlap[\big]{\tilde{\psi}^{1}_{\eta, \sigma \nu}}{\tilde{\psi}^{1}_{\eta, \sigma \nu}} \overlap[\big]{\tilde{\chi}^{1}}{\tilde{\chi}^{1}} \\ + \overlap[\big]{\tilde{\psi}^{1}_{\eta, \sigma \nu}}{\tilde{\psi}^{N}_{\eta, \sigma \nu}} \overlap[\big]{\tilde{\chi}^{1}}{\tilde{\chi}^{N}} + \overlap[\big]{\tilde{\psi}^{N}_{\eta, \sigma \nu}}{\tilde{\psi}^{1}_{\eta, \sigma \nu}} \overlap[\big]{\tilde{\chi}^{N}}{\tilde{\chi}^{1}} + \overlap[\big]{\tilde{\psi}^{N}_{\eta, \sigma \nu}}{\tilde{\psi}^{N}_{\eta, \sigma \nu}} \overlap[\big]{\tilde{\chi}^{N}}{\tilde{\chi}^{N}}\Bigr)
\end{multline}
is the total probability of detection.
Note that we can only speak of an interference,
or fringe pattern, if this expectation value
is nonadditively dependent on the phase of
the field components.
In turn, non-additivity and phase-dependency can only originate
from the cross terms, which
 depend on both transition amplitudes
($T_{1}$ and $T_{N}$)
contained in the respective final material states
$\bigl(\ket[\big]{\tilde{\psi}^{1}_{\eta, \sigma \nu}}$
and $\ket[\big]{\tilde{\psi}^{N}_{\eta, \sigma \nu}}\bigr)$,
as well as the overlaps of both final radiation states
$\bigl(\ket[\big]{\tilde{\chi}^{1}}$
and $\ket[\big]{\tilde{\chi}^{N}}\bigr)$.
If these states are perfectly distinguishable, \ie{}
$\overlap[\big]{\tilde{\psi}^{1}_{\eta, \sigma \nu}}{\tilde{\psi}^{N}_{\eta, \sigma \nu}}$
or $\overlap[\big]{\tilde{\chi}^{1}}{\tilde{\chi}^{N}} = 0$,
the cross terms vanish and the
output statistics are additive and phase-independent.
If, on the contrary,
these overlaps are both finite, then the coherences
$\overlap[\big]{\tilde{\psi}^{1}_{\eta, \sigma \nu}}{\tilde{\psi}^{N}_{\eta, \sigma \nu}}$
and
$\overlap[\big]{\tilde{\chi}^{1}}{\tilde{\chi}^{N}}$
impart the phase difference of the
excitation pathways on the expectation value
$\expval*{\tfrac{1}{2} \left(\operatorone + \eta M\right)}$.
In the CCI, this phenomenon is
akin to interference and is synonymous
with phase control.
These observations will be useful in the next section,
where we define
the particle and the wave property
in the context of the CCI,
a prerequisite for studying
wave-particle complementarity.

\section{Path Distinguishability}
\label{sec:path_distinguishability}

Central to quantum interference is the
complementarity of particles and waves.
In both, the Young double-slit (YDS)
experiment and the MZI,
the distinction between a wave and a particle
is intuitive and elementary.
Moreover, in these standard devices,
interference is tangible in that
it is macroscopic and occurs in free space.
By contrast, it is by no means obvious how
waves and particles manifest in the
abstract context of
the CCI.
In order to show that
wave- and particle-like behavior
can be demonstrated in the CCI, as
in a YDS or an MZI,
it is necessary to clarify the notion of each such behavior.
To this end we consider two particular
CCI configurations,
the \emph{closed} and the \emph{open configuration},
that allow an explicit identification of
wave- and particle-like features.

\subsection[The Open Interferometer Configuration]{The Open Interferometer Configuration: Particle-like Statistics}
\label{sec:path_distinguishability:open_config}

\begin{figure}
  \centering%
  \begingroup%
    \iftikz%
      \tikzexternalenable%
      \input{tikz/mzi_open.tex}%
    \else%
      \includegraphics{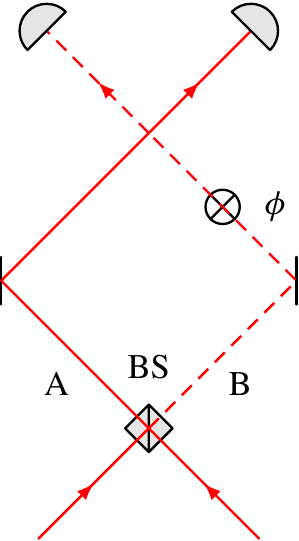}%
    \fi%
  \endgroup%
  \caption{%
    (color online)
    Mach-Zehnder interferometer in the open configuration
    with missing beam merger.
    The beams are mapped one-to-one
    into the detector ports.
    No interference pattern is measured.
  }%
  \label{fig:mzi:open}%
\end{figure}

There exist versions of
every two-path interference experiment
in which it is possible to obtain ``welcher-weg''
(\ie{}, which-way or which-path)
information about the object in the interferometer.
By definition,
complete welcher-weg information (WWI) allows us to unambiguously
predict or trace the path of the object
from input to output.
For example,
in the single-photon YDS experiment,
it is possible to determine which slit a photon
has passed through by replacing the screen with a pair
of telescopes that focus on each slit
\cite{Wheeler:1978pd}.
Similarly,
in a YDS experiment with single electrons,
WWI is obtainable by placing
the electron counter close to one of the slits
\cite{Feynman:1967uq}.
Further, WWI is trivially available
for an MZI with the beam merger removed
\cite{Ionicioiu:2011fk}
as depicted in Fig.~\ref{fig:mzi:open}.
In these alternative experimental arrangements,
interference is absent.
Rather, the welcher-weg measurement assigns classical alternatives
to the object in the interferometer
and the resultant classical alternatives do not interfere.
Interference occurs only if the path of the object through the interferometer
cannot be fully inferred, even in principle.

\begin{figure}
  \centering%
  \begingroup%
    \iftikz%
      \tikzexternalenable%
      \input{tikz/pattern_open.tex}%
    \else%
      \includegraphics{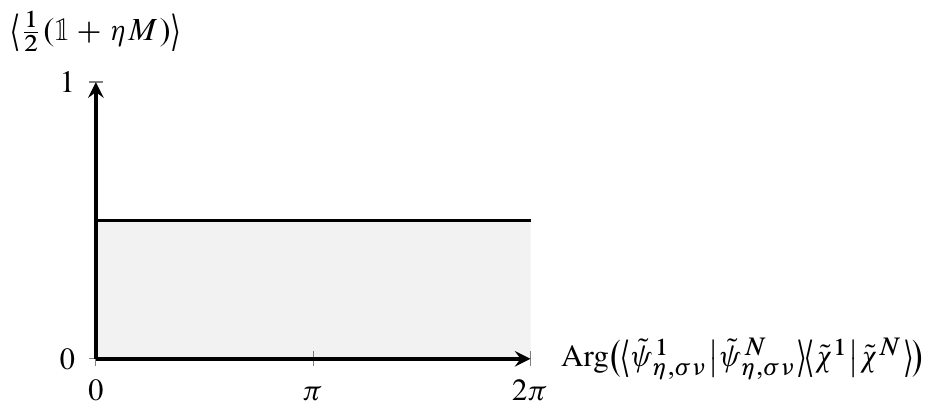}%
    \fi%
  \endgroup%
  \caption{%
    Sketch of the output port populations
    $\expval[\big]{\tfrac{1}{2} \left(\operatorone + \eta M\right)}$,
    $\eta = \pm 1$,
    of the coherent control interferometer (CCI)
    in the open configuration $o$.
    Interference is absent.
    An identical pattern is produced by the
    open Mach-Zehnder interferometer (MZI) shown in
    Fig.~\ref{fig:mzi:open}. 
  }%
  \label{fig:pattern:open}%
\end{figure}

According to the definition of the CCI,
interference fringes are manifest
in the phase sensitivity of the population statistics
of material eigenstates $\ket*{\eta \mu}$
that carry
a particular label, $\eta = M_{\mu} = \pm 1$.
By convention,
the statistical fringe pattern
emerges from counting the fraction of
measurement events with outcome $+1$ or $-1$
in many repeated measurements, and by varying
the phase relationship of the
two incoming wavepackets.
The analysis of the preceding
Sec.~\ref{sec:cci:definition}
shows that the fringe pattern can only be observed
if the states
$\ket[\big]{\tilde{\psi}^{1}_{\eta, \sigma \nu}} \ket[\big]{\tilde{\chi}^{1}}$
and
$\ket[\big]{\tilde{\psi}^{N}_{\eta, \sigma \nu}} \ket[\big]{\tilde{\chi}^{N}}$
are not orthogonal.
Individually,
these quantum states describe
the final state of the system
reaching the output port $\eta$
from the input port $\sigma$
in one of the following two
photoexcitation events:
\begin{enumerate}
  \renewcommand{\theenumi}{\Alph{enumi}}%
  \renewcommand{\labelenumi}{\theenumi,}%
  \item
    a first-order transition process
    fed by the second field component,
    $\left(f_{2}, g_{2}\right)$,
    that is concentrated around
    the frequencies $\Delta_{\mu \nu} / \hbar$,
    where $\mu$ is in the domain
    $\mathcal{D}_{+}$
    and $\nu$ in $\mathcal{D}_{-}$, or
    \label{itm:1photon}
  \item
    an $N^{\text{th}}$-order transition
    driven by the first field component,
    $\left(f_{1}, g_{1}\right)$,
    with central frequency approximating
    $\Delta_{\mu \nu} / \hbar N$,
    where $N = 2, 3, \ldots$~.
    \label{itm:Nphoton}
\end{enumerate}
Therefore, according to this quantum description,
distinguishing the final states
$\ket[\big]{\tilde{\psi}^{1}_{\eta, \sigma \nu}} \ket[\big]{\tilde{\chi}^{1}}$
and
$\ket[\big]{\tilde{\psi}^{N}_{\eta, \sigma \nu}} \ket[\big]{\tilde{\chi}^{N}}$
is equivalent to distinguishing
the excitation processes,
the consequence of which is then the absence
of interference, Fig.~\ref{fig:pattern:open}. 
This conclusion is in full agreement with the
`general principle of coherent control'
as originally stated
\cite{Shapiro:2012ly}:
\begin{quote}
  ``The general principle of coherent control is to
  coherently drive a state with phase coherence
  through multiple, coherent,
  \emph{indistinguishable},
  optical excitation routes to the same final state
  in order to allow for the possibility of control.''
  \label{quot:coherent_control_principle}
\end{quote}
However, for this procedure to be rigorously analogous
to the interference between binary paths in a YDS,
it is necessary that these excitation routes
constitute genuine binary path alternatives.
To show that this is the case,
we first have to verify that, \emph{classically},
the excitation processes are
mutually exclusive.
In the case of the YDS, the mutual exclusivity
of a classical particle passing through
the first or the second slit is obvious.
Further, if we took an electron---%
a quantum particle---%
and let it pass through a YDS,
we could first make, and then falsify,
the classical proposition that the electron
is either going through one
or the other slit \cite{Feynman:1967uq}.
In the case of the CCI this is much less clear.
First, the processes \ref{itm:1photon}
and \ref{itm:Nphoton}
do not constitute a \emph{path qubit},
\ie{}, a two-state quantum path degree of freedom
with a mathematical structure
of a spin-$\frac{1}{2}$.
This is different for the YDS
and most other
two-path interferometers in use,
since these have a path qubit
\cite{Englert:1998kl}.
For instance,
in order to model the kinematics of
interference in a YDS,
the path alternatives can be
described by two orthonormal state vectors,
$\ket*{\text{\ref{itm:1photon}}}$
and $\ket*{\text{\ref{itm:Nphoton}}}$,
that symbolize the amplitudes of
slit \ref{itm:1photon} and slit \ref{itm:Nphoton}
\cite{Englert:1996ul,Englert:2000kh}.
However, such a description does not lend itself
immediately to the case of the CCI.
We address this issue in greater detail
in the next section
(Sec.~\ref{sec:path_distinguishability:closed_config}).

\begin{figure}
  \centering%
  \begingroup%
    \iftikz%
      \tikzexternalenable%
      \input{tikz/cci_open.tex}%
    \else%
      \includegraphics{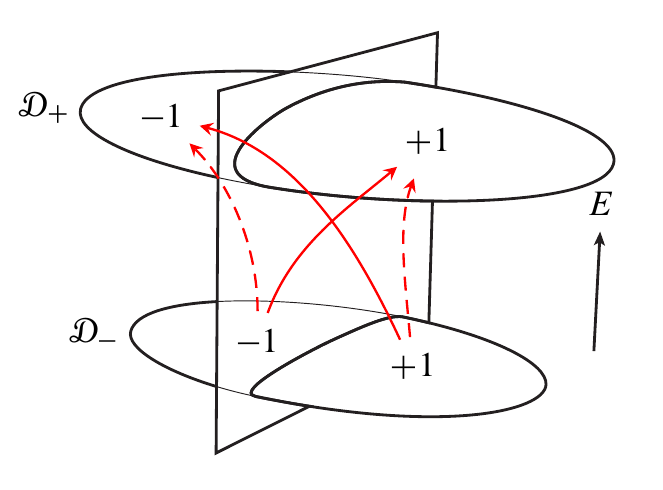}%
    \fi%
  \endgroup%
  \caption{%
    (color online)
    A trivially open CCI.
    Solid: process \ref{itm:1photon}.
    Dashed: process \ref{itm:Nphoton}.
  }%
  \label{fig:cci:open}%
\end{figure}

The second complication of the CCI is
that, due to the finite nature of $\hbar$,
the material system may evolve on paths
for which a classical analog does not exist.
If this is the case, then
there is no classical concept
of what one should expect to occur.
In order to mitigate this latter issue,
we can invoke
the principle of complementarity
that gives meaning to
``classical mutual exclusivity'''.
According to the principle,
every measurement outcome must
be interpreted in classical terms.
Thus, classical mutual exclusivity
means that there has to be a measurement
that distinguishes
the photoexcitation processes
and guarantees that
only one process is executed.
Such measurements indeed exist.
As a trivial example
(non-trivial examples
are discussed in
Sec.~\ref{sec:quantum_erasure})
consider a scenario in which,
starting from the input port $\sigma = +1$,
the output port $\eta = +1$
can only be reached by process \ref{itm:1photon}
and the port $\eta = -1$ only by \ref{itm:Nphoton}
(and vice versa for $\sigma = -1$)
\ie{}
\begin{subequations}%
  \begin{align}%
    T_{1}\bigl[\hat{f}_{2}\bigr]\left(\eta \mu, \sigma \nu\right) & \propto \delta_{\eta, -\sigma}, \\
    T_{N}\bigl[\hat{f}_{1}\bigr]\left(\eta \mu, \sigma \nu\right) & \propto \delta_{\eta, \sigma}
  \end{align}%
  \label{eqs:transition_amplitudes_mutually_exclusive}%
\end{subequations}%
This is the case shown in Fig.~\ref{fig:cci:open}.
It is analogous to the MZI configuration
shown in Fig.~\ref{fig:mzi:open}.
In this highly simplified scenario,
measuring the observable $M$
amounts to discriminating between the transition processes
\ref{itm:1photon} and \ref{itm:Nphoton}.
Moreover, one would always find
that only one of them is chosen,
since
measurements on quantum systems
always have a single outcome.
Classically, it would then
only be conceivable that
the state $\tilde{\varrho}\left(\infty\right)$
is reached by one process at a time,
not by both.
By this paradigm,
\ref{itm:1photon} and \ref{itm:Nphoton}
are indeed mutually exclusive.

Second, we also have to verify the completeness
(\ie{}, there are only these two paths)
as well as the unidirectionality of the processes.
In the cases of the YDS and the MZI,
we have a clear and strict order of events:
there are only two pathways
from injection to detection,
and these are traversed exactly once.
In the CCI,
a violation of unidirectionality
can come about
through higher order effects such as multiphoton processes.
However, these are typically orders of magnitude smaller
than processes
\ref{itm:1photon}
and \ref{itm:Nphoton}.
By neglecting
such higher order effects,
the conditions of completeness and
unidirectionality are met.

The conclusion is that,
from a classical perspective,
processes \ref{itm:1photon}
and \ref{itm:Nphoton} are indeed
binary path alternatives.
Furthermore,
there is only one relevant phase
that occurs between the pathways:
the phase of the coherence
$\overlap[\big]{\tilde{\psi}^{1}_{\eta, \sigma \nu}}{\tilde{\psi}^{N}_{\eta, \sigma \nu}} \overlap[\big]{\tilde{\chi}^{1}}{\tilde{\chi}^{N}}$.
If the
photoexcitation pathways
are fully distinguishable,
then the principle of complementarity dictates
that the controllability of $\eta$
(\ie{} the contrast of the
statistical fringe pattern
in the output port $\eta$)
is zero.
This is the same interpretation
that is used to understand
similar circumstances
with photons in the YDS or an MZI.
If the configuration of the interferometer
is such that it is possible to say
with certainty on which spatial path---%
upper or lower slit, left or right arm---%
the photon wavepacket reaches the detector,
then a fringe pattern does not appear.
Hence, the output port statistics
from repeated measurements would indicate a particle-like
character of the photon.
Indeed, like a particle,
the photon is traveling
only on one path instead of two.
Similarly,
in the absence of interference fringes
in the CCI, we
speak of particle-like statistics
of the output port populations.
That does not imply that
we regard a particle
as being responsible for the statistics;
rather, the conclusion is that the statistics are akin to
those of a classical particle
traversing a YDS.
This is the case since, in a nondeterministic
and probabilistic theory
such as quantum mechanics,
it is only meaningful to base
the notion of a particle
on observed statistics.

Particle-like statistics
can have a number of different causes,
not all of which are profound.
For example,
an interference pattern could simply be
washed out due to imperfections
in the measurement.
It is also easily possible to
falsely identify particle behavior,
since an interference pattern
requires a statistically large number of measurements.
However, of the physical causes of particle-like statistics,
the most fundamental one is
the distinguishability of path alternatives,
the WWI.
In general,
it is enough that WWI
can be acquired \emph{in principle}
for a quantum system to display
particle-like statistics;
it is of no consequence
if this measurement
cannot be done in practice.

We call an interferometer ``open''
when it is possible to trace paths,
and use the symbol $o$ to denote
the open interferometer configuration.

\subsection[The Closed Interferometer Configuration]{The Closed Interferometer Configuration: Wavelike Statistics}
\label{sec:path_distinguishability:closed_config}

\begin{figure}
  \centering%
  \begingroup%
    \iftikz%
      \tikzexternalenable%
      \input{tikz/pattern_closed.tex}%
    \else%
      \includegraphics{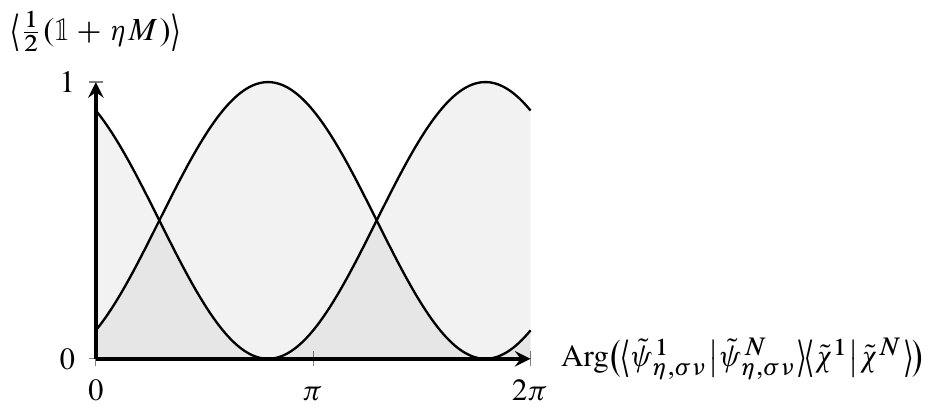}%
    \fi%
  \endgroup%
  \caption{%
    Sketch of the output port populations
    $\expval[\big]{\tfrac{1}{2} \left(\operatorone + \eta M\right)}$,
    $\eta = \pm 1$,
    of the coherent control interferometer (CCI)
    in the closed configuration $c$.
    The interference contrast is maximal
    on both ports.
    An analogous pattern is produced by the
    closed Mach-Zehnder interferometer (MZI) shown in
    Fig.~\ref{fig:mzi:closed}. 
  }%
  \label{fig:pattern:closed}%
\end{figure}

In quantum mechanics,
wave and particle behavior
are classical alternatives
that are realized in
complementary measurements;
they are not intrinsic properties
of a system.
The measurement scheme
that demonstrates the wave property
of a quantum system is provided by
the closed interferometer configuration,
denoted $c$.
It is characterized by
a complete absence of WWI;
it is the standard configuration
of any two-path interferometer,
and it maximizes the amount
of observed interference.
It is also the
the configuration that
adheres most to the
general principle of coherent control,
as stated on page
\pageref{quot:coherent_control_principle}.
That is, in a closed CCI
 it is impossible to identify
the excitation process by which
the output port $\eta$ was reached and,
as such, it displays maximal controllability over $\eta$.

\begin{figure}
  \centering%
  \begingroup%
    \iftikz%
      \tikzexternalenable%
      \input{tikz/cci_closed.tex}%
    \else%
      \includegraphics{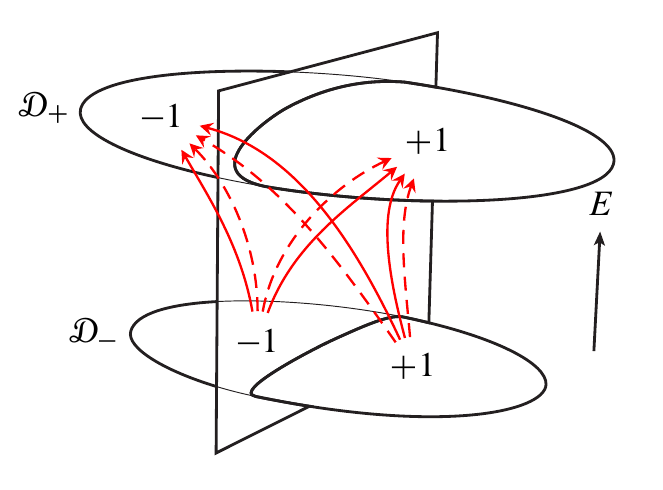}%
    \fi%
  \endgroup%
  \caption{%
    A closed CCI.
    Solid: process \ref{itm:1photon}.
    Dashed: process \ref{itm:Nphoton}.
  }%
  \label{fig:cci:closed}%
\end{figure}

To illustrate,
consider the model case in which the
transition amplitudes
of the processes \ref{itm:1photon}
and \ref{itm:Nphoton}
are each unbiased with respect to the label $\eta$,
that is where
\begin{subequations}%
  \begin{align}%
    T_{1}\bigl[\hat{f}_{2}\bigr]\left(+1 \mu, \sigma \nu\right) & = T_{1}\bigl[\hat{f}_{2}\bigr]\left(-1 \mu, \sigma \nu\right), \\
    T_{N}\bigl[\hat{f}_{1}\bigr]\left(+1 \mu, \sigma \nu\right) & = - T_{N}\bigl[\hat{f}_{1}\bigr]\left(-1 \mu, \sigma \nu\right)
  \end{align}%
  \label{eqs:transition_amplitudes_unbiased_eta}%
\end{subequations}%
for all $\mu \in \mathcal{D}_{+}$,
$\nu \in \mathcal{D}_{-}$,
and $\sigma = \pm 1$.
If Eqs.~\eqref{eqs:transition_amplitudes_unbiased_eta} hold,
then it is straightforward to show
that the final state
$\tilde{\varrho}\left(\infty\right)$
can be written as
\begin{subequations}%
  \begin{multline}%
    \tilde{\varrho}\left(\infty\right) = 2 \, \tilde{p}^{-1} \, \smashoperator[l]{\sum_{\sigma = \pm 1}} \frac{p_{\sigma}}{\sum_{\nu \in \mathcal{D}_{-}} \delta_{M_{\nu} \sigma}} \\
    \begin{aligned}[b]
      \times \sum_{\nu \in \mathcal{D}_{-}} R\left(\tfrac{\pi}{4}\right)^{\dagger} & \left(\ket[\big]{\tilde{\psi}^{1}_{+1, \sigma \nu}} \ket[\big]{\tilde{\chi}^{1}} + \ket[\big]{\tilde{\psi}^{N}_{-1, \sigma \nu}} \ket[\big]{\tilde{\chi}^{N}}\right) \\
      \times & \left(\bra[\big]{\tilde{\psi}^{1}_{+1, \sigma \nu}} \bra[\big]{\tilde{\chi}^{1}} + \bra[\big]{\tilde{\psi}^{N}_{-1, \sigma \nu}} \bra[\big]{\tilde{\chi}^{N}}\right) R\left(\tfrac{\pi}{4}\right)
    \end{aligned}
    \label{eq:tilde_varrho_infty:transition_amplitudes_unbiased_eta}
  \end{multline}%
  with
  \begin{multline}%
    \tilde{p} = 2 \smashoperator[l]{\sum_{\sigma = \pm 1}} \frac{p_{\sigma}}{\sum_{\nu \in \mathcal{D}_{-}} \delta_{M_{\nu} \sigma}} \sum_{\nu \in \mathcal{D}_{-}} \bigl(\overlap[\big]{\tilde{\psi}^{1}_{+1, \sigma \nu}}{\tilde{\psi}^{1}_{+1, \sigma \nu}} \overlap[\big]{\tilde{\chi}^{1}}{\tilde{\chi}^{1}} \\
    + \overlap[\big]{\tilde{\psi}^{N}_{-1, \sigma \nu}}{\tilde{\psi}^{N}_{-1, \sigma \nu}} \overlap[\big]{\tilde{\chi}^{N}}{\tilde{\chi}^{N}}\bigr),
    \label{eq:tilde_p:transition_amplitudes_unbiased_eta}
  \end{multline}%
\end{subequations}%
and where $R\left(\zeta\right)$ denotes
the unitary transformation
\begin{align}
  R\left(\zeta\right) & = \exp\bigl[\textstyle \sumint \mathrm{d}\mu \, \left(\zeta \ket*{+1 \mu}\bra*{-1 \mu} - \zeta^{*} \ket*{-1 \mu}\bra*{+1 \mu}\right)\bigr].
  \label{eq:R_zeta}
\end{align}
Consider, in Eq.~\eqref{eq:tilde_varrho_infty:transition_amplitudes_unbiased_eta},
the terms bracketed by the rotations $R$.
Interestingly,
they take a form analogous to the superposition
$(\ket*{\text{\ref{itm:1photon}}} + \ket*{\text{\ref{itm:Nphoton}}}) / \sqrt{2}$
of \emph{path states}
$\ket*{\text{\ref{itm:1photon}}}$
and $\ket*{\text{\ref{itm:Nphoton}}}$
that corresponds to the intermediate stage
of a two-path interferometer
\cite{Englert:1996ul,Englert:1998kl},
\eg{}, the stage between the beam splitter
and the beam merger in a MZI,
as shown in Fig.~\ref{fig:mzi:closed}.
In fact, from this perspective
the states
$\ket[\big]{\tilde{\psi}^{1}_{+1, \sigma \nu}}$
and $\ket[\big]{\tilde{\psi}^{N}_{-1, \sigma \nu}}$
assume the role of path states for the CCI.
To make the argument clearer, we temporarily assume that
$\ket*{\tilde{\chi}^{1}}$ and $\ket*{\tilde{\chi}^{N}}$
are proportional to one another, \ie{} that
\begin{gather}
  \textstyle \alpha^{-1} \ket*{\tilde{\chi}^{1}} = \beta^{-1} \ket*{\tilde{\chi}^{N}} = \ket*{\tilde{\chi}}
  \label{eq:no_ww_marking}
\end{gather}
with some $\alpha, \beta \in \mathbb{C}$ and
the state $\ket*{\tilde{\chi}}$ for which $\overlap*{\tilde{\chi}}{\tilde{\chi}} = 1$.
We can then say that an intermediate
effect of the light-matter interaction is
to transform the initial state
$\ket*{\sigma \nu} \ket*{\chi}$
(with $\ket*{\chi}$ from Eq.~\eqref{eq:chi}
on page \pageref{eq:chi})
into the superposition
\begin{gather}
  \textstyle \left(\alpha \ket*{\tilde{\psi}^{1}_{+1, \sigma \nu}} + \beta \ket*{\tilde{\psi}^{N}_{-1, \sigma \nu}}\right) \ket*{\tilde{\chi}}.
\end{gather}
This state factorizes into a material and a radiative component.
Here we are particularly interested in the material component.
Looking at the above equation it appears that,
as with common two-path interferometers,
the routes through the CCI
are characterized by
the state of the system
at the intermediate stage,
\cf{} Fig.~\ref{fig:cci:closed}.
In the positively labeled way, characterized by
$\ket[\big]{\tilde{\psi}^{1}_{+1, \sigma \nu}}$,
the excitation happens
through process \ref{itm:1photon},
whereas, in the negatively labeled way, characterized by
$\ket[\big]{\tilde{\psi}^{N}_{-1, \sigma \nu}}$,
the excitation occurs
through process \ref{itm:Nphoton}.
Since
$\overlap[\big]{\tilde{\psi}^{1}_{+1, \sigma \nu}}{\tilde{\psi}^{N}_{-1, \sigma \nu}} = 0$,
these are the binary alternatives
in all CCIs that fulfill
Eqs.~\eqref{eqs:transition_amplitudes_unbiased_eta}.
According to
Eq.~\eqref{eq:tilde_varrho_infty:transition_amplitudes_unbiased_eta},
these alternatives
are subsequently brought into interference
by the rotation $R$---%
that is the effect of the light-matter interaction.
In this view,
this operation plays a role analogous to the
beam merger in a MZI.
In fact, for
\begin{gather}
  \begin{aligned}[b]
    R\left(\tfrac{\pi}{4}\right) = \frac{1}{\sqrt{2}} \sumint \mathrm{d}\mu \, \bigl(\ket*{+1 \mu}\bra*{+1 \mu} & + \ket*{+1 \mu}\bra*{-1 \mu} \\
    - \ket*{-1 \mu}\bra*{+1 \mu} & + \ket*{-1 \mu}\bra*{-1 \mu}\bigr) \, ,
  \end{aligned}
\end{gather}
 the action of the operator
$R\left(\tfrac{\pi}{4}\right)$
is equivalent,
barring phase shifts
introduced by reflections
at the mirror,
to the transformation between
the single-photon input and output modes
of a half-transparent mirror
comprising the MZI's beam merger
(BM in Fig.~\ref{fig:mzi:closed}).

In the perfectly symmetric interferometer [\ie{},
given by Eq.~\eqref{eqs:transition_amplitudes_unbiased_eta} above],
one cannot make an educated guess
about which of the processes,
\ref{itm:1photon} or \ref{itm:Nphoton},
will be realized.
Such a prediction would be
correct in only half of the cases.
By contrast, in asymmetric interferometers,
where there is a bias with respect to the label $\eta$,
the accuracy of a welcher-weg guess
depends on the amount of
\emph{a priori} WWI.
In the following, we assume
that only one CCI input port is populated, \ie{},
we fix $\sigma$ and
set $p_{\sigma} = 1$, $p_{-\sigma} = 0$.
In this case,
\begin{align}
  w_{+1} & = \sumint \mathrm{d}\mu \, \bracket[\big]{+1 \mu}{R\left(\tfrac{\pi}{4}\right) \tilde{\varrho}_{\mathcal{M}}\left(\infty\right) R\left(\tfrac{\pi}{4}\right)^{\dagger}}{\text{$+1 \mu$}}
  \label{eq:wplus1} \\
  & = \frac{\sum_{\nu \in \mathcal{D}_{-}} \abs*{\alpha}^{2} \overlap[\big]{\tilde{\psi}^{1}_{+1, \sigma \nu}}{\tilde{\psi}^{1}_{+1, \sigma \nu}}}{\sum_{\nu \in \mathcal{D}_{-}} \bigl(\abs*{\alpha}^{2} \overlap[\big]{\tilde{\psi}^{1}_{+1, \sigma \nu}}{\tilde{\psi}^{1}_{+1, \sigma \nu}} + \abs*{\beta}^{2} \overlap[\big]{\tilde{\psi}^{N}_{-1, \sigma \nu}}{\tilde{\psi}^{N}_{-1, \sigma \nu}}\bigr)}
\end{align}
is the \emph{a priori} probability
that the CCI executes process \ref{itm:1photon},
and $w_{-1} = 1 - w_{+1}$
is the \emph{a priori} probability of process \ref{itm:Nphoton}.
In Eq.~\eqref{eq:wplus1},
$\tilde{\varrho}_{\mathcal{M}}\left(\infty\right)$
is the reduced density matrix of the material system,
$\tilde{\varrho}_{\mathcal{M}}\left(\infty\right) = \Tr_{\mathcal{R}} \tilde{\varrho}\left(\infty\right)$.
The amount of \emph{a priori} WWI
is given by the
\emph{predictability} $\mathfrak{P}$ of the alternatives:
\begin{align}
  \mathfrak{P} & = \abs*{w_{+1} - w_{-1}}.
  \label{eq:predictability}
\end{align}
It states that, of all educated guesses made,
(at least)
the fraction $(1+\mathfrak{P})/2$ will be correct.
If the CCI is biased completely
towards one excitation process---%
as it is the case in the open configuration discussed above---,
then $\mathfrak{P} = 1$,
and all educated guesses will be $100 \%$ accurate.

Maximum interference contrast requires
a symmetric CCI with $\mathfrak{P} = 0$,
\ie{} a CCI with equally likely \emph{a priori}
path alternatives.
To see this, we compute
the output port statistics,
which reduce to
\begin{align}
  \expval*{\tfrac{1}{2} \left(\operatorone + \eta M\right)} = \frac{1}{2} + \frac{\eta \, \sum_{\nu \in \mathcal{D}_{-}} \Re\bigl[\alpha \beta^{*} \overlap[\big]{\tilde{\psi}^{1}_{+1, \sigma \nu}}{\tilde{\psi}^{N}_{+1, \sigma \nu}}\bigr]}{\sum_{\nu \in \mathcal{D}_{-}} \bigl(\abs*{\alpha}^{2} \overlap[\big]{\tilde{\psi}^{1}_{+1, \sigma \nu}}{\tilde{\psi}^{1}_{+1, \sigma \nu}} + \abs*{\beta}^{2} \overlap[\big]{\tilde{\psi}^{N}_{-1, \sigma \nu}}{\tilde{\psi}^{N}_{-1, \sigma \nu}}\bigr)}.
  \label{eq:expval_eta_M_projector:transition_amplitudes_unbiased_eta}
\end{align}
If we additionally examine the case in which
\begin{subequations}%
  \begin{gather}%
    \abs*{\alpha} \, T_{1}\bigl[\hat{f}_{2}\bigr]\left(+1 \mu, \sigma \nu\right) = \exp\bigl\{- \ii \gamma\bigl[\hat{f}_{1}, \hat{f}_{2}\bigr]\left(+1, \sigma\right)\bigr\} \, \abs*{\beta} \, T_{N}\bigl[\hat{f}_{1}\bigr]\left(+1 \mu, \sigma \nu\right)
    \label{eq:symmetric_conditions:gamma}%
    \shortintertext{and}%
    \alpha \, \abs[\big]{T_{1}\bigl[\hat{f}_{2}\bigr]\left(+1 \mu, \sigma \nu\right)} = \ee^{- \ii \phi} \, \beta \, \abs[\big]{T_{N}\bigl[\hat{f}_{1}\bigr]\left(+1 \mu, \sigma \nu\right)}
    \label{eq:symmetric_conditions:phi}%
  \end{gather}%
  \label{eqs:symmetric_conditions}%
\end{subequations}%
for all $\mu \in \mathcal{D}_{+}$
and $\nu \in \mathcal{D}_{-}$, then
the predictability vanishes, $\mathfrak{P} = 0$, and
Eq.~\eqref{eq:expval_eta_M_projector:transition_amplitudes_unbiased_eta}
becomes
\begin{align}
  \expval*{\tfrac{1}{2} \left(\operatorone + \eta M\right)} & = \tfrac{1}{2} \bigl[1 + \eta \cos\left(\phi + \gamma\right)].
  \label{eq:expval_eta_M_projector:transition_amplitudes_unbiased_eta:symmetric}
\end{align}
This simplification results because,
by assumption,
Eqs.~\eqref{eqs:symmetric_conditions}
hold for all transition amplitudes,
regardless of the values of $\mu$ and $\nu$.
This allows us 
in Eq.~\eqref{eq:expval_eta_M_projector:transition_amplitudes_unbiased_eta:symmetric}
to eliminate
the incoherent sum over $\nu$.

The phases $\gamma$ and $\phi$
in Eq.~\eqref{eq:expval_eta_M_projector:transition_amplitudes_unbiased_eta:symmetric}
are assumed to be independent of both $\mu$ and $\nu$.
Indeed, $\gamma$ and $\phi$ are the primary control parameters of the CCI
and, in the following, are assumed to be arbitrarily variable.

The first phase,
\begin{align}
  \gamma & = \gamma\bigl[\hat{f}_{1}, \hat{f}_{2}\bigr]\left(+1, \sigma\right),
  \label{eq:gamma}
\end{align}
is sensitive to both material and radiative properties.
For example,
if the amplitude functions of the incident radiation
were to gain additional phase factors, \ie{}, if
we applied the map
\begin{align}
  \Phi_{i}\colon \hat{f}_{i} \mapsto \ee^{\ii \varphi_{i}} \hat{f}_{i}
\end{align}
with $-\pi < \varphi_{i} \le \pi$,
then $\gamma$ would transform as
$\gamma \mapsto \gamma + N \varphi_{1} - \varphi_{2}$.

On the other hand,
the second phase, $\phi$, is sensitive only to
the properties of the radiation field.
It characterizes the phase relationship
between the two outgoing radiation states
$\ket*{\tilde{\chi}^{1}}$ and $\ket*{\tilde{\chi}^{N}}$,
\begin{subequations}%
  \begin{align}%
    \phi & = \textstyle \Arg \overlap*{\tilde{\chi}^{1}}{\tilde{\chi}^{N}} \\
    & = \Arg \left\{\sum_{n = 0}^{\infty} \frac{1}{n!} \left[\cramped{g_{1}^{\left(n\right)}}\left(0\right)\right]^{*} g_{1}^{\left(N + n\right)}\left(0\right) \, \sum_{m = 0}^{\infty} \frac{1}{m!} \left[\cramped{g_{2}^{\left(1 + m\right)}}\left(0\right)\right]^{*} g_{2}^{\left(m\right)}\left(0\right)\right\} \\
    & = \textstyle \phi\left[g_{1}, g_{2}\right].
  \end{align}%
  \label{eq:phi}%
\end{subequations}%
Thus $\phi$ can have a value different from
$0$ or $\pi$ only if $g_{1}$ or $g_{2}$ are complex.
For example, for given real $g_{i}$,
we can make the transformation
$g_{i} \mapsto g_{i} \circ \Phi_{i}$.
Then $\phi$ transforms according to
$\phi \mapsto \phi + N \varphi_{1} - \varphi_{2}$.

Let us now go back to
Eq.~\eqref{eq:expval_eta_M_projector:transition_amplitudes_unbiased_eta:symmetric}.
This expectation value describes
interference fringes with maximum contrast,
as can be seen from
the value of the fringe visibility $\mathfrak{V}$,
that, following Michelson's definition \cite{Mandel:1995zt}, is
\begin{align}
  \mathfrak{V} & = \frac{\max_{\gamma, \phi} \expval*{\tfrac{1}{2} \left(\operatorone + \eta M\right)} - \min_{\gamma, \phi} \expval*{\tfrac{1}{2} \left(\operatorone + \eta M\right)}}{\max_{\gamma, \phi} \expval*{\tfrac{1}{2} \left(\operatorone + \eta M\right)} + \min_{\gamma, \phi} \expval*{\tfrac{1}{2} \left(\operatorone + \eta M\right)}} = 1.
\end{align}
For general, unbiased CCIs
[those for which Eqs.~\eqref{eqs:transition_amplitudes_unbiased_eta} hold],
the fringe visibility can also be computed directly
from Eq.~\eqref{eq:tilde_varrho_infty:transition_amplitudes_unbiased_eta}
by using
\begin{align}
  \mathfrak{V} & = 2 \abs*{\sumint \mathrm{d}\mu \, \bracket[\big]{+1 \mu}{R\left(\tfrac{\pi}{4}\right) \tilde{\varrho}_{\mathcal{M}}\left(\infty\right) R\left(\tfrac{\pi}{4}\right)^{\dagger}}{\text{$-1 \mu$}}}.
\end{align}
For any two-path interferometer,
it can further be shown
\cite{Wootters:1979pb,Glauber:1986lh,Greenberger:1988fu,Mandel:1991ye,Jaeger:1995kx}
that the complementarity between
wave and particle behavior
is quantitatively manifest
in the \emph{duality relation}
\begin{align}
  \mathfrak{V}^{2} + \mathfrak{P}^{2} \le 1.
  \label{eq:duality_relation:VP}
\end{align}
That is,
the contrast of the interference pattern
(given by $ \mathfrak{V}$)
is bounded by the amount of WWI
(given by $\mathfrak{P}$)
that is \emph{a priori} available, and
solely due to the
asymmetry of the interferometer.
The inequality also states
that full particle statistics
with absolute welcher-weg knowledge
(as in the open configuration $o$)
as well as wavelike statistics
with truly indistinguishable
path alternatives
(as in the closed configuration $c$)
are idealized cases.
Quantum mechanics
allows a system to display
either one of those ideal properties;
however, the standard case is a blend of the two
that fits into neither classical concept
of wave or particle \cite{Englert:1996ul}.
Hence, as quantified in the duality relation,
it is possible to have limited \emph{a priori} WWI
(increasing the accuracy of welcher-weg predictions)
and simultaneously observe interference
with a reduced contrast.

This concludes the introduction of
the closed CCI configuration,
in which we can fully control
the population of the CCI's output ports.
It is the control scheme
that allows one to demonstrate
wavelike behavior of the output statistics;
they are fully analogous
to the measurement statistics of a classical wave
passing through a YDS.

One comment is in order.
The open and the closed configuration
(and thus the wave and the particle property)
are defined here
with respect to measurements
in the eigenbasis of
the material Hamiltonian $H_{\mathcal{M}}$,
particularly, a measurement of
the dichotomous observable $M$
as specified by Eq.~\eqref{eq:M}.
In general, however,
wave- and particle-like statistics
are basis dependent.
It is easy to show
that there exists another basis
in which the wave- and
the particle-like behavior
are interchanged,
\ie{}, the eigenbasis of the observable
$R\left(\tfrac{\pi}{4}\right)^{\dagger} M R\left(\tfrac{\pi}{4}\right)$.
Notwithstanding,
in the setting of coherent control,
the material eigenbasis
is the preferred basis.

\section{Quantum Erasure Coherent Control}
\label{sec:quantum_erasure}

Our goal is to create scenarios in which
genuine, nontrivial features of quantum interference
are the origin of coherent phase control.
As a first scenario of this kind
we propose a generalized coherent phase control scenario
based on quantum erasure,
in which quantum entanglement and nonclassical correlations play a role.
This provides
a measurement scheme by which an open configuration
can be transformed into a closed configuration in a nontrivial way
in order to probe and verify
the nonclassicality of the scenario.
We follow up on this aspect
in Sec.~\ref{sec:quantum_erasure:relationship_to_quantum_correlations}.

\subsection{Welcher-Weg Marking And Nontrivial CCI Configurations}
\label{sec:quantum_erasure:ww_marking}

The duality relation
$\mathfrak{V}^{2} + \mathfrak{P}^{2} \le 1$
provides a quantitative measure of the fact that
any \emph{a priori} WWI
(as specified by the asymmetry $\mathfrak{P}$ of the CCI)
must cause a loss of interference contrast $\mathfrak{V}$
in the measurement statistics of the output ports.
A second source of WWI
may be found in other degrees of freedom
that are part of the material system or of the radiation field.
We will now deal with them in greater detail---%
with a special focus on the role of the radiation.

The process by which
degrees of freedom acquire WWI
is called ``welcher-weg marking'' (WWM).
The degrees of freedom involved in WWM,
commonly referred to as the
welcher-weg marker, can be understood in terms of
observation and measurement, and
 is linked to an outward flow of information
from the path degree of freedom to the marker.
In general, this flow is mediated by
an entangling interaction,
that is, the monitoring
of the path degree of freedom by the marker
creates an entangled state,
in which the
path degree of freedom and marker
are in a quantum superposition.
This being the case, one cannot ascribe
realistic properties
to \emph{either} of the two parties;
the properties of the first system depend on
the choice of measurement performed on the second
inasmuch as the state of the second system is determined
by the observable being measured on the first.
In the context of two-path interference,
these quantum correlations can result in
the distinguishability of path alternatives,
if they correspond to a recording of WWI.
According to the complementarity principle,
this would lead to
a reduction of interference contrast.

Consider then the role of entanglement and WWM in our agenda.
A coherent control scenario in which
both entanglement and WWM participate
could be properly classified as
quantum coherent control.
Below we utilize the CCI to
develop such a model scenario, starting with
a CCI with
unbiased transition amplitudes
[with respect to the output ports, \cf{}
Eq.~\eqref{eqs:transition_amplitudes_unbiased_eta}]
and populating just one of the two input ports.
That is, we set $p_{\sigma} = 1$
for $\sigma = +1$ or $\sigma = -1$.
The final state of the total system is then
\begin{multline}%
  \tilde{\varrho}\left(\infty\right) = \frac{1}{\sum_{\nu \in \mathcal{D}_{-}} \bigl(\overlap[\big]{\tilde{\psi}^{1}_{+1, \sigma \nu}}{\tilde{\psi}^{1}_{+1, \sigma \nu}} \overlap[\big]{\tilde{\chi}^{1}}{\tilde{\chi}^{1}} + \overlap[\big]{\tilde{\psi}^{N}_{-1, \sigma \nu}}{\tilde{\psi}^{N}_{-1, \sigma \nu}} \overlap[\big]{\tilde{\chi}^{N}}{\tilde{\chi}^{N}}\bigr)} \\
  \times \smashoperator{\sum_{\nu \in \mathcal{D}_{-}}} \; \Bigl[R\left(\tfrac{\pi}{4}\right)^{\dagger} \bigl(\ket[\big]{\tilde{\psi}^{1}_{+1, \sigma \nu}} \ket[\big]{\tilde{\chi}^{1}} + \ket[\big]{\tilde{\psi}^{N}_{-1, \sigma \nu}} \ket[\big]{\tilde{\chi}^{N}}\bigr) \\
  \times \bigl(\bra[\big]{\tilde{\psi}^{1}_{+1, \sigma \nu}} \bra[\big]{\tilde{\chi}^{1}} + \bra[\big]{\tilde{\psi}^{N}_{-1, \sigma \nu}} \bra[\big]{\tilde{\chi}^{N}}\bigr) R\left(\tfrac{\pi}{4}\right)\Bigr],
  \label{eq:tilde_varrho_infty:transition_amplitudes_unbiased_eta:sigma_fixed}
\end{multline}%
where the sums over the quantum number $\nu$
are over the input domain $\mathcal{D}_{-}$.
In this expression,
the path and radiation states
are such
that $\ket[\big]{\tilde{\chi}^{1}}$
correlates with the path labeled $+1$,
and $\ket[\big]{\tilde{\chi}^{N}}$
with the path $-1$.
This situation introduces a complication:
the radiation field is
not only driving the dynamics,
but it can additionally act as
a welcher-weg marker.
In the setting of coherent control,
however, these two roles conflict.
Phase-sensitive control relies on
the indistinguishability of the pathways,
but WWM can inhibit the control
by making the pathways distinguishable.

In conventional coherent control,
we would simply measure $M$ and determine the output port populations
$\expval*{\tfrac{1}{2} \left(\operatorone + \eta M\right)}$
with $\eta = \pm 1$.
Such an experiment would not provide
any insight into the correlated nature
of the paths $\pm 1$ and the radiation,
and hence into this quantum feature.
In what can be called an extended control experiment,
we join the measurement of $M$
with a measurement of a radiation field observable $O_{\mathcal{R}}$.
In this case, correlations can have an impact,
and they may be exploited to
gain such insight.
In fact,
given sufficient control over the field,%
\footnote{We will not concern ourselves
with the very details and
practical experimental requirements of
such control, which are well enough
discussed elsewhere
\cite{Mandel:1995zt,Scully:1997il,Walls:2008tw}.
However, we will discuss a realistic case of control
in Sec.~\ref{sec:quantum_erasure:photon_threshold}.}
two antithetic measurement strategies are conceivable
\cite{Englert:2000kh}:
the ``optimum welcher-weg measurement''
and quantum erasure.
Both are useful for the agenda of developing a fully quantum control scenario,
and both are explained in detail below.
In the first, one tries to make the interferometer
as open as possible
and to extract as much WWI as possible.
Although this minimizes the phase control---%
the strategy is in fact contrary to
the general principle of coherent control
as quoted on page \pageref{quot:coherent_control_principle}---,
the procedure may still be beneficial
since the goal is to introduce quantum correlations.
The second strategy is quantum erasure,
where one attempts to close the interferometer
as much as possible,
which also gives as much phase control as possible.
The logic behind the choice between the two
antithetic strategies is
provided by the nonclassical principle of complementarity.
The existence of that choice is
a distinctive feature of
the extended coherent control scheme.

We discuss these strategies below,
emphasizing once again that
our goal is not to enhance the extent of control.
Indeed, the closed configuration is known
to provide maximum control in this scenario,
\ie{}, maximum sensitivity to the value of the phase.

\subsubsection{Welcher-Weg Measurements}

In accord with
Eq.~\eqref{eq:tilde_varrho_infty:transition_amplitudes_unbiased_eta:sigma_fixed},
WWI may be acquired
via a measurement $O_\mathcal{R}$
on the radiation field $\mathcal{R}$
that can distinguish
$\ket*{\tilde{\chi}^{1}}$ and
$\ket*{\tilde{\chi}^{N}}$---%
provided that they are distinguishable
to some degree.
A measure
of the acquired amount of WWI is
the welcher-weg \emph{knowledge}
$\mathfrak{K}\left(O_{\mathcal{R}}\right)$
which is always greater than or equal to
the predictability [Eq.~\eqref{eq:predictability}]
$\mathfrak{P} = \mathfrak{K}\left(\operatorone_{\mathcal{R}}\right)$.
Significantly,
Ref.~\citealp{Englert:2000kh}
proves that the duality relation
$\mathfrak{V}^{2} + \mathfrak{P}^{2} \le 1$
can be generalized to the duality relation
\begin{align}
  \mathfrak{V}\left(O_{\mathcal{R}}\right)^{2} + \mathfrak{K}\left(O_{\mathcal{R}}\right)^{2} \le 1,
  \label{eq:duality_relation:VORKOR}
\end{align}
where $\mathfrak{V}\left(O_{\mathcal{R}}\right)$
denotes the fringe visibility
in the CCI's output ports
under joint measurement of $O_{\mathcal{R}}$.%
\footnote{Note that, since there cannot be
any measurement $O_{\mathcal{R}}$
that reduces the interference contrast,
we have
$\mathfrak{V}\left(O_{\mathcal{R}}\right) \ge \mathfrak{V}\left(\operatorone_{\mathcal{R}}\right) = \mathfrak{V}$
and, further, an additional duality relation,
$\mathfrak{V}^{2} + \mathfrak{K}\left(O_{\mathcal{R}}\right)^{2} \le 1$,
which is less strong than
Eq.~\eqref{eq:duality_relation:VORKOR},
but stronger than $\mathfrak{V}^{2} + \mathfrak{P}^{2} \le 1$.}
For projective measurements,
$O_{\mathcal{R}}^{2} = O_{\mathcal{R}}$,
$\mathfrak{K}$ is computable from
\begin{align}
  \mathfrak{K}\left(O_{\mathcal{R}}\right) & = \abs[\big]{\expval[\big]{R\left(\tfrac{\pi}{4}\right){}^{\dagger} M R\left(\tfrac{\pi}{4}\right) \, O_{\mathcal{R}}}} + \abs[\big]{\expval[\big]{R\left(\tfrac{\pi}{4}\right){}^{\dagger} M R\left(\tfrac{\pi}{4}\right) \, \left(\operatorone_{\mathcal{R}} - O_{\mathcal{R}}\right)}}.
  \label{eq:knowledge}
\end{align}
The first term quantifies the knowledge
$\mathfrak{K}_{\mathrm{click}}$
obtained from measurement events
with the outcome $1$
(\ie{}, detector clicks),
whereas the second term specifies
the knowledge
$\mathfrak{K}_{\mathrm{no-click}}$
gained from events
with outcome $0$ (no click).
For the state
\eqref{eq:tilde_varrho_infty:transition_amplitudes_unbiased_eta:sigma_fixed},
the knowledge amounts to
\begin{multline}%
  \mathfrak{K}\left(O_{\mathcal{R}}\right) = \abs*{w_{+1} \frac{\bracket{\tilde{\chi}^{1}}{O_{\mathcal{R}}}{\tilde{\chi}^{1}}}{\overlap{\tilde{\chi}^{1}}{\tilde{\chi}^{1}}} - w_{-1} \frac{\bracket{\tilde{\chi}^{N}}{O_{\mathcal{R}}}{\tilde{\chi}^{N}}}{\overlap{\tilde{\chi}^{N}}{\tilde{\chi}^{N}}}} \\ + \abs*{w_{+1} \left(1 - \frac{\bracket{\tilde{\chi}^{1}}{O_{\mathcal{R}}}{\tilde{\chi}^{1}}}{\overlap{\tilde{\chi}^{1}}{\tilde{\chi}^{1}}}\right) - w_{-1} \left(1 - \frac{\bracket{\tilde{\chi}^{N}}{O_{\mathcal{R}}}{\tilde{\chi}^{N}}}{\overlap{\tilde{\chi}^{N}}{\tilde{\chi}^{N}}}\right)}.
  \label{eq:knowledge:tilde_varrho_infty:transition_amplitudes_unbiased_eta:sigma_fixed}%
\end{multline}%
If we want to acquire a maximum of WWI
(in order to realize a maximally open CCI),
we should not, however,
restrict ourselves to projective measurements,
which are, in general, not optimal.
However,
even without explicitly finding the optimal measurement,
we can still calculate the welcher-weg knowledge
it would provide.
Specifically, the highest amount of WWI acquirable
in any measurement of the field
is determined by the radiative \emph{distinguishability}
$\mathfrak{D}_{\mathcal{R}} = \sup_{O_{\mathcal{R}}} \mathfrak{K}\left(O_{\mathcal{R}}\right)$
\cite{Englert:2000kh},
given by \cite{Englert:1998kl}
\begin{multline}
  \mathfrak{D}_{\mathcal{R}} = \Tr\biggl|\sumint \! \mathrm{d}\mu \, \Bigl(\bracket[\big]{+1 \mu}{R\left(\tfrac{\pi}{4}\right) \tilde{\varrho}\left(\infty\right) R\left(\tfrac{\pi}{4}\right)^{\dagger}}{\text{$+1 \mu$}} \\
  - \bracket[\big]{-1 \mu}{R\left(\tfrac{\pi}{4}\right) \tilde{\varrho}\left(\infty\right) R\left(\tfrac{\pi}{4}\right)^{\dagger}}{\text{$-1 \mu$}}\Bigr)\biggr|.
  \label{eq:radiative_distinguishability}
\end{multline}
where the trace of the norm is defined as
$\textrm{Tr}|X| = \textrm{Tr}\sqrt{X^\dagger X}$.

For the state
\eqref{eq:tilde_varrho_infty:transition_amplitudes_unbiased_eta:sigma_fixed},
we find from Eq.~\eqref{eq:radiative_distinguishability}
that the radiative distinguishability is
\begin{align}
  \mathfrak{D}_{\mathcal{R}} & = \sqrt{1 - 4 w_{+1} w_{-1} \frac{\abs{\overlap{\tilde{\chi}^{1}}{\tilde{\chi}^{N}}}^{2}}{\overlap{\tilde{\chi}^{1}}{\tilde{\chi}^{1}} \overlap{\tilde{\chi}^{N}}{\tilde{\chi}^{N}}}},
\end{align}
which becomes equal to $\mathfrak{P}$
for indistinguishable,
and unity for orthogonal, radiative out-states,
\ie{}, $\mathfrak{K}\left(O_{\mathcal{R}}\right) = \mathfrak{D}_{\mathcal{R}} = 1$.
With the latter, the paths are fully distinguished,%
\footnote{If we failed to optimize the measurement,
then, due to
$\mathfrak{K}\left(O_{\mathcal{R}}\right) \ge \mathfrak{P}$,
we would be at most as ignorant
about the paths as
without any welcher-weg measurement.}
the duality relation
Eq.~\eqref{eq:duality_relation:VORKOR}
indicates the absence of interference,
$\mathfrak{V}\left(O_{\mathcal{R}}\right) = 0$,
giving the fully particle-like statistics
of the open CCI configuration $o$
(Sec.~\ref{sec:path_distinguishability:open_config}).
The optimum welcher-weg measurement thus
demonstrates the significance of
correlations between
the path and the radiation field.
The extent to which they are nonclassical is addressed below.

\subsubsection{Quantum Erasure}

With the quantum erasure measurement scheme,
we select a statistical subensembles for which
the path alternatives are indistinguishable.
In doing so, we can recover maximum fringe contrast
\cite{Englert:1998kl,Kim:2000th,Walborn:2002gd,Bramon:2004yf,Aharonov:2005cs}
and thus a maximum of phase control.
What is remarkable about this is that
the phase control can originate entirely from
a nontrivial quantum interference effect.
For quantum erasure to work,
we again have to join
the result of a measurement of $M$
with that of an optimally chosen $O_{\mathcal{R}}$.
The largest attainable fringe contrast---%
and thus phase controllability---%
in such a procedure is quantified by
the radiative \emph{coherence}
$\mathfrak{C}_{\mathcal{R}} = \sup_{O_{\mathcal{R}}} \mathfrak{V}\left(O_{\mathcal{R}}\right)$
\cite{Englert:2000kh}, which is
the counterpart of
the radiative distinguishability,
$\mathfrak{D}_{\mathcal{R}}$,
introduced above.

For an unbiased CCI,
the radiative coherence is given by
\begin{subequations}%
  \begin{align}%
    \mathfrak{C}_{\mathcal{R}} & = 2 \Tr\abs*{\sumint \mathrm{d}\mu \, \bracket[\big]{+1 \mu}{R\left(\tfrac{\pi}{4}\right) \tilde{\varrho}\left(\infty\right) R\left(\tfrac{\pi}{4}\right)^{\dagger}}{\text{$-1 \mu$}}}
    \label{eq:radiative_coherence} \\
    \intertext{which, for the final state
    \eqref{eq:tilde_varrho_infty:transition_amplitudes_unbiased_eta:sigma_fixed},
    reduces to}
    \mathfrak{C}_{\mathcal{R}} & = 2 \, \frac{\abs[\big]{\sum_{\nu} \overlap[\big]{\tilde{\psi}^{1}_{+1, \sigma \nu}}{\tilde{\psi}^{N}_{+1, \sigma \nu}}} \, \Tr\abs[\big]{\ket[\big]{\tilde{\chi}^{1}} \bra[\big]{\tilde{\chi}^{N}}}}{\sum_{\nu} \bigl(\overlap[\big]{\tilde{\psi}^{1}_{+1, \sigma \nu}}{\tilde{\psi}^{1}_{+1, \sigma \nu}} \overlap[\big]{\tilde{\chi}^{1}}{\tilde{\chi}^{1}} \! + \! \overlap[\big]{\tilde{\psi}^{N}_{-1, \sigma \nu}}{\tilde{\psi}^{N}_{-1, \sigma \nu}} \overlap[\big]{\tilde{\chi}^{N}}{\tilde{\chi}^{N}}\bigr)} \\
    \intertext{and, in the case of (at least slightly)
    distinguishable final radiative states,
    $\abs{\overlap{\tilde{\chi}^{1}}{\tilde{\chi}^{N}}}$ > 0,
    to}
    \mathfrak{C}_{\mathcal{R}} & = \mathfrak{V} \, \frac{\sqrt{\overlap{\tilde{\chi}^{1}}{\tilde{\chi}^{1}} \overlap{\tilde{\chi}^{N}}{\tilde{\chi}^{N}}}}{\abs{\overlap{\tilde{\chi}^{1}}{\tilde{\chi}^{N}}}}.
  \end{align}%
\end{subequations}
If the final radiative states,
$\ket*{\tilde{\chi}^{1}}$ and
$\ket*{\tilde{\chi}^{N}}$,
are indistinguishable
(\ie{}, proportional to one another),
then
$\mathfrak{C}_{\mathcal{R}} = \mathfrak{V}$,
and we cannot improve the control;
otherwise, we have
$\mathfrak{C}_{\mathcal{R}} > \mathfrak{V}$.
According to wave-particle complementarity
as quantified by the duality relation
\eqref{eq:duality_relation:VORKOR},
succeeding to
completely restore the fringe contrast,
$\mathfrak{V}\left(O_{\mathcal{R}}\right) = \mathfrak{C}_{\mathcal{R}} = 1$,
means to fully reverse the WWM,
$\mathfrak{K}\left(O_{\mathcal{R}}\right) = 0$.
This situation realizes the
closed CCI configuration
and thus full phase control
in an \emph{entirely nontrivial} way.

The measurement strategies disussed here---%
optimum welcher-weg measurement,
quantum erasure, and nonoptimum mixes thereof---%
will in general deliver a blend of
particle- and wavelike statistics,
which goes beyond Bohr's original version of
complementarity.
Interestingly, it turns out that
the experimenter has a choice between measurement strategies even
after detecting the output port event
(\ie{} after the ``click'').
In fact, \emph{when} the choice is made
has no impact on the measurement statistics.
This is the essence of the
delayed choice gedanken experiment
\cite{Wheeler:1984cq,Zeilinger2016},
whose relevance for the demonstration of
quantum coherent control
is discussed in Sec.~\ref{sec:delayed_choice}.

\subsection{The Special Case of Symmetric CCIs}
\label{sec:quantum_erasure:symmetric_CCIs}

Below we derive
optimum welcher-weg and quantum erasure measurements
for the case of a symmetric CCI.
In this example, WWM involves only
the radiative degrees of freedom.
For simplicity,
we populate only one input port
(\ie{}, $p_{\sigma} = 1$ for either $\sigma = -1$ or $1$),
and assume unbiased transition amplitudes,
as defined in
Eq.~\eqref{eqs:transition_amplitudes_unbiased_eta}.
Additionally, we stipulate that
\begin{multline}
  \sqrt{\overlap{\tilde{\chi}^{1}}{\tilde{\chi}^{1}}} \, T_{1}\bigl[\hat{f}_{2}\bigr]\left(+1 \mu, \sigma \nu\right) = \exp\bigl\{- \ii \gamma\bigl[\hat{f}_{1}, \hat{f}_{2}\bigr]\left(+1, \sigma\right)\bigr\} \\
  \times \sqrt{\overlap{\tilde{\chi}^{\smash{N}}}{\tilde{\chi}^{\smash{N}}}} \, T_{N}\bigl[\hat{f}_{1}\bigr]\left(+1 \mu, \sigma \nu\right)
  \label{eq:symmetric_conditions:gamma:again}
\end{multline}
for all $\mu \in \mathcal{D}_{+}$
and $\nu \in \mathcal{D}_{-}$,
the same symmetry condition as in
Eq.~\eqref{eq:symmetric_conditions:gamma}.
Note that, once again, the phase
$\gamma = \gamma\bigl[\hat{f}_{1}, \hat{f}_{2}\bigr]\left(+1, \sigma\right)$
is independent of both $\mu$ and $\nu$.
The CCI's output port labels and
the remaining material degrees of freedom
are therefore not correlated,
excluding the latter from WWM.

With the above restrictions,
the final state of the system becomes
\begin{multline}
  \tilde{\varrho}\left(\infty\right) = \smashoperator{\sumint_{\mathcal{D}_{+}}} \mathrm{d}\mu \smashoperator{\sumint_{\mathcal{D}_{+}}} \mathrm{d}\mu' \, \frac{\sum_{\nu} T_{1}^{\vphantom{*}}\bigl[\hat{f}_{2}\bigr]\left(+1 \mu, \sigma \nu\right) T_{1}^{*}\bigl[\hat{f}_{2}\bigr]\left(+1 \mu', \sigma \nu\right)}{\sum_{\nu} \overlap[\big]{\tilde{\psi}^{1}_{+1, \sigma \nu}}{\tilde{\psi}^{1}_{+1, \sigma \nu}}} \\ \times R\left(\tfrac{\pi}{4}\right)^{\dagger} \frac{1}{\sqrt{2}} \left(\ket*{+1 \mu} \frac{\ket*{\tilde{\chi}^{1}}}{\sqrt{\overlap{\tilde{\chi}^{1}}{\tilde{\chi}^{1}}}} - \ee^{\ii \gamma} \ket*{-1 \mu} \frac{\ket*{\tilde{\chi}^{N}}}{\sqrt{\overlap{\tilde{\chi}^{N}}{\tilde{\chi}^{N}}}}\right) \\ \times \frac{1}{\sqrt{2}} \left(\bra*{+1 \mu'} \frac{\bra*{\tilde{\chi}^{1}}}{\sqrt{\overlap{\tilde{\chi}^{1}}{\tilde{\chi}^{1}}}} - \ee^{- \ii \gamma} \bra*{-1 \mu'} \frac{\bra*{\tilde{\chi}^{N}}}{\sqrt{\overlap{\tilde{\chi}^{N}}{\tilde{\chi}^{N}}}}\right) R\left(\tfrac{\pi}{4}\right).
  \label{eq:tilde_varrho_infty:transition_amplitudes_unbiased_eta:sigma_fixed:symmetric}
\end{multline}
From this we find
\begin{subequations}%
  \begin{align}%
    \mathfrak{P} & = 0, &
    \mathfrak{D}_{\mathcal{R}} & = \sqrt{1 - \frac{\textstyle \abs*{\overlap*{\tilde{\chi}^{1}}{\tilde{\chi}^{N}}}^{2}}{\overlap{\tilde{\chi}^{1}}{\tilde{\chi}^{1}} \overlap{\tilde{\chi}^{N}}{\tilde{\chi}^{N}}}}, \\
    \mathfrak{V} & = \frac{\textstyle \abs*{\overlap*{\tilde{\chi}^{1}}{\tilde{\chi}^{N}}}}{\sqrt{\overlap{\tilde{\chi}^{1}}{\tilde{\chi}^{1}} \overlap{\tilde{\chi}^{N}}{\tilde{\chi}^{N}}}}, &
    \mathfrak{C}_{\mathcal{R}} & = 1.
    \label{eq:visibility_coherence:transition_amplitudes_unbiased_eta:sigma_fixed:symmetric}
  \end{align}%
\end{subequations}%
Hence, these quantities ``saturate'' the duality relations with
$\mathfrak{V}^{2} + \mathfrak{D}_{\mathcal{R}}^{2} = 1$
and $\mathfrak{C}_{\mathcal{R}}^{2} + \mathfrak{P}^{2} = 1$.
That is, the duality relations now provide a direct relationship,
as opposed to an inequality, between the quantities in
Eq.~\eqref{eq:visibility_coherence:transition_amplitudes_unbiased_eta:sigma_fixed:symmetric}.
Note that since the duality relation
\eqref{eq:duality_relation:VORKOR} holds
for all measurements $O_\mathcal{R}$,
it specifically holds for the measurements
that maximize the acquired WWI,
$\mathfrak{K}\left(O_\mathcal{R}\right) = \mathfrak{D}_{\mathcal{R}}$,
and restore the most fringe contrast,
$\mathfrak{V}\left(O_\mathcal{R}\right) = \mathfrak{C}_{\mathcal{R}}$,
respectively
\cite{Jaeger:1995kx,Englert:2000kh}.

In order to find
optimal measurements $O_\mathcal{R}$
(for which either a maximum of WWI is obtained,
or full interference contrast is restored, respectively),
consider the ket
in the second line of
Eq.~\eqref{eq:tilde_varrho_infty:transition_amplitudes_unbiased_eta:sigma_fixed:symmetric}.
This state \emph{almost} looks like
a standard example
of an entangled state
\cite{Horne:1989qf},
but the radiation states
$\ket*{\tilde{\chi}^{1}}$
and $\ket*{\tilde{\chi}^{N}}$
are in general nonorthogonal.
However,
there is an alternative form
of the ket
that involves only orthogonal states
\cite{Mann:1995jl},
and is of great value
below. It reads
\begin{align}
  \textstyle \sum_{\pm } \sqrt{\lambda_{\pm}} \, \ket*{\pm_{\mathcal{M}}} \ket*{\pm_{\mathcal{R}}},
  \label{eq:schmidt_decomposition}
\end{align}
where
\begin{subequations}%
  \begin{align}%
    \lambda_{\pm} & = \frac{1 \pm \mathfrak{V}}{2},
    \label{eq:schmidt_decomposition:coefficients} \\
    \ket*{\pm_{\mathcal{M}}} & = R\left(\tfrac{\pi}{4}\right)^{\dagger} \frac{1}{\sqrt{2}} \left(\pm \ee^{- \ii \phi} \ket*{+1 \mu} - \ee^{\ii \gamma} \ket*{-1 \mu}\right) \notag \\
    & = \pm \ee^{- \ii \left(\phi - \gamma\right) / 2} \left(\cos\tfrac{\phi + \gamma}{2} \ket*{\pm 1 \mu} - \ii \sin\tfrac{\phi + \gamma}{2} \ket*{\mp 1 \mu}\right),
    \label{eq:schmidt_decomposition:material_states}
    \shortintertext{and}
    \ket*{\pm_{\mathcal{R}}} & = \frac{1}{2 \sqrt{\lambda_{\pm}}} \left(\pm \ee^{\ii \phi} \frac{\textstyle \ket*{\tilde{\chi}^{1}}}{\sqrt{\overlap{\tilde{\chi}^{1}}{\tilde{\chi}^{1}}}} + \frac{\textstyle \ket*{\tilde{\chi}^{N}}}{\sqrt{\overlap{\tilde{\chi}^{N}}{\tilde{\chi}^{N}}}}\right)
    \label{eq:schmidt_decomposition:radiative_states}
  \end{align}%
  \label{eqs:schmidt_decomposition:results}%
\end{subequations}%
with phase
$\phi = \Arg \overlap*{\tilde{\chi}^{1}}{\tilde{\chi}^{N}}$
as in Eq.~\eqref{eq:phi}.
Here $\overlap*{\pm_{\mathcal{R}}}{\pm_{\mathcal{R}}} = 1$
and $\overlap*{\pm_{\mathcal{R}}}{\mp_{\mathcal{R}}} = 0$,
whereas the states $\ket*{+_{\mathcal{M}}}$
and $\ket*{-_{\mathcal{M}}}$
are orthogonal,
but in general (\eg{}, continuum states)
not normalized.
Their dependence on $\mu$ and $\sigma$ is understood,
though not explicitly stated.

Equation \eqref{eq:schmidt_decomposition}
is similar to a \emph{Schmidt decomposition}
\cite{Nielsen:2000qq,Guhne:2009vn},
in which the factors $\sqrt{\lambda_{\pm}}$
would be referred to as the \emph{Schmidt coefficients}.
The difference to a standard Schmidt decomposition is
that the $\ket*{\pm_{\mathcal{R}}}$ are improper (unnormalizable) states.
Despite this difference,
we will refer to Eq.~\eqref{eq:schmidt_decomposition}
as a Schmidt decomposition
and to $\sqrt{\lambda_{\pm}}$ as the Schmidt coefficients.

If $\ket*{\tilde{\chi}^{1}}$ and $\ket*{\tilde{\chi}^{N}}$
are linearly independent,
then both Schmidt coefficients are finite
and the state \eqref{eq:schmidt_decomposition} is entangled.
In general, however, that does not imply
that $\tilde{\varrho}\left(\infty\right)$ in
Eq.~\eqref{eq:tilde_varrho_infty:transition_amplitudes_unbiased_eta:sigma_fixed:symmetric}
is entangled.
That is, since it is a mixed state,
correlations may be classical.
Here, however,
$\tilde{\varrho}\left(\infty\right)$
contains the same entanglement as
the state \eqref{eq:schmidt_decomposition},
if the remaining degrees of freedom
of the material system
(everything except the labels $\pm 1$)
are ignored.%
\footnote{This is easily justifiable
in the case in which
the material Hilbert space
has the structure of a tensor product
between a two-dimensional Hilbert space,
$\mathcal{H}_{2}$,
for the CCI port labels $\pm 1$
and another Hilbert space,
$\mathcal{H}_{\perp 2}$,
for the remaining degrees of freedom,
such that
$\mathcal{H}_{\mathcal{M}} \otimes \mathcal{H}_{\mathcal{R}} = \mathcal{H}_{\perp 2} \otimes \mathcal{H}_{2} \otimes \mathcal{H}_{\mathcal{R}}$.
Then the state
\eqref{eq:tilde_varrho_infty:transition_amplitudes_unbiased_eta:sigma_fixed:symmetric}
admits a similar decomposition into
$\tilde{\varrho}_{\perp 2}\left(\infty\right) \otimes \tilde{\varrho}_{2 \otimes \mathcal{R}}\left(\infty\right)$,
where the mixed state
$\tilde{\varrho}_{\perp 2}\left(\infty\right)$
is not correlated with anything and
$\tilde{\varrho}_{2 \otimes \mathcal{R}}\left(\infty\right)$
is pure and entangled.}
A quantitative measure of the
entanglement strength is provided
by the \emph{concurrence} $\mathfrak{c}$
\cite{Horodecki:2009la,Guhne:2009vn}.
For the case described here,
it is given by
\begin{subequations}%
  \begin{align}%
    \mathfrak{c} & = \sqrt{\textstyle 2 \left(1 - \sum_{\pm} \lambda_{\pm}^{2}\right)}
    \label{eq:concurrence:definition} \\
    & = \sqrt{1 - \mathfrak{V}^{2}},
    \label{eq:concurrence:calculated}
  \end{align}%
  \label{eqs:concurrence}%
\end{subequations}%
which varies smoothly from $0$
for a product state, to $1$,
for a maximally entangled state
for which the Schmidt coefficients are equal.
This can only happen for fully distinguishable
radiative out-states
and hence vanishing interference.

Here, in accord with
Eqs.~\eqref{eq:visibility_coherence:transition_amplitudes_unbiased_eta:sigma_fixed:symmetric}      and \eqref{eqs:concurrence},
the distinguishability
and the concurrence seem to be equal,
$\mathfrak{D}_{\mathcal{R}} = \mathfrak{c}$,
which is an explicit statement that,
in this example,
the distinguishability
is quantum
and not the result of
classical correlations
\cite{Jakob:2010gd}.
Below we discuss whether
this provides sufficient leverage
for probing the nonclassicality of the scenario.
To do so, we begin by designing an
optimum welcher-weg measurement.

\subsubsection{An Optimum Welcher-Weg Measurement}

It is only possible to obtain additional WWI
through a measurement of the radiation field
if $\mathfrak{D}_{\mathcal{R}} > 0$.
The acquired welcher-weg knowledge,
$\mathfrak{K}\left(O_{\mathcal{R}}\right)$,
depends on the observable $O_{\mathcal{R}}$,
and we can derive an optimum welcher-weg observable for which
$\mathfrak{K}\left(O_{\mathcal{R}}\right) = \mathfrak{D}_{\mathcal{R}}$.
For the symmetric CCI with
$w_{+1} = w_{-1} = \tfrac{1}{2}$ examined here, it is
\begin{gather}
  O_{\mathcal{R}} = \frac{1}{\sqrt{1 - \mathfrak{V}^{2}}} \, \left(\frac{\textstyle \ket*{\tilde{\chi}^{N}} \bra*{\tilde{\chi}^{N}}}{\textstyle \overlap*{\tilde{\chi}^{N}}{\tilde{\chi}^{N}}} - \frac{\textstyle \ket*{\tilde{\chi}^{1}} \bra*{\tilde{\chi}^{1}}}{\textstyle \overlap*{\tilde{\chi}^{1}}{\tilde{\chi}^{1}}}\right)
  \label{eq:OR:optimum_ww_measurement}
\end{gather}
which has an eigendecomposition
$O_{\mathcal{R}} = \Pi^{+}_{\mathcal{R}} - \Pi^{-}_{\mathcal{R}}$,
where
$\Pi^{\pm}_{\mathcal{R}} = (\ket*{+_{\mathcal{R}}} \pm \ket*{-_{\mathcal{R}}}) (\bra*{+_{\mathcal{R}}} \pm \bra*{-_{\mathcal{R}}}) / 2$.
Consequentially,
the measurement of the $\pm 1$ eigenvalue reduces the state
$\tilde{\varrho}\left(\infty\right)$ to
\begin{subequations}
  \begin{gather}
    \tilde{\varrho}_{\mathcal{M}}^{\pm}\left(\infty\right) = \sum_{\nu} R\left(\tfrac{\pi}{4}\right)^{\dagger} \ket*{\tilde{\psi}_{\sigma \nu}^{\pm}} \bra*{\tilde{\psi}_{\sigma \nu}^{\pm}} R\left(\tfrac{\pi}{4}\right)
  \end{gather}
  with
  \begin{multline}
    \ket*{\tilde{\psi}_{\sigma \nu}^{\pm}} = \sqrt{\frac{1 \mp \sqrt{1 - \mathfrak{V}^{2}}}{2 \sum_{\nu'} \smash[b]{\overlap[\big]{\tilde{\psi}^{1}_{+1, \sigma \nu'}}{\tilde{\psi}^{1}_{+1, \sigma \nu'}}}}} \, \ee^{- \ii \phi} \, \ket[\big]{\tilde{\psi}^{1}_{+1, \sigma \nu}} \\
    + \sqrt{\frac{1 \pm \sqrt{1 - \mathfrak{V}^{2}}}{2 \sum_{\nu'} \smash[b]{\overlap[\big]{\tilde{\psi}^{N}_{-1, \sigma \nu'}}{\tilde{\psi}^{N}_{-1, \sigma \nu'}}}}} \, \ket[\big]{\tilde{\psi}^{N}_{-1, \sigma \nu}} \, .
  \end{multline}
\end{subequations}
For $\mathfrak{V} \to 0$, these states becomes fully particle-like:
$\tilde{\varrho}_{\mathcal{M}}^{-}\left(\infty\right)$
describes the state after excitation in a one-photon process,
whereas $\tilde{\varrho}_{\mathcal{M}}^{+}\left(\infty\right)$
describes the result of $N$-photon excitation.
According to Eq.~\eqref{eq:concurrence:calculated},
$\mathfrak{V} = 0$ also means $\mathfrak{c} = 1$,
\ie{} the quantum correlations described
by the state $\tilde{\varrho}\left(\infty\right)$
are maximal. Successful extraction of full WWI
therefore suggests a strong nonclassicality to the scenario.
However, as we shall see, that is not sufficient to prove it.

\subsubsection{An Optimum What-Phase Measurement}

Consider the intricacies of
the complementary quantum erasure scheme,
where phase control can emerge
as a nontrivial quantum interference effect.
In Eq.~\eqref{eq:tilde_varrho_infty:transition_amplitudes_unbiased_eta:sigma_fixed:symmetric},
the final state $\tilde{\varrho}\left(\infty\right)$
is written in a way that puts emphasis
on the particle-like property of the CCI labels.
However, as shown above,
an equally valid description
(or ``as-if reality'' \cite{Englert:1998kl,Englert:2013eu}) is
in terms of the Schmidt decomposition
\eqref{eq:schmidt_decomposition},
in which the material system is in a superposition of
the two wavelike states $\ket*{\pm_{\mathcal{M}}}$,
one corresponding to fringes
[$\ket*{+_{\mathcal{M}}}$, in conjunction with
the fringe marker state $\ket*{+_{\mathcal{R}}}$]
and the other to antifringes
[$\ket*{-_{\mathcal{M}}}$, with
the orthogonal antifringe marker state $\ket*{-_{\mathcal{R}}}$].
The goal of quantum erasure is to
realize one of these wavelike alternatives.

As an illustration,
consider the joint measurement statistics
of $\tfrac{1}{2} \left(\operatorone + \eta M\right)$
and some radiative projector
$O_{\mathcal{R}}$ with
$\cramped{O_{\mathcal{R}}^{2}} = O_{\mathcal{R}}$.
Using the Schmidt decomposition
\eqref{eq:schmidt_decomposition},
the statistics can be immediately obtained.
On the one hand,
within the subensemble
for which we measure the
$O_{\mathcal{R}}$ eigenvalue $+1$, \ie{},
 the subensemble of click events,
we observe the output port statistics
\begin{subequations}%
  \begin{multline}%
    \frac{\expval*{\tfrac{1}{2} \left(\operatorone_{\mathcal{M}} + \eta M\right) O_{\mathcal{R}}}}{\expval*{O_{\mathcal{R}}}} = \frac{1}{\expval*{O_{\mathcal{R}}}} \biggl\{\frac{\lambda_{+}}{2} \left[1 + \eta \cos\left(\phi + \gamma\right)\right] \bracket{+_{\mathcal{R}}}{O_{\mathcal{R}}}{+_{\mathcal{R}}} \\
    - \vphantom{\biggl\{} \sqrt{\lambda_{+} \lambda_{-}} \, \eta \sin\left(\phi + \gamma\right) \Im\bracket{+_{\mathcal{R}}}{O_{\mathcal{R}}}{-_{\mathcal{R}}} \\
    + \frac{\lambda_{-}}{2} \left[1 - \eta \cos\left(\phi + \gamma\right)\right] \bracket{-_{\mathcal{R}}}{O_{\mathcal{R}}}{-_{\mathcal{R}}}\biggr\},
    \label{eq:expval_eta_M_O_projector:schmidt_decomposition:click}%
  \end{multline}
  where
  $\expval*{O_{\mathcal{R}}} = \sum_{\pm} \lambda_{\pm} \bracket{\pm_{\mathcal{R}}}{O_{\mathcal{R}}}{\pm_{\mathcal{R}}}$
  is the weight of the subensemble.
  On the other hand,
  for the subensemble to the eigenvalue $0$---%
  the no-click subensemble---%
  with
  \begin{multline}
    \frac{\expval*{\tfrac{1}{2} \left(\operatorone_{\mathcal{M}} + \eta M\right) \left(\operatorone_{\mathcal{R}} - O_{\mathcal{R}}\right)}}{1 - \expval*{O_{\mathcal{R}}}} = \frac{1}{1 - \expval*{O_{\mathcal{R}}}} \biggl\{\frac{1}{2} \left[1 + \eta \, \mathfrak{V} \cos\left(\phi + \gamma\right)\right] \\
    - \expval*{\tfrac{1}{2} \left(\operatorone_{\mathcal{M}} + \eta M\right) O_{\mathcal{R}}}\biggr\}.
    \label{eq:expval_eta_M_O_projector:schmidt_decomposition:no_click}%
  \end{multline}%
  \label{eqs:expval_eta_M_O_projector:schmidt_decomposition}%
\end{subequations}%
Using
Eq.~\eqref{eq:expval_eta_M_O_projector:schmidt_decomposition:click},
it is easy to construct
a projector $O_{\mathcal{R}}$
that restores full fringe visibility,
$\mathfrak{V}\left(O_{\mathcal{R}}\right) = 1$.
Formally, a perfect choice is
\begin{align}
  O_{\mathcal{R}} & = \ket*{+_{\mathcal{R}}} \bra*{+_{\mathcal{R}}}.
  \label{eq:OR:optimum_quantum_erasure}
\end{align}
With this, the statistics of the click
subensemble becomes
\begin{align}
  \frac{\expval*{\tfrac{1}{2} \left(\operatorone_{\mathcal{M}} + \eta M\right) O_{\mathcal{R}}}}{\expval*{O_{\mathcal{R}}}} & = \frac{1}{2} \left[1 + \eta \cos\left(\phi + \gamma\right)\right]
\end{align}
with weight
$\expval*{O_{\mathcal{R}}} = \lambda_{+}$.
Indeed, this expression describes a fringe pattern
with maximal contrast.
The choice \eqref{eq:OR:optimum_quantum_erasure} works,
because the projection onto $\ket*{+_{\mathcal{R}}}$
reduces the decomposition
\eqref{eq:schmidt_decomposition}
to a product state.
Since WWM relies here fully on the
entanglement between
the final state of the marker
and the path alternatives,
the interference can be fully recovered.
A less abstract explanation is
that the field is detected in a state
for which a distinction
between the two excitation processes
is not possible.
The WWI has been \emph{erased}.

With $O_{\mathcal{R}}$ as in
Eq.~\eqref{eq:OR:optimum_quantum_erasure},
a fully saturated interference pattern
also shows in the subensemble
for which the detector does not click,
\begin{align}
  \frac{\expval*{\tfrac{1}{2} \left(\operatorone_{\mathcal{M}} + \eta M\right) \left(\operatorone_{\mathcal{R}} - O_{\mathcal{R}}\right)}}{1 - \expval*{O_{\mathcal{R}}}} & = \frac{1}{2} \left[1 - \eta \cos\left(\phi + \gamma\right)\right],
\end{align}
where $1 - \expval*{O_{\mathcal{R}}} = \lambda_{-}$.
At first glance,
this result seems at odds with
the limited visibility $\mathfrak{V}$
when no measurement on the field is carried out.
However, there is no contradiction,
since the weighted average
\begin{align}
  \expval*{O_{\mathcal{R}}} \, \frac{\expval*{\tfrac{1}{2} \left(\operatorone_{\mathcal{M}} + \eta M\right) O_{\mathcal{R}}}}{\expval*{O_{\mathcal{R}}}} + \left(1 - \expval*{O_{\mathcal{R}}}\right) \, \frac{\expval*{\tfrac{1}{2} \left(\operatorone_{\mathcal{M}} + \eta M\right) \left(\operatorone_{\mathcal{R}} - O_{\mathcal{R}}\right)}}{1 - \expval*{O_{\mathcal{R}}}}
\end{align}
of both interference patterns
is a sum of fringes and antifringes,
which, in general,
has less contrast than either
pattern ($\mathfrak{V}$ in this case).

In contrast to
the optimum welcher-weg measurement scheme,
a successful demonstration
of quantum erasure would not only suggest
nonclassicality,
but also imply a genuinely quantum interference origin
for phase control in the experiment.
Sec.~\ref{sec:quantum_erasure:relationship_to_quantum_correlations}
discusses whether or not
this would also be a valid rigorous proof.
Before that, however,
we address in
Sec.~\ref{sec:quantum_erasure:photon_threshold}
how ``quantum erasure coherent control''
could be done under more realistic conditions,
\ie{}, conditions under which a measurement of
$O_{\mathcal{R}} = \ket*{+_{\mathcal{R}}} \bra*{+_{\mathcal{R}}}$
[Eq.~\eqref{eq:OR:optimum_quantum_erasure}]
is not available.

\subsection{Quantum Erasure with Displaced Photon Threshold Measurements}
\label{sec:quantum_erasure:photon_threshold}

The observable $O_{\mathcal{R}}$
in Eq.~\eqref{eq:OR:optimum_quantum_erasure}
is formally the best choice
for quantum erasure,
but it may not be the most practical.
For an arbitrary
radiative in-state $\ket*{\chi}$,
it may not be possible to
realize a projection onto
the state $\ket*{+_{\mathcal{R}}}$.
We can illustrate and resolve this issue
via an example of perfect WWM.
Consider the case where the radiation field
is initialized in the nonclassical
($n_{1} \! + \! n_{2}$)-photon
wavepacket state
\begin{align}
  \ket*{\chi} & = \frac{1}{\sqrt{n_{1}!}} \, \bigl(a^{\dagger}\bigl[\hat{f}_{1}\bigr]\bigr)^{n_{1}} \, \frac{1}{\sqrt{n_{2}!}} \, \bigl(a^{\dagger}\bigl[\hat{f}_{2}\bigr]\bigr)^{n_{2}} \ket*{\mathrm{vac}},
  \label{eq:n1n2photon_wavepacket}
\end{align}
where $n_{1} \ge N$,
$n_{2} \ge 1$, and where
the amplitude functions
$f_{1}$, $f_{2}$
have disjoint support
and meet the requirements
of the $1$ \vs{} $N$ coherent phase control
as stated in Sec.~\ref{sec:cci:coherent_control_theory}.
Then, according to
Eqs.~\eqref{eqs:tilde_chi},
the final (path conditioned) radiation states are
\begin{subequations}%
  \begin{align}%
    \ket*{\textstyle \tilde{\chi}^{1}} & = \frac{1}{\sqrt{n_{1}!}} \, \bigl(a^{\dagger}\bigl[\hat{f}_{1}\bigr]\bigr)^{n_{1}} \, \frac{\sqrt{n_{2}}}{\sqrt{\left(n_{2} - 1\right)!}} \, \bigl(a^{\dagger}\bigl[\hat{f}_{2}\bigr]\bigr)^{n_{2} - 1} \ket*{\mathrm{vac}}\
    \shortintertext{and}
    \ket*{\textstyle \tilde{\chi}^{N}} & = \frac{\sqrt{n_{1} \left(n_{1} - 1\right) \cdots \left(n_{1} - N + 1\right)}}{\sqrt{\left(n_{1} - N\right)!}} \, \bigl(a^{\dagger}\bigl[\hat{f}_{1}\bigr]\bigr)^{n_{1} - N} \, \frac{1}{\sqrt{n_{2}!}} \, \bigl(a^{\dagger}\bigl[\hat{f}_{2}\bigr]\bigr)^{n_{2}} \ket*{\mathrm{vac}}.
  \end{align}%
  \label{eq:final_radiation_states}%
\end{subequations}%
These states are orthogonal,
$\overlap*{\tilde{\chi}^{1}}{\tilde{\chi}^{N}} = 0$,
and therefore $\phi = 0$, $\mathfrak{V} = 0$,
and $\mathfrak{D} = 1$.
Full WWI is thus available.

Complete WWI can be acquired
by measuring the total light intensity
$I \sim Tr\left[H_{\mathcal{R}} \tilde{\varrho}\left(\infty\right)\right]$
\cite{Loudon:2000yg},
which can be understood as counting
the final numbers of photons in the modes \cite{Gong:2010dq}.
In principle, as described in the previous section,
this WWI can be removed by a projection of the field
into $\ket*{+_{\mathcal{R}}}$ or $\ket*{-_{\mathcal{R}}}$.
Yet, by definition
[\cf{} Eq.~\eqref{eq:schmidt_decomposition:radiative_states}],
these states are coherent superpositions of
$\ket*{\tilde{\chi}^{1}}$ and $\ket*{\tilde{\chi}^{N}}$
and thus are \emph{entangled} multimode Fock states.
Hence,
it is very difficult to implement a projection into one of them.
However, there is an experimentally feasible alternative,
a so-called ``displaced photon threshold measurement''
which can be defined as
\begin{gather}
  \textstyle O_{\mathcal{R}} = D^{\dagger}\left[- f\right] \left(2 \ket*{\mathrm{vac}} \bra*{\mathrm{vac}} - \operatorone_{\mathcal{R}}\right) D\left[- f\right],
  \label{eq:displaced_photon_threshold_measurement}
\end{gather}
where
\begin{gather}
  \textstyle D\left[- f\right] = \exp\left(a\left[f\right] - a^{\dagger}\left[f\right] \right) = D^{\dagger}\left[f\right]
\end{gather}
is a displacement operator
in continuous Fock space.
Such a displaced photon threshold measurement can be realized
in a homodyne detection scheme
that employs a click/no-click detector,%
\footnote{A click/no-click detector,
also referred to as a threshold detector,
is a photodetector that
does not resolve the number of photons.
It responds
if it detects one or more photons,
but its signal will not differentiate between
one photon and two or more photons.
A threshold detector
with unit efficiency for all frequencies
and polarizations
is described
by the dichotomous observable
$\left[\ket*{\mathrm{vac}} \bra*{\mathrm{vac}} - \left(\operatorone_{\mathcal{R}} - \ket*{\mathrm{vac}} \bra*{\mathrm{vac}}\right)\right]$.
A click corresponds to an event with
one or more photons
and the eigenvalue $-1$.
The eigenvalue $1$ is assigned
to the photon-less no-click event.}
an auxiliary coherent light source,
and a beam splitter
with an almost ideal transmission.
In the beam splitter,
prior to feeding it
into the detector,
the light is superposed
with a coherent state
\begin{align}
  \ket{- f} & = D\left[- f\right] \ket*{\mathrm{vac}} = \ee^{-\norm{f}^{2}/2} \sum_{n=0}^{\infty} \frac{1}{n!} \left(\textstyle a^{\dagger}\left[-f\right]\right)^{n} \ket*{\mathrm{vac}}
\end{align}
from the auxiliary source
whose radiation is characterized by the amplitude function $- f$.
In those cases where
the eigenvalue $1$ is found (\ie{}, the absence
of a click in the detector),
the radiation field is projected into the
coherent state
$\ket*{f} = D\left[f\right] \ket*{\mathrm{vac}}$.
The visibility
of the fringes observed in this subensemble is then:
\begin{multline}
  \mathfrak{V}_{\mathrm{no-click}} = \frac{1}{2 \expval*{\ket*{f} \bra*{f}}} \, \Bigl|\abs*{\overlap*{+_{\mathcal{R}}}{f}}^{2} - \overlap*{+_{\mathcal{R}}}{f} \overlap*{f}{-_{\mathcal{R}}} \\
  + \overlap*{-_{\mathcal{R}}}{f} \overlap*{f}{+_{\mathcal{R}}} - \abs*{\overlap*{-_{\mathcal{R}}}{f}}^{2}\Bigr|,
\end{multline}
where
$\expval*{\ket*{f} \bra*{f}} = \bigl(\cramped{\abs*{\overlap*{+_{\mathcal{R}}}{f}}^{2}} + \cramped{\abs*{\overlap*{-_{\mathcal{R}}}{f}}^{2}}\bigr) / 2$
is the probability to not detect a click.
For the radiation states \eqref{eq:final_radiation_states},
these expressions become
\begin{gather}
  \mathfrak{V}_{\mathrm{no-click}} = \frac{2 \sqrt{n_{1}! \, \left(n_{1} - N\right)! \, n_{2}} \; \abs[\big]{\sproduct[\big]{\hat{f}_{1}}{f}}^{N} \abs[\big]{\sproduct[\big]{\hat{f}_{2}}{f}}}{\left(n_{1} - N\right)! \, n_{2} \, \abs[\big]{\sproduct[\big]{\hat{f}_{1}}{f}}^{2 N} + n_{1}! \, \abs[\big]{\sproduct[\big]{\hat{f}_{2}}{f}}^{2}}
\end{gather}
and
\begin{multline}
  \expval*{\ket*{f} \bra*{f}} = \frac{\abs[\big]{\sproduct[\big]{\hat{f}_{1}}{f}}^{2 \left(n_{1} - N\right)} \abs[\big]{\sproduct[\big]{\hat{f}_{2}}{f}}^{2 \left(n_{2} - 1\right)}}{4 \, \ee^{\norm{f}^{2}} n_{1}! \, \left(n_{1} - N\right)! \, n_{2}!} \\
  \times \left[\left(n_{1} - N\right)! \, n_{2} \, \abs[\big]{\sproduct[\big]{\hat{f}_{1}}{f}}^{2 N} + n_{1}! \, \abs[\big]{\sproduct[\big]{\hat{f}_{2}}{f}}^{2}\right].
\end{multline}
The visibility $\mathfrak{V}_{\mathrm{no-click}}$
can then be optimized by tuning the
intensity, proportional to the norm $\norm{f}$ of
the amplitude function $f$,
of the auxiliary coherent light field described by
the state $\ket{-f}$.
The visibility becomes maximal,
$\mathfrak{V}_{\mathrm{no-click}} = 1$, for
\begin{align}
  \norm{f} = \left(\frac{\abs[\big]{\sproduct[\big]{\hat{f}_{1}}{\hat{f}}}^{N}}{\abs[\big]{\sproduct[\big]{\hat{f}_{2}}{\hat{f}}}} \sqrt{n_{2} \frac{\left(n_{1} - N\right)!}{n_{1}!}}\right)^{\frac{1}{1-N}}
  \label{eq:normf_optimal}
\end{align}
with $\hat{f} = f / \norm{f}$.%
\footnote{With $\norm{f}$ as in Eq.~\eqref{eq:normf_optimal},
$\mathfrak{K}\left(O_{\mathcal{R}}\right) = \mathfrak{K}_{\mathrm{no-click}} = \mathfrak{K}_{\mathrm{click}} = 0$.
It is thus impossible to identify the excitation process by
which the output port is reached.}
That means,
given the right intensity of the auxiliary field
(which, according to the above equation, only depends on
the number of photons in the incident radiation
and the absolute overlaps
$\abs[\big]{\sproduct[\big]{\hat{f}_{1}}{\hat{f}}}$,
$\abs[\big]{\sproduct[\big]{\hat{f}_{2}}{\hat{f}}}$
of the amplitude functions),
one achieves full coherent control over the label $\eta$
of the material system,
\begin{align}
  \frac{\expval*{\tfrac{1}{2} \left(\operatorone_{\mathcal{M}} + \eta M\right) \ket*{f} \bra*{f}}}{\expval*{\ket*{f} \bra*{f}}} & = \frac{1}{2} \bigl\{1 + \eta \cos\bigl[\gamma + N \Arg \sproduct[\big]{\hat{f}_{1}}{\hat{f}} - \Arg \sproduct[\big]{\hat{f}_{2}}{\hat{f}}\bigr]\bigr\},
\end{align}
in the event that the detector does not fire.%
\footnote{Note that the phase relationship between the
incoming radiation and auxiliary field,
$\Arg \sproduct[\big]{\hat{f}_{i}}{\hat{f}}$,
offers an additional control opportunity.}
Subject to the constraint \eqref{eq:normf_optimal},
one now optimizes this probability,
by finding $\max_{\hat{f}} \expval*{\ket*{f} \bra*{f}}$.
For convenience, we restrict our attention to $N = 2$.
In this case,
the expectation value $\expval*{\ket*{f} \bra*{f}}$
is maximized by
\begin{subequations}%
  \begin{align}%
    \hat{f} & = \textstyle \alpha \hat{f}_{1} + \beta \hat{f}_{2},
  \end{align}%
  where
  \begin{align}%
    \abs*{\alpha} & = \textstyle \abs[\big]{\sproduct[\big]{\hat{f}_{1}}{\hat{f}}} = \frac{1}{\sqrt{u}}, &
    \abs*{\beta} & = \textstyle \abs[\big]{\sproduct[\big]{\hat{f}_{2}}{\hat{f}}} = \sqrt{1 - \frac{1}{u}},
  \end{align}
  and
  \begin{align}
    u & = \textstyle \frac{1}{4} \left(3 + \sqrt{1 + \frac{8 n_{2} \left(n_{1} + 2 n_{2} - 2\right)}{n_{1} \left(n_{1} - 1\right)}}\right).
  \end{align}%
\end{subequations}%
The maximum is given by
\begin{align}
  \expval*{\ket*{f} \bra*{f}} & = \frac{\left(u - 1\right)^{n_{1} + 2 n_{2} - 2}}{n_{1}! \left(n_{2} - 1\right)!} \left[\frac{n_{1} \left(n_{1} - 1\right)}{n_{2}}\right]^{n_{1} + n_{2} - 1} \ee^{- \frac{n_{1} \left(n_{1} - 1\right)}{n_{2}} u \left(u - 1\right)}.
\end{align}
For two photons in the first
and one photon in the second incident wavepacket,
$n_{1} = 2$, $n_{2} = 1$,
and
$\expval*{\ket*{f} \bra*{f}} = \ee^{-3/2}/2 \simeq 0.11$,
which is also the largest possible value for
any valid choice of $n_{1}$ and $n_{2}$.
That means that the statistical weight of the no-click subensemble
is $11 \%$ at best.
The remaining $89 \%$ contribute to the click subensemble
where one observes interference fringes with contrast
\begin{align}
  \mathfrak{V}_{\mathrm{click}} & = \frac{\expval*{\ket*{f} \bra*{f}}}{1 - \expval*{\ket*{f} \bra*{f}}} \, \mathfrak{V}_{\mathrm{no-click}} = \frac{1}{2 \ee^{3/2} - 1} \simeq 0.13,
\end{align}
much less than for the no click subensemble,
but still better than without quantum erasure.
Note that we were unable to improve over these numbers,
with or without condition \eqref{eq:normf_optimal}.

\subsection[Can Welcher-Weg Or What-Phase Measurements Probe Nonclassicality?]{Can Welcher-Weg Or What-Phase Measurements Rigorously Probe Nonclassicality?}
\label{sec:quantum_erasure:relationship_to_quantum_correlations}

The question addresed in this section is whether or not
the measurement strategies discussed above
can serve as rigorous verifications
of the nonclassical principle of complementarity,
of true quantum coherent control,
and, in the case of quantum erasure,
of genuinely quantum interference.
For the example of quantum erasure,
let us reiterate why this might be the case.
In the above examples,
the WWI is accompanied by quantum entanglement between
the material system and the radiation field:
measuring the former in one of the states
$R\left(\tfrac{\pi}{4}\right)^{\dagger} \ket*{\tilde{\psi}^{1}_{+1, \sigma \nu}}$ or
$R\left(\tfrac{\pi}{4}\right)^{\dagger} \ket*{\tilde{\psi}^{1}_{-1, \sigma \nu}}$
is equivalent to detecting the latter in either
$\ket{\tilde{\chi}^{1}}$ or $\ket{\tilde{\chi}^{N}}$,
respectively.
The projection on a path ignorant marker state
($\ket*{+_{\mathcal{R}}}$ or $\ket*{-_{\mathcal{R}}}$)
results in the interference reappearing;
quantum erasure is complete.
Based on this finding,
one can speculate that
an erasure of WWI generally involves
the dissolution of quantum entanglement.
If that were indeed true, then successful quantum erasure,
as indicated by the recovery of interference,
would be able to certify the presence of entanglement.
Thus, we may ask whether or not quantum erasure
necessarily implies quantum entanglement

The answer is, in general, unfortunately, no.
The example studied in
Sec.~\ref{sec:quantum_erasure:symmetric_CCIs} is
an exceptional, \emph{pure-state case} where
quantum entanglement is prerequisite to WWI,
and WWI is prerequisite to the recovery of interference.
However, as we discuss in the following,
that causal chain breaks
when mixed states are considered.
However, if we can guarantee that the system is in a pure
state, then quantum erasure ensures quantum entanglement
and rigorously verifies quantumness in the control scenario.

Considering, however, the difficulty of preparing the pure
state case, we examine the general case, below. In that
case:
\begin{description}
  \item[Interference recovery does not require the presence of WWI.]
    In fact, the term ``quantum erasure'' is a misnomer.
    Sometimes, interference can be restored in cases
    for which there is no WWI (and no quantum entanglement)
    to begin with.
    This is known as ``nonerasing quantum erasure''
    \cite{Englert:2000kh}.
    The function of quantum erasure is solely to prepare subensembles
    within which the interference contrast is maximal
    \cite{Englert:1998kl,Kwiat:2004yu}.
    This can include the removal of WWI, but it also may not.
  \item[WWI does not rely on quantum entanglement.]
    For example, if the state $\tilde{\varrho}\left(\infty\right)$ of
    Eq.~\eqref{eq:tilde_varrho_infty:transition_amplitudes_unbiased_eta:sigma_fixed:symmetric}
    were to decohere into a mixture with respect to the
    path labels $\pm 1$,
    then the WWI would remain intact, but would exist as classical,
    rather than quantum, correlations.
    These correlations could still be removed by postselection,
    yet the recovery of an interference pattern would not be possible
    (\ie{}, WWI can be erased without
    regaining any interference contrast).
    That is, the procedure would not be able to bring back
    the cross-term contributions
    that are necessary for interference,
    but that were lost due to decoherence.
  \item[Interference recovery does not certify quantum entanglement.]
    In some cases and
    contrary to the situation just described,
    interference contrast can be restored
    from a fully classical, mixed state.
    Below we give such an example
    (adapted from Ref.~\citealp{Englert:2000kh}).
    It counters the proposition
    that quantum erasure is a reliable test
    for nonclassical correlations
    (or any correlations for that matter) unless the state is pure.
    To see this, consider the case in which condition
    \eqref{eqs:transition_amplitudes_unbiased_eta}
    is replaced by
    \begin{subequations}%
      \begin{align}%
        T_{1}\bigl[\hat{f}_{2}\bigr]\left(+1 \mu, \sigma \nu\right) & = \sigma \, T_{1}\bigl[\hat{f}_{2}\bigr]\left(-1 \mu, \sigma \nu\right), \\
        T_{N}\bigl[\hat{f}_{1}\bigr]\left(+1 \mu, \sigma \nu\right) & = - \sigma \, T_{N}\bigl[\hat{f}_{1}\bigr]\left(-1 \mu, \sigma \nu\right).
      \end{align}%
      \label{eqs:transition_amplitudes_unbiased_eta:alternative}%
    \end{subequations}%
    We further require the validity of
    Eq.~\eqref{eq:symmetric_conditions:gamma:again}
    and demand that
    \begin{subequations}%
      \begin{align}%
        T_{1}\bigl[\hat{f}_{2}\bigr]\left(+1 \mu, \sigma \nu\right) & = \exp\bigl\{- \ii \gamma\bigl[\hat{f}_{1}, \hat{f}_{2}\bigr]\left(+1, \sigma\right)\bigr\} \, T_{1}\bigl[\hat{f}_{2}\bigr]\left(+1 \mu, -\sigma \nu\right), \\
        T_{N}\bigl[\hat{f}_{2}\bigr]\left(+1 \mu, \sigma \nu\right) & = \exp\bigl\{\ii \gamma\bigl[\hat{f}_{1}, \hat{f}_{2}\bigr]\left(+1, \sigma\right)\bigr\} \, T_{N}\bigl[\hat{f}_{2}\bigr]\left(+1 \mu, -\sigma \nu\right),
      \end{align}%
      as well as
      \begin{gather}%
        \gamma\bigl[\hat{f}_{1}, \hat{f}_{2}\bigr]\left(+1, +1\right) = - \gamma\bigl[\hat{f}_{1}, \hat{f}_{2}\bigr]\left(+1, -1\right).
      \end{gather}%
    \end{subequations}%
    Finally, let us assume that the material system
    starts in the state \eqref{eq:varrhoM:sigma} with
    \begin{gather}
      \frac{p_{+1}}{\sum_{\nu \in \mathcal{D}_{-}} \delta_{M_{\nu}, +1}} = \frac{p_{-1}}{\sum_{\nu \in \mathcal{D}_{-}} \delta_{M_{\nu}, -1}},
    \end{gather}
    \ie{}, in an incoherent (classical) mixture with equal population
    of the CCI input ports $\pm 1$.
    In this situation, the final state will be
    \begin{multline}
      \tilde{\varrho}\left(\infty\right) = \smashoperator{\sumint_{\mathcal{D}_{+}}} \mathrm{d}\mu \smashoperator{\sumint_{\mathcal{D}_{+}}} \mathrm{d}\mu' \, \frac{\sum_{\nu} T_{1}^{\vphantom{*}}\bigl[\hat{f}_{2}\bigr]\left(+1 \mu, +1 \nu\right) T_{1}^{*}\bigl[\hat{f}_{2}\bigr]\left(+1 \mu', +1 \nu\right)}{\sum_{\nu} \overlap[\big]{\tilde{\psi}^{1}_{+1, +1 \nu}}{\tilde{\psi}^{1}_{+1, +1 \nu}}} \\
      \begin{aligned}[b]
        \times R\left(\tfrac{\pi}{4}\right)^{\dagger} \frac{1}{2} \Biggl[ & \left(\frac{\ket*{+1 \mu} \ket*{\tilde{\chi}^{1}}}{\sqrt{2 \overlap{\tilde{\chi}^{1}}{\tilde{\chi}^{1}}}} - \ee^{\ii \gamma} \frac{\ket*{-1 \mu} \ket*{\tilde{\chi}^{N}}}{\sqrt{2 \overlap{\tilde{\chi}^{N}}{\tilde{\chi}^{N}}}}\right) \\
        \times & \left(\frac{\bra*{+1 \mu'} \bra*{\tilde{\chi}^{1}}}{\sqrt{2 \overlap{\tilde{\chi}^{1}}{\tilde{\chi}^{1}}}} - \ee^{- \ii \gamma} \frac{\bra*{-1 \mu'} \bra*{\tilde{\chi}^{N}}}{\sqrt{2 \overlap{\tilde{\chi}^{N}}{\tilde{\chi}^{N}}}}\right) \\
        + & \left(\frac{\ket*{+1 \mu} \ket*{\tilde{\chi}^{N}}}{\sqrt{2 \overlap{\tilde{\chi}^{N}}{\tilde{\chi}^{N}}}} - \ee^{\ii \gamma} \frac{\ket*{-1 \mu} \ket*{\tilde{\chi}^{1}}}{\sqrt{2 \overlap{\tilde{\chi}^{1}}{\tilde{\chi}^{1}}}}\right) \\
        \times & \left(\frac{\bra*{+1 \mu'} \bra*{\tilde{\chi}^{N}}}{\sqrt{2 \overlap{\tilde{\chi}^{N}}{\tilde{\chi}^{N}}}} - \ee^{- \ii \gamma} \frac{\bra*{-1 \mu'} \bra*{\tilde{\chi}^{1}}}{\sqrt{2 \overlap{\tilde{\chi}^{1}}{\tilde{\chi}^{1}}}}\right)\Biggr] R\left(\tfrac{\pi}{4}\right).
      \end{aligned}
      \label{eq:tilde_varrho_infty:transition_amplitudes_unbiased_eta:sigma_not_fixed:symmetric:altnernative}
    \end{multline}
    where
    $\gamma = \gamma\bigl[\hat{f}_{1}, \hat{f}_{2}\bigr]\left(+1, +1\right)$.
    That is an average of
    Eq.~\eqref{eq:tilde_varrho_infty:transition_amplitudes_unbiased_eta:sigma_fixed:symmetric}
    and Eq.~\eqref{eq:tilde_varrho_infty:transition_amplitudes_unbiased_eta:sigma_fixed:symmetric}
    with the path labels $+1$ and $-1$ interchanged.
    Therefore,
    the state \eqref{eq:tilde_varrho_infty:transition_amplitudes_unbiased_eta:sigma_not_fixed:symmetric:altnernative}
    does not represent any correlations---%
    neither classical nor quantum---%
    between the path labels and the radiation field.
    One calculates
    \begin{subequations}%
      \begin{align}%
        \mathfrak{P} & = 0, &
        \mathfrak{D}_{\mathcal{R}} & = 0, \\
        \mathfrak{V} & = \frac{\textstyle \abs*{\overlap*{\tilde{\chi}^{1}}{\tilde{\chi}^{N}}}}{\textstyle \sqrt{\overlap{\tilde{\chi}^{1}}{\tilde{\chi}^{1}} \overlap{\tilde{\chi}^{N}}{\tilde{\chi}^{N}}}} \, \abs*{\cos \phi}, &
        \mathfrak{C}_{\mathcal{R}} & = \abs*{\cos \phi}
      \end{align}%
    \end{subequations}%
    with $\phi = \Arg \overlap*{\tilde{\chi}^{1}}{\tilde{\chi}^{N}}$.
    Thus, if $\overlap*{\tilde{\chi}^{1}}{\tilde{\chi}^{N}} = 0$,
    then the fringe visibility can be fully restored
    because the state
    \eqref{eq:tilde_varrho_infty:transition_amplitudes_unbiased_eta:sigma_not_fixed:symmetric:altnernative}
    contains cross terms
    \begin{align}
      \frac{\expval*{\tfrac{1}{2} \left(\operatorone_{\mathcal{M}} + \eta M\right) \ket*{+_{\mathcal{R}}} \bra*{+_{\mathcal{R}}}}}{\expval*{\ket*{+_{\mathcal{R}}} \bra*{+_{\mathcal{R}}}}} & = \frac{1}{2} \left(1 + \eta \cos \gamma\right),
    \end{align}
    although $\mathfrak{D}_{\mathcal{R}} = 0$.
    We thus have a case of nonerasing quantum erasure.
\end{description}
To conclude,
welcher-weg and quantum erasure measurement schemes
can, in general, neither demonstrate the recording of WWI,
nor can they unambiguously certify
the presence of quantum entanglement
\cite{Englert:1998kl}.
Although, in our description of the scenario
of Sec.~\ref{sec:quantum_erasure:symmetric_CCIs},
it is quantum entanglement that mediates path knowledge
and quantum interference that establishes control,
the question about the nonclassicality of
an actual coherent phase control experiment
cannot be answered on the grounds of
those measurement schemes alone.
The presence of quantum entanglement,
nonclassical correlations,
and true quantum coherent control
has to be certified by additional,
more stringent means
(as discussed in
Sec.~\ref{sec:certification_of_nonclassicality}
below).
However, this is not to say that the proposed
quantum erasure coherent phase control experiment
is useless.
First, it is of practical utility to restore
a maximum amount of phase controllability.
Second, in combination with an optimum welcher-weg measurement,
quantum erasure allows for demonstrating
wave-particle complementarity.
And third,
if an experiment managed to demonstrably produce the state
\eqref{eq:tilde_varrho_infty:transition_amplitudes_unbiased_eta:sigma_fixed:symmetric}
and if quantum erasure succeeded,
then the scenario would be indeed be quantum (even nonclassical),
and phase control would originate from
genuinely quantum interference.
The problem is that the state
\eqref{eq:tilde_varrho_infty:transition_amplitudes_unbiased_eta:sigma_fixed:symmetric}
may not be the only possible description.
From a foundational perspective,
one must show that there is no other,
classical explanation for the measurement results.

\section{Delayed Choice Coherent Control}
\label{sec:delayed_choice}

Below we continue the search for unconventional,
demonstrably quantum phase control scenarios.
The first scenario, the quantum erasure control scheme discussed above
(Sec.~\ref{sec:quantum_erasure}),
was based on the principle of complementarity.
The second scenario below will be based on
another hallmark of quantum mechanics: delayed choice.
As explained below,
the alternative scenario
solves a conceptual issue of quantum erasure phase control.

Complementarity in quantum mechanics implies
that the experimenter
can select from a set of measurement strategies
in order to observe
particle-like, wavelike,
or any arbitrary intermediate statistics
\cite{Bohr:1984fu,Englert:1998kl}.
This is in stark contrast to classical physics,
which invariably classifies every object
as either particle or wave;
never can an entity subscribe to
either property ad libitum.
Complementarity also applies in the setting of quantum erasure,
where a subensemble with wavelike statistics is chosen and where the timing of that choice is irrelevant
for the outcome of the experiment
\cite{Kim:2000th,Zeilinger2016}.
Quantum erasure can therefore be a particular realization of
the so-called delayed-choice gedanken experiment
\cite{Wheeler:1984cq,Kwiat:2004yu}.

Delayed-choice experiments
were proposed by Wheeler
to display the fundamental difference
that the principle of complementarity places
between classical and quantum physics.
In order to render it impossible
for an object to ``know'' \emph{a priori}
whether or not to display the wave or the particle property,
delayed-choice experiments aim to
sever any causal link between
the object and the interferometer until it has entered it.

Quantum erasure adheres to the aims
of delayed choice
by conditioning the interference property
on the outcome of a measurement performed on an auxiliary system,
the welcher-weg marker.
This allows for choosing \emph{a posteriori} between
the open interferometer configuration $o$
with associated particle-like statistics---%
in which case the tracing of paths may be possible---,
to the closed configuration $c$
and associated wavelike statistics---%
for which the path alternatives are
at least partially indistinguishable.
This also applies to the examples studied in
Sec.~\ref{sec:quantum_erasure:symmetric_CCIs},
where the choice of, say, the wave property,
could be postponed to a time
after the measurement of the output port label.
According to quantum theory,
this can occur in direct violation of classical realism
and thus suggests the nonclassicality of coherent control.
That is a significant step towards the goal to
create a demonstrably quantum control scenario.

Yet, one might deem that particular demonstration
as unsatisfactory,
since the preparation of the wave property---%
and thus phase control---%
is \emph{detached} from
the light-matter scattering processes of coherent control.
Rather, an additional process, quantum erasure,
has to be performed on top of
the original coherent control experiment.
For the scope of this article,
we would prefer a situation
in which the wave property emerges more directly
from the scattering processes.

As a solution we propose another
generalized coherent control experiment
in which the CCI delivers
the wave and the particle property directly
and in superposition \cite{Ionicioiu:2011fk}.
This experiment will allow one to
select either the open or the closed CCI configuration
on demand (and at any time)
by flipping a ``quantum switch''.
Prior to turning that switch,
it will be unknowable
whether the CCI output port statistics are that
of a wave or a particle,
\ie{}, whether phase control is possible or not,
a violation of classical realism.
In this regard, this proposed scenario
is entitled to be called quantum coherent control.

As we shall see,
the positions of the quantum switch are represented by
two (nearly) orthogonal states of the radiation field.
That means that,
in contrast to the quantum eraser,
the choice between wave and particle
is not a choice between two complementary,
noncommuting field observables,
but rather between two orthogonal projections.

\subsection{The Simultaneously Both Open And Closed Configuration}
\label{sec:delayed_choice:open_closed_config}

In the following,
we consider two field configurations,
\begin{subequations}%
  \begin{align}
    \ket*{\chi_{o}} & = g_{1 o}\bigl(a^{\dagger}\bigl[\hat{f}_{1 o}\bigr]\bigr) g_{2 o}\bigl(a^{\dagger}\bigl[\hat{f}_{2 o}\bigr]\bigr) \ket*{\mathrm{vac}}
    \intertext{and}
    \ket*{\chi_{c}} & = g_{1 c}\bigl(a^{\dagger}\bigl[\hat{f}_{1 c}\bigr]\bigr) g_{2 c}\bigl(a^{\dagger}\bigl[\hat{f}_{2 c}\bigr]\bigr) \ket*{\mathrm{vac}}.
  \end{align}
\end{subequations}%
\begin{subequations}%
  The former, $\ket*{\chi_{o}}$,
  is assumed to implement
  an open CCI $o$,
  \begin{align}%
    T_{1}\bigl[\hat{f}_{2 o}\bigr]\left(\eta \mu, \sigma \nu\right) & \propto \delta_{\eta, -\sigma}, \\
    T_{N}\bigl[\hat{f}_{1 o}\bigr]\left(\eta \mu, \sigma \nu\right) & \propto \delta_{\eta, \sigma}
    \intertext{whereas the latter, $\ket*{\chi_{c}}$,
    implements a closed CCI $c$,}
    T_{1}\bigl[\hat{f}_{2 c}\bigr]\left(+1 \mu, \sigma \nu\right) & = T_{1}\bigl[\hat{f}_{2 c}\bigr]\left(-1 \mu, \sigma \nu\right), \\
    T_{N}\bigl[\hat{f}_{1 c}\bigr]\left(+1 \mu, \sigma \nu\right) & = - T_{N}\bigl[\hat{f}_{1 c}\bigr]\left(-1 \mu, \sigma \nu\right).
  \end{align}%
\end{subequations}%
These relations are adopted from
Eqs.~\eqref{eqs:transition_amplitudes_mutually_exclusive} and
\eqref{eqs:transition_amplitudes_unbiased_eta},
respectively.
In both cases,
the incoming light
is prepared in a semi-classical
coherent state:
\begin{subequations}%
  \begin{align}
    g_{i o}\left(x\right) & = \exp\bigl(\norm*{f_{i o}} x - \norm*{f_{i o}}{}^{2} / 2\bigr), \\
    g_{i c}\left(x\right) & = \exp\bigl(\norm*{f_{i c}} x - \norm*{f_{i c}}{}^{2} / 2\bigr),
    \label{eqs:gi:coherent_state}
  \end{align}
\end{subequations}%
$i = 1,2$.
These states are eigenstates of
the photon annihilation operator,
and thus WWM does not occur \cite{Gong:2010dq}.
We have
\begin{subequations}%
  \begin{align}
    \textstyle \norm*{f_{2 o}}^{-1} \ket*{\tilde{\chi}^{1}_{o}} = \norm*{f_{1 o}}^{-N} \ket*{\tilde{\chi}^{N}_{o}} & = \textstyle \ket*{\chi_{o}}, \\
    \textstyle \norm*{f_{2 c}}^{-1} \ket*{\tilde{\chi}^{1}_{c}} = \norm*{f_{1 c}}^{-N} \ket*{\tilde{\chi}^{N}_{c}} & = \textstyle \ket*{\chi_{c}},
  \end{align}
\end{subequations}%
similar to Eq.~\eqref{eq:no_ww_marking}.
Coherent states are
explicitly chosen with the intention of
excluding postselection procedures
like optimum welcher-weg measurements or quantum erasure
from the experiment.
That is, we have to realize delayed choice
in a different way, for reasons explained below.

In order to carry out
Wheeler's original delayed choice experiment
in the setting of coherent phase control,
we would have to
both \emph{randomize} and \emph{postpone}
the choice between an open and closed CCI.
We could certainly do the former,
\ie{}, randomize the configuration by
deciding at random between
either $\ket*{\chi_{c}}$ or $\ket*{\chi_{o}}$
as the initial state of the radiation field.
However, since we decided to exclude WWM,
the CCI does not allow us to postpone that decision.
This is a major stumbling block,
since randomization alone would not suffice
if the alternatives ``phase control'' and ``no control''
were chosen in advance.%
\footnote{Mind that it is irrelevant whether or not
the choice is known to us---%
it has to be unknowable \emph{physically}.}

The CCI is thus \emph{not} compatible with
Wheeler's delayed choice experiment.
The only way of physically delaying the choice
is to deviate both
from the traditional coherent control schemes
and from Wheeler's original idea.
The goal is to guarantee the uncertainty
and the unknowability
of the CCI's interference property.
One solution is to quantize the choice
between open and closed \cite{Ionicioiu:2011fk}
by preparing the field
in a quantum-coherent superposition
between the complementary configurations,
\ie{} in
\begin{gather}
  \ket*{\chi} = \frac{\ket*{\chi_{o}} + \ket*{\chi_{c}}}{\sqrt{2 \left(1 + \overlap*{\chi_{o}}{\chi_{c}}\right)}},
  \label{eq:delayed_choice_superposition}
  \intertext{where}
  \overlap*{\chi_{o}}{\chi_{c}} = \textstyle \exp\left[-\frac{1}{2} \left(\norm*{f_{1 o}}^{2} + \norm*{f_{2 o}}^{2} + \norm*{f_{1 c}}^{2} + \norm*{f_{2 c}}^{2}\right)\right].
  \label{eq:chio_chic_overlap}
\end{gather}
This state is quite unusual.
It is composed out of
two semi-classical, distinct states,
each of which a direct product of
two coherent states in, in total,
four mutually orthogonal modes, \ie{},
$\ket*{\chi}$ is a ``cat state''
\cite{Nielsen:2000qq}.
It is an entangled coherent state
with nonclassical correlations of the GHZ type
\cite{Sanders:1992zr,Jeong:2006kx,Greenberger:1990oq,Guhne:2009vn}.
From an experimental point of view,
handling $\ket*{\chi}$ may prove difficult:
the state will be very sensitive to decoherence,
and its creation requires a special,
highly nonlinear MZI
that, at the time of this writing,
is not readily available.%
\footnote{Specifically,
an entangled coherent state with a single frequency, $\omega$,
is produced by letting laser light interfere with itself
in a MZI in which one arm
has been replaced with a strong
Kerr nonlinearity \cite{Sanders:1992zr}.
Yet another nonlinear optical process is required
to transform this state into an
entangled coherent states with two frequencies,
$\omega$ and $N \omega$,
as required for $\ket*{\chi}$.}

\subsection{Quantum Delayed Choice}
\label{sec:delayed_choice:quantum_delayed_choice}

However, assuming a system prepared in the state
\begin{align}
  \varrho\left(-\infty\right) & = \frac{1}{\sum_{\nu \in \mathcal{D}_{-}} \delta_{M_{\nu} \sigma}} \sum_{\nu \in \mathcal{D}_{-}} \ket*{\sigma \nu} \bra*{\sigma \nu} \otimes \ket*{\chi} \bra*{\chi}
  \label{eq:varrho:qdc}
\end{align}
with $\sigma$ either equal to $+1$ or to $-1$,
we can expect
(following a similar calculation as in
Appendix~\ref{app:derivation_total_density_operator})
the final state
to become approximately equal to
\begin{multline}
  \tilde{\varrho}\left(\infty\right) = \textstyle \Bigl[\sum_{\nu} \left(\ket*{\tilde{\psi}_{o, \sigma \nu}} \ket*{\chi_{o}} + \ket*{\tilde{\psi}_{c, \sigma \nu}} \ket*{\chi_{c}}\right) \left(\bra*{\tilde{\psi}_{o, \sigma \nu}} \bra*{\chi_{o}} + \bra*{\tilde{\psi}_{c, \sigma \nu}} \bra*{\chi_{c}}\right)\Bigr] \\
  \textstyle \times \Bigl[\sum_{\nu} \bigl(\overlap*{\tilde{\psi}_{o, \sigma \nu}}{\tilde{\psi}_{o, \sigma \nu}} + \overlap*{\tilde{\psi}_{o, \sigma \nu}}{\tilde{\psi}_{c, \sigma \nu}} \overlap*{\chi_{o}}{\chi_{c}} \\
  + \overlap*{\tilde{\psi}_{c, \sigma \nu}}{\tilde{\psi}_{o, \sigma \nu}} \overlap*{\chi_{c}}{\chi_{o}} + \overlap*{\tilde{\psi}_{c, \sigma \nu}}{\tilde{\psi}_{c, \sigma \nu}}\bigr)\Bigr]^{-1} ,
  \label{eq:tilde_varrho_infty:qdc}
\end{multline}
where the sums
run over $\nu \in \mathcal{D}_{-}$.
With the equivalent of
Eq.~\eqref{eq:symmetric_conditions:gamma:again},
\begin{align}
  \norm{f_{2 c}} \, T_{1}\bigl[\hat{f}_{2 c}\bigr]\left(+1 \mu, \sigma \nu\right) = \ee^{- \ii \gamma_{c}} \, \norm{f_{1 c}}^{N} \, T_{N}\bigl[\hat{f}_{1 c}\bigr]\left(+1 \mu, \sigma \nu\right)
\end{align}
and
$\gamma_{c} = \gamma\bigl[\hat{f}_{1 c}, \hat{f}_{2 c}\bigr]\left(+1, \sigma\right)$
for all $\mu \in \mathcal{D}_{+}$
and $\nu \in \mathcal{D}_{-}$,
the wavelike state $\ket*{\tilde{\psi}_{c, \sigma \nu}}$
becomes
\begin{subequations}%
  \begin{multline}%
    \ket*{\tilde{\psi}_{c, \sigma \nu}} = 2 \norm{f_{2 c}} \, \smashoperator{\sumint_{\mathcal{D}_{+}}} \mathrm{d}\mu \, T_{1}^{\vphantom{*}}\bigl[\hat{f}_{2 c}\bigr]\left(+1 \mu, \sigma \nu\right) \\
    \times \ee^{\ii \gamma_{c} / 2} \left(\cos \frac{\gamma_{c}}{2} \ket*{+ 1 \mu} - \ii \sin \frac{\gamma_{c}}{2} \ket*{-1 \mu}\right).
  \end{multline}
  For the particle-like state, on the other hand,
  we arrive at
  \begin{gather}
    \ket*{\tilde{\psi}_{o, \sigma \nu}} = \norm{f_{2 o}} \, \smashoperator{\sumint_{\mathcal{D}_{+}}} \mathrm{d}\mu \, T_{1}^{\vphantom{*}}\bigl[\hat{f}_{2 o}\bigr]\left(-\sigma \mu, \sigma \nu\right) \, \frac{1}{\sqrt{2}} \left(\ket*{-\sigma \mu} - \ee^{\ii \gamma_{o}}\ket*{\sigma \mu}\right),
  \end{gather}%
\end{subequations}%
if we impose the requirement that
\begin{align}
  \norm{f_{2 o}} \, T_{1}\bigl[\hat{f}_{2 o}\bigr]\left(-\sigma \mu, \sigma \nu\right) = - \ee^{- \ii \gamma_{o}} \, \norm{f_{1 o}}^{N} \, T_{N}\bigl[\hat{f}_{1 o}\bigr]\left(\sigma \mu, \sigma \nu\right)
\end{align}
with
$\gamma_{o} = \gamma\bigl[\hat{f}_{1 o}, \hat{f}_{2 o}\bigr]\left(\sigma\right)$
for all $\mu \in \mathcal{D}_{+}$
and $\nu \in \mathcal{D}_{-}$.

If an experiment succeeded in creating
the state $\tilde{\varrho}\left(\infty\right)$
of Eq.~\eqref{eq:tilde_varrho_infty:qdc},
then, according to quantum theory,
the interference property would be
physically guaranteed to be random:
Eq.~\eqref{eq:tilde_varrho_infty:qdc} describes
a \emph{coherent} ``blend'' \cite{Englert:1998kl}
or ``morphing'' \cite{Ionicioiu:2011fk}
between particle- and wavelike output port statistics
that is conditioned on the state of the radiation field.
Therefore,
a measurement of the field
allows for a delayed-choice determination of whether
the system is in
$\ket*{\tilde{\psi}_{c, \sigma \nu}}$
or $\ket*{\tilde{\psi}_{o, \sigma \nu}}$,
thus whether the CCI displays phase control
or no control.%
\footnote{Note that $\ket*{\chi_{o}}$
and $\ket*{\chi_{c}}$ are not orthogonal,
\cf{} Eq.~\eqref{eq:chio_chic_overlap}.}
Such a coherent control experiment would be
an example of ``quantum delayed-choice''
\cite{Ionicioiu:2011fk,%
Tang:2012kx,%
Peruzzo:2012mw,Kaiser:2012fk,%
Roy:2012uq,Auccaise:2012kx,%
Stassi:2012fk}.

The effect of the coherent control process
is reminiscent of ``entanglement swapping''
\cite{Zukowski:1993kq,Bose:1998mz}.
Whereas, initially,
the quantum correlations reside
exclusively within the field
(between coherent states in very different field modes),
the CCI redistributes them such that,
in the final state \eqref{eq:tilde_varrho_infty:qdc},
quantum entanglement appears instead
between the matter and the radiation.
The similarity to
the entangling swapping protocol means that the
quantum delayed choice coherent control scenario
relies on quantum entanglement as a resource.

\subsection{A Comparison between Quantum Erasure And Quantum Delayed Choice}
\label{sec:delayed_choice:comparison_with_quantum_erasure}

It is worthwhile to compare
these proposed coherent control implementations of
quantum erasure and quantum delayed choice.
In the following, we list
a number of differences and similarities:
\begin{enumerate}[label=(\roman{enumi})]
  \item
    In both cases,
    the interference property
    (\ie{}, phase control or no control)
    can be prepared after the
    output port of the CCI has been measured \ie{},
    both approaches implement
    Wheeler's delayed choice experiment.
  \item
    The quantum erasure coherent control scenario
    produces entanglement,
    whereas the quantum delayed choice coherent control scenario
    relies on quantum entanglement as a resource.
    However, this entanglement is not consumed,
    but is transformed and
    used to deliver the wave and the particle property
    in superposition.
    In doing so,
    quantum delayed choice coherent control can
    demonstrate both complementarity and delayed choice
    in a fashion that is closer to the spirit of coherent control
    than it would be possible in a quantum eraser experiment.
  \item
    In the quantum erasure case,
    it is the choice between
    two complementary measurements
    on the light field
    that decides between phase control or no control.
    In quantum delayed choice,
    the choice between these alternative
    is \emph{random} and
    dictated by the outcome of
    the measurement of a \emph{single} observable.%
    \footnote{As originally discussed
    in Ref.~\citealp{Ionicioiu:2011fk},
    this result is at variance with Bohr's original view on
    complementarity, according to which
    the wave and the particle property manifest themselves
    only in complementary measurements
    corresponding to complementary experiments.
    In quantum delayed choice, however,
    both properties can be observed by
    measuring a single observable
    in a single experiment.
    The authors of Ref.~\citealp{Ionicioiu:2011fk}
    therefore propose to revise the phrasing of the principle
    and speak of complementarity of experimental data
    rather than complementary experiments.}
    This is due to the fact that
    the interference property
    is set by a quantum switch
    whose position
    is unknowable and nondeterministic, 
    a consequence of
    the fundamental randomness
    of nature as manifest in quantum theory.
  \item
    Yet, how can a quantum delayed choice coherent control
    experiment demonstrate that its outcome is fundamentally random?
    It cannot, that is, not on its own.
    Like quantum erasure coherent control,
    quantum delayed choice coherent control alone is
    insufficient to rigorously certify nonclassicality.
    For illustration,
    consider the case in which the amplitudes
    $\norm{f_{1 o}}$, $\norm{f_{2 o}}$,
    $\norm{f_{1 c}}$, and $\norm{f_{2 c}}$
    are macroscopic.
    Then $\ket*{\chi_{o}}$ and $\ket*{\chi_{c}}$
    are readily distinguishable.
    Let us predict the outcome of
    a joint measurement of
    the CCI output ports and
    the population of a selection of modes of the radiation field.
    If these modes correspond to the closed configuration,
    \ie{}, if they are associated with
    the amplitude functions $f_{1 c}$ or $f_{2 c}$,
    then the detection of one or more photons
    will be perfectly correlated with wavelike
    output port statistics and the absence of photons will
    be perfectly correlated with particle-like
    output port statistics.
    One will find that
    the case that applies is unpredictable.
    However, an experiment that establishes just that
    is not able to discriminate the observed statistics
    against those of the corresponding classical situation:
    a simple Bernoulli trial (\ie{}, essentially a coin flip)
    between the open and the closed CCI configuration.
    The experiment would not even demonstrate Wheeler's original
    version of delayed choice,
    because the statistics of the output port populations
    could as well have been fully determined from the moment
    of the field's creation,
    without any change to the results.
\end{enumerate}
The last item on this list
illustrates that,
for a valid, credible,
and meaningful demonstration of nonclassicality,
and from a foundational perspective, 
\emph{it is generally insufficient to acknowledge that
observations are consistent with quantum theory.
Rather, it is necessary to show that they are inconsistent
with all classical descriptions.}
The next section deals with this issue.

\section{Towards Rigorous Experimental Certification of
The Nonclassicality of Coherent Phase Control}
\label{sec:certification_of_nonclassicality}

The above discussion makes clear that neither the quantum erasure
nor the quantum delayed choice scenario
(Secs.~\ref{sec:quantum_erasure} and
\ref{sec:delayed_choice}, respectively)
are stringent enough to ensure
truly nonclassical control scenarios.
In this section, we show how the nonclassicality
of these scenarios can be certified,
\ie{}, we clarify to what extent classical explanations
of the experimental observations can be excluded.
This is the final step in demonstrating
quantum coherent control.

Such a certification of nonclassicality and related problems
are of course not new.
Rather, these issues have led to
major advances in understanding the foundations
and basic principles of quantum mechanics.
Specifically, Bell test experiments have been devised and executed
in order to show the untenability of local realism
\cite{Bell:1964zt,Bell:1966zr,Clauser:1969dq,Guhne:2009vn},
and Leggett-Garg test experiments have been performed to refute
classical theories combining macrorealism and
noninvasive measurability
\cite{Leggett:1985eu,Leggett:2008wd,Emary:2014rr,Robens:2015aa}.
These (and most other) tests assume the existence of
classical probability distributions
for the measurement statistics,
in particular, joint distributions
that account for all inquired marginal probabilities,
independent of the measurement.
From this assumption,
constraints on the statistics of certain
measurement outcomes can be derived
that are usually expressed
in terms of an inequality
(\cf{} Bell-CHSH inequalities
and Leggett-Garg inequalities),
violations of which are then
rigorous indicators of nonclassicality
 (barring loopholes).
In these derivations,
the formalism of quantum mechanics
(or any other nonclassical theory)
is nowhere used.
This is crucial,
because bounds on classical behavior
can only be acquired from within classical physics.
Moreover, in a discussion about the necessity of
nonclassical theories to describe certain aspects of nature,
arguments originating from within quantum mechanics
(or other nonclassical theories)
are not allowed.
Nobody who refutes nonclassicality would accept them.

\subsection{A Bell Inequality Test for Coherent Control}
\label{sec:certification_of_nonclassicality:bell_test}

The ultimate purpose of Bell inequality experiments
has been to settle the question of
whether nature obeys \emph{local realism}.
\emph{Realism} is mentioned at the beginning of Sec.~\ref{sec:CCI};
it is the idea that all properties
and values of a system exist prior to,
and independent of, their measurement.
Quantum theory refutes realism,
classical physics does not.
\emph{Locality}, on the other hand,
refers to the assumption
that interaction between
spatially distant physical systems is not possible.
Although classical and quantum mechanics are local theories,
both of them can predict nonlocal correlations
that stem from local processes like
coupling, interaction, and scattering.
If nature satisfied local realism, then
two space-like separated measurement events would never
be able to influence one another, since a signal heralding the
measurement outcome from one event to the other
would have to travel faster than the speed of light.
Such influence is excluded by special relativity.

Numerous Bell test experiments have been conducted
\cite{Aspect:1982rr,Weihs:1998ul,Rowe:2001qq,Groblacher:2007la,Matsukevich:2008nx,Hensen:2015lf,Giustina:2015sl,Shalm:2015sl},
some with considerable violations of the inequality. Experiments
usually implement tests based on the 
Clauser-Horne-Shimony-Holt (CHSH) inequality \cite{Clauser:1969dq,Guhne:2009vn}
that is an extension of Bell's original inequality.
Both inequalities provide bounds for
the correlations between measurement outcomes
obtained from a bipartite system.
The CHSH inequality, in particular,
is derived for the case where
two observables, each with two possible outcomes,
are measured on both subsystems.
Here we use this inequality where
the first subsystem is the pair of output ports of the CCI
and the second subsystem is a collection of relevant modes
of the radiation field.

\subsection{The Case of Quantum Erasure Coherent Control}
\label{sec:certification_of_nonclassicality:quantum_erasure}

For the certification of nonclassicality
in the quantum erasure coherent control scenario
of Sec.~\ref{sec:quantum_erasure:symmetric_CCIs},
we propose the following CHSH inequality:
\begin{gather}
  \abs*{\expval*{B}} = \left\lvert\expval*{O_{\mathcal{M}}^{\vphantom{\prime}} \otimes O_{\mathcal{R}}^{\vphantom{\prime}}} + \expval*{O_{\mathcal{M}}' \otimes O_{\mathcal{R}}^{\vphantom{\prime}}} + \expval*{O_{\mathcal{M}}^{\vphantom{\prime}} \otimes O_{\mathcal{R}}'} - \expval*{O_{\mathcal{M}}' \otimes O_{\mathcal{R}}'}\right\rvert \le 2,
  \label{eq:bell_test:quantum_erasure}
\end{gather}
where we introduced the Bell operator $B$
and the four dichotomous observables
$O_{\mathcal{M}}^{\vphantom{\prime}}$, $O_{\mathcal{M}}'$,
$O_{\mathcal{R}}^{\vphantom{\prime}}$, and $O_{\mathcal{R}}'$, defined below.
One can then easily verify that the inequality
\eqref{eq:bell_test:quantum_erasure} holds for classical physics.
If we denote
$o_{\mathcal{M}}^{\vphantom{\prime}}$, $o_{\mathcal{M}}'$,
$o_{\mathcal{R}}^{\vphantom{\prime}}$, and $o_{\mathcal{R}}'$
the possible outcomes ($\pm 1$)
of the observables of the Bell operator, then clearly
$-2 \le o_{\mathcal{M}}^{\vphantom{\prime}} o_{\mathcal{R}}^{\vphantom{\prime}} + o_{\mathcal{M}}' o_{\mathcal{R}}^{\vphantom{\prime}} + o_{\mathcal{M}}^{\vphantom{\prime}} o_{\mathcal{R}}' - o_{\mathcal{M}}' o_{\mathcal{R}}' \le 2$.
However, in quantum mechanics,
the largest possible expectation value of a Bell-CHSH operator
is the Tsirelson bound, $2 \sqrt{2}$
\cite{Tsirelson:1980fk,Barrett:2005uq}.
We shall see that,
in quantum erasure coherent control,
the inequality \eqref{eq:bell_test:quantum_erasure}
can be violated
and even equality,
$\abs*{\expval*{B}} = 2 \sqrt{2}$,
be achieved.
In accord with Bell's theorem,
this implies that the physics does not abide by local realism.

\begin{figure}
  \centering%
  \begingroup%
    \iftikz%
      \tikzexternalenable%
      \input{tikz/belltest_quantumerasure_expvalB.tex}%
    \else%
      \includegraphics{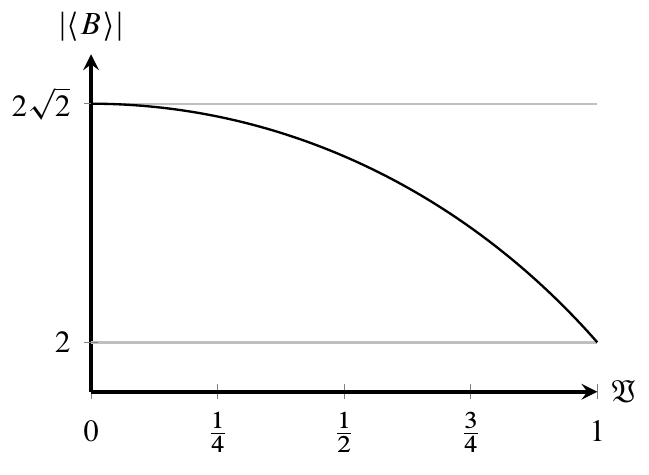}%
    \fi%
  \endgroup%
  \caption{%
    Absolute Bell expectation value $\abs*{\expval*{B}}$
    with respect to the state
    \eqref{eq:tilde_varrho_infty:transition_amplitudes_unbiased_eta:sigma_fixed:symmetric}
    as a function of the visibility $\mathfrak{V}$.
    Maximum violation, $\abs*{\expval*{B}} = 2 \sqrt{2}$,
    occurs for $\mathfrak{V} = 0$.
  }%
  \label{fig:bell_test:quantum_erasure:expvalB}%
\end{figure}

Consider now the four observables
$O_{\mathcal{M}}^{\vphantom{\prime}}$,
$O_{\mathcal{M}}'$,
$O_{\mathcal{R}}^{\vphantom{\prime}}$,
and $O_{\mathcal{R}}'$.
The first two are sensitive only to the CCI output labels:
\begin{subequations}%
  \begin{align}%
    O_{\mathcal{M}}^{\vphantom{\prime}} & = \frac{1}{\sqrt{2 - \mathfrak{V}^{2}}} \, \left[- \sqrt{1 - \mathfrak{V}^{2}} \, \sigma_{\mathcal{M}}^{1} - \sin\left(\phi + \gamma\right) \, \sigma_{\mathcal{M}}^{2} + \cos\left(\phi + \gamma\right) \, \sigma_{\mathcal{M}}^{3}\right], \\
    O_{\mathcal{M}}' & = \frac{1}{\sqrt{2 - \mathfrak{V}^{2}}} \, \left[\sqrt{1 - \mathfrak{V}^{2}} \, \sigma_{\mathcal{M}}^{1} - \sin\left(\phi + \gamma\right) \, \sigma_{\mathcal{M}}^{2} + \cos\left(\phi + \gamma\right) \, \sigma_{\mathcal{M}}^{3}\right],
  \end{align}%
  \label{eqs:bell_test:quantum_erasure:OM_OMp}%
\end{subequations}%
where
\begin{subequations}%
  \begin{align}%
    \sigma_{\mathcal{M}}^{1} & = \textstyle \sumint \mathrm{d}\mu \, \left(\ket*{+1 \mu}\bra*{-1 \mu} + \ket*{-1 \mu}\bra*{+1 \mu}\right), \\
    \sigma_{\mathcal{M}}^{2} & = \textstyle \sumint \mathrm{d}\mu \, \left(-\ii \ket*{+1 \mu}\bra*{-1 \mu} + \ii \ket*{-1 \mu}\bra*{+1 \mu}\right), \\
    \sigma_{\mathcal{M}}^{3} & = M = \textstyle \sumint \mathrm{d}\mu \, \left(\ket*{+1 \mu}\bra*{+1 \mu} - \ket*{-1 \mu}\bra*{-1 \mu}\right)
  \end{align}%
  \label{eqs:pseudospin_operators}%
\end{subequations}%
are Hermitian, unitary pseudospin operators
(the equivalents of the Pauli spin matrices).
The visibility $\mathfrak{V}$ is the same as in
Eq.~\eqref{eq:visibility_coherence:transition_amplitudes_unbiased_eta:sigma_fixed:symmetric}
(page \pageref{eq:visibility_coherence:transition_amplitudes_unbiased_eta:sigma_fixed:symmetric}),
and the phases $\phi$ and $\gamma$
are defined in Eqs.~\eqref{eq:phi}
and \eqref{eq:symmetric_conditions:gamma:again}
(pages \pageref{eq:phi} and \pageref{eq:symmetric_conditions:gamma:again},
respectively).
The second two observables are sensitive only to the radiation:
\begin{subequations}%
  \begin{align}%
    O_{\mathcal{R}}^{\vphantom{\prime}} & = \frac{1}{\sqrt{1 - \mathfrak{V}^{2}}} \, \left(\frac{\textstyle \ket*{\tilde{\chi}^{N}} \bra*{\tilde{\chi}^{N}}}{\textstyle \overlap*{\tilde{\chi}^{N}}{\tilde{\chi}^{N}}} - \frac{\textstyle \ket*{\tilde{\chi}^{1}} \bra*{\tilde{\chi}^{1}}}{\textstyle \overlap*{\tilde{\chi}^{1}}{\tilde{\chi}^{1}}}\right)
    \shortintertext{and}
    O_{\mathcal{R}}' & = \ket*{+_{\mathcal{R}}} \bra*{+_{\mathcal{R}}} - \ket*{-_{\mathcal{R}}} \bra*{-_{\mathcal{R}}},
  \end{align}%
  \label{eqs:bell_test:quantum_erasure:OR_ORp}%
\end{subequations}%
where $\ket*{\pm_{\mathcal{R}}}$ are
the orthonormal radiative states derived
in Sec.~\ref{sec:quantum_erasure:symmetric_CCIs}
[\cf{} Eq.~\eqref{eq:schmidt_decomposition:radiative_states}].
With these operators and
the final state $\tilde{\varrho}\left(\infty\right)$
of the quantum erasure coherent control scenario
given in Eq.~\eqref{eq:tilde_varrho_infty:transition_amplitudes_unbiased_eta:sigma_fixed:symmetric},
we have
\begin{gather}
  \abs*{\expval*{B}} = \abs*{\Tr\left[B \tilde{\varrho}\left(\infty\right)\right]} = 2 \sqrt{2 - \mathfrak{V}^2} \, ,
  \label{eq:bell_test:quantum_erasure:expvalB}%
\end{gather}
see Fig.~\ref{fig:bell_test:quantum_erasure:expvalB}.
Clearly $\abs*{\expval*{B}}$ reaches the Tsirelson bound $2 \sqrt{2}$
for vanishing visibility,
\ie{}, $\mathfrak{V} = 0$,
a result that is impossible classically.

Several points are worth noting.
First, the field observables
Eq.~\eqref{eqs:bell_test:quantum_erasure:OR_ORp}
are known from Sec.~\ref{sec:quantum_erasure:symmetric_CCIs}.
In particular, $O_{\mathcal{R}}^{\vphantom{\prime}}$
is an optimal observable
for a welcher-weg measurement
[\cp{} Eq.~\eqref{eq:OR:optimum_ww_measurement}],
whereas $O_{\mathcal{R}}'$ is optimal
for quantum erasure
[\cp{} Eq.~\eqref{eq:OR:optimum_quantum_erasure}].
In this sense,
$O_{\mathcal{R}}^{\vphantom{\prime}}$
and $O_{\mathcal{R}}'$
are maximally complementary.
Second,
although there is an intimate relation,
testing the Bell inequality \eqref{eq:bell_test:quantum_erasure}
is \emph{not} the same as combining the results from
an optimum welcher-weg measurement and
from quantum erasure.
For a proper Bell test,
the field observables have to be joint
in a particular way with material observables,
\eg{} $O_{\mathcal{M}}^{\vphantom{\prime}}$
and $O_{\mathcal{M}}'$ from above,
that are different from $M$
and also probe coherences between the
CCI output ports.
Third, the maximum Bell violation
is achieved for vanishing visibility, $\mathfrak{V} = 0$,
which, according to the discussion in
Sec.~\ref{sec:quantum_erasure:symmetric_CCIs},
is equivalent to
a maximum of radiative distinguishability $\mathfrak{D}_{\mathcal{R}}$
and a maximum of quantum entanglement $\mathfrak{c}$.%
\footnote{Recall that, for $\mathfrak{V} = 0$,
the final states of the radiation field,
$\ket*{\tilde{\chi}^{1}}$ and $\ket*{\tilde{\chi}^{N}}$,
are orthogonal and thus fully distinguishable.}
This is in agreement with
the fact that all states with maximum Bell violation
are the equivalent of a direct sum of
maximally quantum entangled singlet states
\cite{Popescu:1992kq}.
Finally,
due to the presence of coherences between
$\ket*{\tilde{\chi}^{1}}$
and $\ket*{\tilde{\chi}^{N}}$,
a measurement of
the field observable $O_{\mathcal{R}}'$
may not be possible.
Therefore, 
in a manner similar to what is done in
Sec.~\ref{sec:quantum_erasure:photon_threshold},
one could instead consider another set of operators
that substitute $O_{\mathcal{R}}^{\vphantom{\prime}}$ and
$O_{\mathcal{R}}'$ and that are given
in terms of displaced photon threshold measurement observables.
We leave this as an exercise to the interested reader.

\subsection{The Case of Quantum Delayed Choice Coherent Control}

Consider now 
the quantum delayed choice coherent control scenario
introduced in Sec.~\ref{sec:delayed_choice:quantum_delayed_choice}.
As in the case of quantum erasure coherent control above,
we propose a Bell-CHSH inequality test for
certifying nonclassicality.
The Bell operator $B$ is, however,
different from that used in the previous section;
it needs to be adapted to the physical situation
described by the state \eqref{eq:tilde_varrho_infty:qdc}.
We restrict attention, as an example,
to a special case where particle-like and wavelike statistics occur
with equal probabilities,
\ie{} we add the requirement that
\begin{align}
  2 \norm{f_{2 c}} \, T_{1}\bigl[\hat{f}_{2 c}\bigr]\left(+1 \mu, \sigma \nu\right) = \ee^{\ii \kappa} \, \norm{f_{2 o}} \, T_{1}\bigl[\hat{f}_{2 o}\bigr]\left(-\sigma \mu, \sigma \nu\right)
\end{align}
with
$\kappa = \kappa\bigl[\hat{f}_{2 o}, \hat{f}_{2 c}\bigr]\left(\sigma\right)$
for all $\mu \in \mathcal{D}_{+}$
and $\nu \in \mathcal{D}_{-}$.
Further, we fix the relationship between
the relative phases of the transition amplitudes
by setting
\begin{align}
  \gamma_{o} & = - \phi - \frac{\pi}{2}, &
  \sigma & = -1, \\
  \gamma_{c} & = \phi, &
  \kappa & = - 2 \phi - \pi,
\end{align}
independently of $\nu \in \mathcal{D}_{-}$.
Here, the new free phase parameter $\phi$
replaces all other phase parameters,
and the input port has been set to $-1$.
With these assumptions,
the state \eqref{eq:tilde_varrho_infty:qdc} becomes
\begin{multline}
  \tilde{\varrho}\left(\infty\right) = \left(\textstyle \norm{f_{2 o}}^{2} \sumint_{\mathcal{D}_{+}} \mathrm{d}\mu \sum_{\nu \in \mathcal{D}_{-}} \abs[\big]{T_{1}^{\vphantom{*}}\bigl[\hat{f}_{2 o}\bigr]\left(+1 \mu, -1 \nu\right)}^{2}\right)^{-1} \\ \times \sum_{\nu \in \mathcal{D}_{-}} \frac{\left(\ket*{\tilde{\psi}_{o, -1 \nu}} \ket*{\chi_{o}} + \ket*{\tilde{\psi}_{c, -1 \nu}} \ket*{\chi_{c}}\right) \left(\bra*{\tilde{\psi}_{o, -1 \nu}} \bra*{\chi_{o}} + \bra*{\tilde{\psi}_{c, -1 \nu}} \bra*{\chi_{c}}\right)}{2 - \frac{1}{\sqrt{2}} \overlap*{\chi_{o}}{\chi_{c}} \left[\cos\phi + \cos\left(2 \phi\right) - \sin\phi\right]},
  \label{eq:tilde_varrho_infty:qdc:simple}
\end{multline}
where
\begin{subequations}%
  \begin{gather}%
    \ket*{\tilde{\psi}_{o, -1 \nu}} = \norm{f_{2 o}} \, \smashoperator{\sumint_{\mathcal{D}_{+}}} \mathrm{d}\mu \, T_{1}^{\vphantom{*}}\bigl[\hat{f}_{2 o}\bigr]\left(+1 \mu, -1 \nu\right) \, \frac{1}{\sqrt{2}} \Bigl(\ket*{+1 \mu} + \ii \ee^{- \ii \phi} \ket*{-1 \mu}\Bigr)
  \end{gather}%
  and
  \begin{multline}%
    \ket*{\tilde{\psi}_{c, -1 \nu}} = \norm{f_{2 o}} \, \smashoperator{\sumint_{\mathcal{D}_{+}}} \mathrm{d}\mu \, T_{1}^{\vphantom{*}}\bigl[\hat{f}_{2 o}\bigr]\left(+1 \mu, -1 \nu\right) \\
    \times \ee^{- 3 \ii \phi / 2} \Bigl(- \cos \frac{\phi}{2} \ket*{+ 1 \mu} + \ii \sin \frac{\phi}{2} \ket*{-1 \mu}\Bigr)
  \end{multline}%
\end{subequations}%
are the particle-like and the wavelike CCI state, respectively.
The overlap $\overlap*{\chi_{o}}{\chi_{c}}$ is real
and the same as in Eq.~\eqref{eq:chio_chic_overlap}.

\begin{figure}
  \centering%
  \begingroup%
    (a)
    \iftikz%
      \tikzexternalenable%
      \input{tikz/belltest_quantumdelayedchoice_expvalB_full.tex}%
    \else%
      \includegraphics{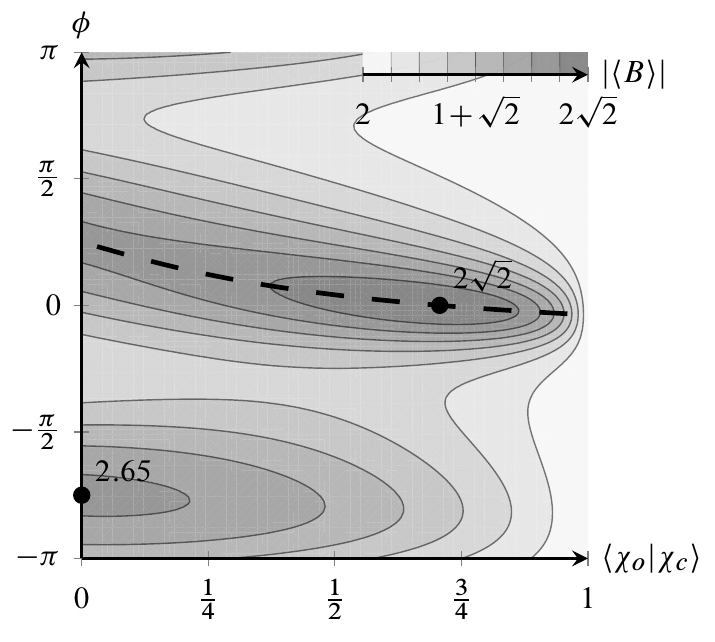}%
    \fi%
  \endgroup%
  \\%
  \centering%
  \begingroup%
    (b)
    \iftikz%
      \tikzexternalenable%
      \input{tikz/belltest_quantumdelayedchoice_expvalB_simple.tex}%
    \else%
      \includegraphics{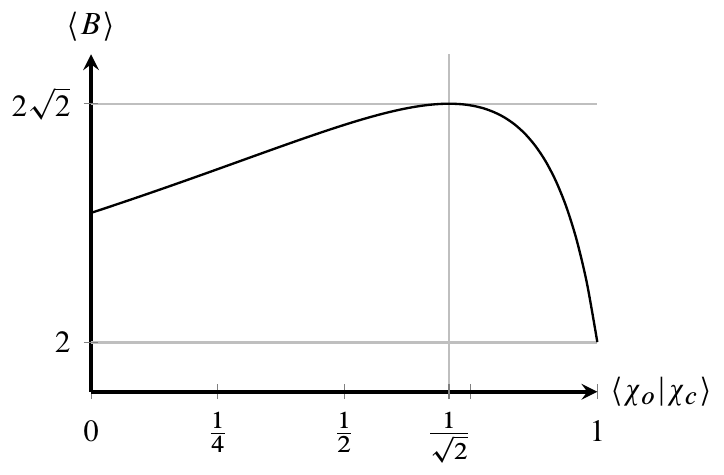}%
    \fi%
  \endgroup%
  \caption{%
    (a) Contour plot of the absolute Bell expectation value
    $\abs*{\expval*{B}}$ with respect to the state
    \eqref{eq:tilde_varrho_infty:qdc:simple}
    [from
    Eq.~\eqref{eq:bell_test:quantum_delayed_choice:expvalB:full}]
    as a function of the phase $\phi$ and
    the overlap $\overlap*{\chi_{o}}{\chi_{c}}$.
    Higher values have a darker shade (see scale in upper right corner).
    The dashed black line is the position of the global maximum
    of $\abs*{\expval*{B}}$ with respect to $\phi$,
    but for fixed $\overlap*{\chi_{o}}{\chi_{c}}$.
    This shows how, for given $\overlap*{\chi_{o}}{\chi_{c}}$,
    one would have to choose $\phi$
    to achieve the highest possible Bell violation.
    Local maxima are labeled and indicated with a dot.
    Especially noteworthy is the saturation
    of the Bell violation at the Tsirelson bound $2 \sqrt{2}$
    for $\phi = 0$ and
    $\overlap*{\chi_{o}}{\chi_{c}} = 1/\sqrt{2}$.
    (b) Plot of
    Eq.~\eqref{eq:bell_test:quantum_delayed_choice:expvalB:simple},
    which is a cut through panel (a) at $\phi = 0$.
  }%
  \label{fig:bell_test:quantum_delayed_choice:expvalB}%
\end{figure}

Since most phases
and interferometer imbalances have been eliminated,
the only remaining variables are
the overlap $\overlap*{\chi_{o}}{\chi_{c}}$
and the phase $\phi$
that, unless stated otherwise,
are treated below
as arbitrary and controllable.%
\footnote{The overlap $\overlap*{\chi_{o}}{\chi_{c}}$
is always between $0$ and $1$,
since, as we explain in what follows,
the bounds of that interval are not achievable practically.
On the one hand, the value $0$ is not physical.
Since it would imply an infinite population of the field.
For macroscopic populations, $\overlap*{\chi_{o}}{\chi_{c}}$
can, however, come indistinguishably close to $0$.
The value $1$, on the other hand,
is equivalent to the absence of all incoming light,
where there would be no dynamics, no coherent control,
and nothing to measure.}
Since there is no single Bell test that would be applicable
to all settings of these two variables.
One has to construct a family of Bell tests
that take these settings into account.
This can be done
using a method similar to that
outlined in Ref.~\citealp{Mann:1995jl}.
Due to the complexity of the expressions,
we only give the result
for $\expval*{B}$ with respect to
the state \eqref{eq:tilde_varrho_infty:qdc:simple}:
\begin{align}
  \expval*{B} & = 2 \sqrt{1 + \frac{\left(1 - \overlap*{\chi_{o}}{\chi_{c}}{}^{2}\right) \left[2 + \sin\left(2 \phi\right)\right]}{\left\{2 - \frac{1}{\sqrt{2}} \overlap*{\chi_{o}}{\chi_{c}} \left[\cos\phi + \cos\left(2 \phi\right) - \sin\phi\right]\right\}^{2}}}.
  \label{eq:bell_test:quantum_delayed_choice:expvalB:full}
\end{align}
This expression is plotted
in Fig.~\ref{fig:bell_test:quantum_delayed_choice:expvalB}(a).
The expectation value is seen to be larger than $2$ for all
$\overlap*{\chi_{o}}{\chi_{c}} < 1$.
In other words, Bell tests can be found and violated for
practically all relevant values of $\phi$
and $\overlap*{\chi_{o}}{\chi_{c}}$.
Moreover,
$\expval*{B}$ can reach
the global maximum Tsirelson bound of $2 \sqrt{2}$,
for the particular pair of values:
$\phi = 0$ and
$\overlap*{\chi_{o}}{\chi_{c}} = 1 / \sqrt{2}$.
The latter value corresponds to a miniscule field amplitude
with a significant population of the vacuum mode.

In the following, we explore the
neighborhood of the global maximum of $\expval*{B}$ and identify
the specific Bell test that gives rise to it.
We expect that, in an experiment,
$\phi$ can be controlled better than
$\overlap*{\chi_{o}}{\chi_{c}}$.
Therefore, we fix $\phi$ at $0$ below
and derive a Bell operator family that
adapts to any given value of $\overlap*{\chi_{o}}{\chi_{c}}$.
As in
Sec.~\ref{sec:certification_of_nonclassicality:quantum_erasure},
each Bell operator $B$ is comprised of
four dichotomous observables,
two of which are sensitive to the matter $\mathcal{M}$
and two to the radiation $\mathcal{R}$.
They are:
\begin{subequations}%
  \begin{multline}%
    O_{\mathcal{M}}^{\vphantom{\prime}} = \frac{1}{\sqrt{\vphantom{\sqrt{2}}\smash{6 - 4 \sqrt{2} \overlap*{\chi_{o}}{\chi_{c}}}}} \, \Bigl[\left(-\sqrt{2} + \overlap*{\chi_{o}}{\chi_{c}} - \sqrt{1 - \overlap*{\chi_{o}}{\chi_{c}}{}^{2}}\right) \sigma_{\mathcal{M}}^{2} \\
    + \left(-\sqrt{2} + \overlap*{\chi_{o}}{\chi_{c}} + \sqrt{1 - \overlap*{\chi_{o}}{\chi_{c}}{}^{2}}\right) \sigma_{\mathcal{M}}^{3}\Bigr],
  \end{multline}
  \begin{multline}
    O_{\mathcal{M}}' = \frac{1}{\sqrt{\vphantom{\sqrt{2}}\smash{6 - 4 \sqrt{2} \overlap*{\chi_{o}}{\chi_{c}}}}} \, \Bigl[\left(-\sqrt{2} + \overlap*{\chi_{o}}{\chi_{c}} + \sqrt{1 - \overlap*{\chi_{o}}{\chi_{c}}{}^{2}}\right) \sigma_{\mathcal{M}}^{2} \\
    + \left(-\sqrt{2} + \overlap*{\chi_{o}}{\chi_{c}} - \sqrt{1 - \overlap*{\chi_{o}}{\chi_{c}}{}^{2}}\right) \sigma_{\mathcal{M}}^{3}\Bigr],
  \end{multline}%
  \label{eqs:bell_test:quantum_delayed_choice:OM_OMp:simple}%
\end{subequations}%
with $\sigma_{\mathcal{M}}^{2}$, $\sigma_{\mathcal{M}}^{3}$
from Eqs.~\eqref{eqs:pseudospin_operators},
and
\begin{subequations}%
  \begin{align}%
    O_{\mathcal{R}}^{\vphantom{\prime}} & = \frac{1}{1 - \overlap*{\chi_{o}}{\chi_{c}}^{2}} \, \left[-\overlap*{\chi_{o}}{\chi_{c}} \, \left(\ket*{\chi_{c}} \bra*{\chi_{c}} + \ket*{\chi_{o}} \bra*{\chi_{o}}\right) + \ket*{\chi_{c}} \bra*{\chi_{o}} + \ket*{\chi_{o}} \bra*{\chi_{c}}\right], \\
    O_{\mathcal{R}}' & = \frac{1}{\sqrt{1 - \overlap*{\chi_{o}}{\chi_{c}}{}^{2}}} \, \left(\ket*{\chi_{c}} \bra*{\chi_{c}} - \ket*{\chi_{o}} \bra*{\chi_{o}}\right).
  \end{align}%
  \label{eqs:bell_test:quantum_delayed_choice:OR_ORp:simple}%
\end{subequations}%
The expectation value of the Bell operator
$B = O_{\mathcal{M}}^{\vphantom{\prime}} \otimes O_{\mathcal{R}}^{\vphantom{\prime}} + O_{\mathcal{M}}' \otimes O_{\mathcal{R}}^{\vphantom{\prime}} + O_{\mathcal{M}}^{\vphantom{\prime}} \otimes O_{\mathcal{R}}' - O_{\mathcal{M}}' \otimes O_{\mathcal{R}}'$
with respect to the state \eqref{eq:tilde_varrho_infty:qdc:simple}
with $\phi = 0$ is
\begin{align}
  \expval*{B} & = \frac{2 \sqrt{\vphantom{\sqrt{2}}\smash{6 - 4 \sqrt{2} \overlap*{\chi_{o}}{\chi_{c}}}}}{2 - \sqrt{2} \overlap*{\chi_{o}}{\chi_{c}}}.
  \label{eq:bell_test:quantum_delayed_choice:expvalB:simple}
\end{align}
A plot of that is shown in
Fig.~\ref{fig:bell_test:quantum_delayed_choice:expvalB}(b).
At the maximum, where $\langle B \rangle = 2\sqrt{2}$, 
\ie{}, at $\overlap*{\chi_{o}}{\chi_{c}} = 1 / \sqrt{2}$,
the material observables become particularly simple.
They are
$O_{\mathcal{M}}^{\vphantom{\prime}} = - \sigma_{\mathcal{M}}^{2}$
and
$O_{\mathcal{M}}' = - \sigma_{\mathcal{M}}^{3} = -M$.

As before,
measuring the observables $O_{\mathcal{R}}^{\vphantom{\prime}}$
and $O_{\mathcal{R}}'$ pose a serious experimental challenge.
Therefore, a switch to displaced photon measurements
of a kind similar to what was discussed in
Sec.~\ref{sec:quantum_erasure:photon_threshold}
may be necessary.
See Ref.~\citealp{Scholak:2014aa}
for further information.

\section{Application To Photoelectron Spin Polarization Control in Alkali Photoionization}
\label{sec:application}

Here,
we apply this formalism to a specific example
used throughout the remainder of the article,
phase-coherent control of
the spin polarization of an electron emitted in
an $1$ \vs{} $2$ photoionization process from
a heavy alkali atom \cite{Scholak:2014aa}.
This control scenario is an extension
of Elliott's two-color photoionization experiments on alkali atoms
\cite{Chen:1990kx,Chen:1990ve,Yin:1992fu,Yin:1993fk},
which demonstrated that
the angular distribution of an emitted photoelectron
can be controlled by the relative phase between
two ionizing laser beams.
The sensitivity to phase was attributed to
quantum interference between competing ionization processes.
However, as discussed in Sec.~\ref{sec:introduction},
phase sensitivity by itself does not suffice to guarantee quantum interference.

The control scenario below deviates from
the original experiment in two ways:
\begin{enumerate*}[label=(\roman{enumi})]
  \item
    the control target is the polarization of
    the spin of the emitted photoelectron,
    rather than its angular distribution.
    Hence, the binary alternatives of
    the electron's spin-$\frac{1}{2}$
    allow for a straightforward definition of
    the dichotomous observable $M$, and
  \item
    the framework of
    the controlherent control interferometer (CCI)
    allows us to demonstrate true quantum features.
\end{enumerate*}
For this section,
the goal is to show that
the spin polarization control example can implement a CCI
as defined in Sec.~\ref{sec:CCI}.
Hence, although some of the details of this example are well known,
casting the problem as a CCI immediately allows us to apply all the results 
in Sec.~\ref{sec:CCI} to this scenario.

\subsection{A Model for The Material Degrees of Freedom}
\label{sec:application:model}

Below we introduce a simple photoionization model for the alkali atom.
Note first, that in the weak-field limit,
the scattering recoil of the atom will be small,
so if the atom is tightly confined 
\cite{Grimm:2000tg,Eschner:2003kl},
then we can neglect its motional degrees of freedom
which would otherwise result in distinguishable pathways.
Further we can neglect all but the
outermost electron of the atom.
The atomic Hamiltonian, hence, reduces to an
effective single-electron Hamiltonian with bound and continuum contributions:
\begin{align}
  H_{\mathcal{M}} & = \smashoperator{\sum_{n l j m_{j}}} E_{n l j} \, \ket*{n l j m_{j}} \bra*{n l j m_{j}} + \smashoperator{\int_{0}^{\infty}} \mathrm{d}E \, E \, \smashoperator{\sum_{l j m_{j}}} \, \ket*{E l j m_{j}{}^{+}} \bra*{E l j m_{j}{}^{+}},
  \label{eq:HM:alkali_atom}
\end{align}
where the states are described by the principle quantum number ($n$),
the orbital ($l$) and the total angular momenta ($j$),
and the projection ($m_{j}$) of the total angular momentum
onto the quantization axis, chosen as the laboratory's
$\uvec{z}$-axis below.
The bound eigenstates $\ket*{n l j m_{j}}$ are
hydrogen-like, with energies $E_{n l j}$ that
depend on $l$ and $j$ due to spin-orbit coupling.
The bound state sum in Eq.~\eqref{eq:HM:alkali_atom},
begins at the electron's ground state which,
for alkali metals, is spin-$\frac{1}{2}$
and is denoted $\ket*{n 0 \frac{1}{2} m_{j}}$,
where $n$ equals the metal's atomic number.
Using Clebsch-Gordan coefficients,
we can expand the bound eigenstates in the position basis:
\begin{align}
  \overlap*{\vec{X}}{n l j m_{j}} & = \smashoperator{\sum_{m_{l} m_{s}}} \, \clebschgordan*{l m_{l}}{\tfrac{1}{2} m_{s}}{j m_{j}} \, R_{n l j}\left(X\right) \, Y_{l m_{l}}\bigl(\uvec{X}\bigr) \, \ket*{m_{s}}.
  \label{eq:x_nljmj}
\end{align}
Here, $R_{n l j}$ denotes a solution
of the radial Schr\"{o}dinger equation,
$Y_{l m_{l}}$ is a spherical harmonic,
$m_{l}$ is the $\uvec{z}$-axis projection
of the orbital angular momentum,
and $\ket{m_{s}}$ describes
the electron spin, also in $\uvec{z}$ direction.
The continuum states are, to good approximation,
the asymptotic scattering out-states of energy $E$
in the partial wave expansion for
a screened Coulomb potential
\cite{Taylor:1972nx,Bethe:2008oq}.
In this article,
we use two different, but equivalent sets of continuum states,
$\ket*{E l j m_{j} {}^{+}}$ and
\begin{align}
  \ket*{\vec{K} m_{s} {}^{+}} & = \frac{\hbar}{\sqrt{m_{\mathrm{e}} K}} \smashoperator{\sum_{l m_{l}}} Y_{l m_{l}}\bigl(\uvec{K}\bigr) \smashoperator{\sum_{j m_{j}}} \, \clebschgordan*{l m_{l}}{\tfrac{1}{2} m_{s}}{j m_{j}} \, \ket*{E(K) l j m_{j} {}^{+}} \, .
\end{align}
The latter states are eigenstates
of $H_{\mathcal{M}}$ with energy
$E\left(K\right) = \hbar^{2} K^{2} / 2 m_{\mathrm{e}}$.
Here $\vec{K}$ is the electron's asymptotic outgoing wavevector,
$\uvec{K}$ its direction and $K$ its length.
An expansion analogous to Eq.~\eqref{eq:x_nljmj}
also exists for the continuum wave functions
$\overlap*{\vec{X}}{E l j m_{j} {}^{+}}$.
For an alkali atom,
the radial part in that expansion
can be taken as a Coulomb function
that obeys
\begin{multline}
  R_{E l j} \left(X\right) \underset{X \to \infty}{\sim} \ii^l \ee^{\ii \varpi_{K l j}} \sqrt{\frac{2 m}{\pi \hbar^2 K}} \\
  \times \frac{1}{X} \sin \left[K X + \tfrac{1}{K} \ln \left(2 K X\right) - \tfrac{1}{2} l \pi + \varpi_{K l j} + \delta_{K}\right] \, ,
\end{multline}
where $\varpi_{K l j}$ is the partial wave Coulomb phase shift
and $\delta_{K}$ is the quantum defect.

\subsection{The Proposed Experiment}
\label{sec:application:experiment}

Consider then
the setup for the experiment,
and the associated implementation of the CCI.
We assume an ideal, spherically shaped
electron detector centered around the atom,
all in the vacuum.
The detector is large and can measure both the electron position on the surface,
and its spin polarization with respect to a constant global axis.
The atom is initially in its ground state,
in an incoherent mixture of spin up and down,
\begin{align}
  \varrho_{\mathcal{M}}\left(-\infty\right) & = \sum_{m_{j}} p_{m_{j}} \ket*{n 0 \tfrac{1}{2} m_{j}} \bra*{n 0 \tfrac{1}{2} m_{j}}.
  \label{eq:varrhoM:mj}
\end{align}
It is ionized
with a pair of weak light pulses with
$\varrho_{\mathcal{R}}\left(-\infty\right) = \ket*{\chi}\bra*{\chi}$,
whereupon a photoelectron is emitted.
The outgoing matter and light waves
are described by the final state $\varrho\left(\infty\right)$.

The control target defines the input and output ports of the CCI
as the electron's spin polarization.
The input ports are thus associated with the
spin polarization $m_{j}$ of the electronic ground state
$\ket*{n 0 \frac{1}{2} m_{j}}$,
and the output ports are identified with
the spin projection $m_{s}$
of the outgoing matter wave $\ket*{\vec{K} m_{s}{}^{+}}$.
Hence, the material observable $M$ is defined as
\begin{align}
  M & = \smashoperator{\sum_{n l m_{l} m_{s}}} 2 m_{s} \ket*{n l m_{l} m_{s}} \bra*{n l m_{l} m_{s}} + \smashoperator{\int} \mathrm{d}\vec{K} \smashoperator{\sum_{m_{s}}} 2 m_{s} \ket*{\vec{K} m_{s}{}^{+}} \bra*{\vec{K} m_{s}{}^{+}},
  \label{eq:M:spin}
\end{align}
where the factor of $2$ ensures that $M$
has the eigenvalues $+1$ and $-1$.%
\footnote{Since $M$ need not be sensitive to
anything but the electron's spin,
we assemble $M$'s point spectrum
[the first term in Eq.~\eqref{eq:M:spin}]
from the spin-orbit product states $\ket*{n l m_{l} m_{s}}$.
In alkali metals,
the electron's spin and orbit are strongly coupled,
and these states are in general not eigenstates
of the Hamiltonian $H_{\mathcal{M}}$.
This is in violation of the provisions of
Sec.~\ref{sec:cci:dichotomous_material_observables},
according to which $M$ and $H_{\mathcal{M}}$ are required to commute.
However, this will not be an issue below,
since the properties of $M$ matter only
with respect to the initial and
the final states in the scattering problem.
There is no issue with respect to $\mathcal{D}_{+}$, since
$M$ is diagonal in all continuum states of $H_{\mathcal{M}}$,
including those in $\mathcal{D}_{+}$.
We only have to concern ourselves with $\mathcal{D}_{-}$.
The set $\mathcal{D}_{-}$ represents the ground state
for which spin-orbit coupling is absent.
Therefore, the state $\ket*{n 0 \frac{1}{2} m_{j}}$
is identical to the
spin-orbit product state $\ket*{n 0 0 m_{s}}$
with $m_{s} = m_{j}$.
Hence, $M$ is also diagonal with respect to all states
in $\mathcal{D}_{-}$.}

The incoming light is
quasi-monochromatic around frequencies
$2 \omega$ and $\omega$,
and the detection is limited to photoelectrons
with kinetic energy
$E\left(K\right) \approx E_- + 2\hbar\omega$,
where $E_{-} = E_{n 0 \frac{1}{2}}$,
hence $\mathcal{D}_{-}$ is a region in $\vec{K}$-space
clustered around $E\left(K\right)$.
We make further adjustments to $\mathcal{D}_{+}$ below.

Using the approach 
in Sec.~\ref{sec:cci:coherent_control_theory},
we project 
the total density operator of electron and radiation
onto $\mathcal{D}_{+}$ to obtain:
\begin{multline}
  \tilde{\varrho}\left(\infty\right) = \tilde{p}^{-1} \, \smashoperator[l]{\sum_{m_{s}, m_{s}' = \pm \frac{1}{2}}} \smashoperator[r]{\sum_{m_{j} = \pm \frac{1}{2}}} \, p_{m_{j}} \left(\ket[\big]{\tilde{\psi}^{1}_{m_{s}, m_{j}}} \ket[\big]{\tilde{\chi}^{1}} + \ket[\big]{\tilde{\psi}^{2}_{m_{s}, m_{j}}} \ket[\big]{\tilde{\chi}^{2}}\right) \\ \times \left(\bra[\big]{\tilde{\psi}^{1}_{m_{s}', m_{j}}} \bra[\big]{\tilde{\chi}^{1}} + \bra[\big]{\tilde{\psi}^{2}_{m_{s}', m_{j}}} \bra[\big]{\tilde{\chi}^{2}}\right)
\end{multline}
with $\tilde{p}$ the normalization.
The ket
\begin{align}
  \ket[\big]{\tilde{\psi}^{1}_{m_{s}, m_{j}}} & = \smashoperator{\int_{\mathcal{D}_{+}}} \mathrm{d}\vec{K} \, T_{1}\bigl[\hat{f}_{2}\bigr]\left(\vec{K} m_{s}, m_{j}\right) \ket*{\vec{K} m_{s}{}^{+}},
\end{align}
is the final scattering state of a single ionization event, 
and $\ket[\big]{\tilde{\psi}^{2}_{m_{s}, m_{j}}}$
is an analogous expression for the
final state of a two-photon photon ionization event.
Note that these states are defined in terms
of the one- and two-photon scattering probability amplitudes,
$T_{1}$ and $T_{2}$.
For monochromatic cw light,
these amplitudes are well known \cite{Yin:1992fu,Yin:1993fk}.
However, here we deal with quasi-monochromatic light pulses.
The approach in
Sec.~\ref{sec:cci:coherent_control_theory},
specifically Eq.~\eqref{eq:T1},
gives
\begin{multline}
  T_{1}\bigl[\hat{f}_{2}\bigr]\left(\vec{K} m_{s}, m_{j}\right) = 2 \pi \ii \int \mathrm{d}\vec{k} \sum_{\lambda} \delta\left[\textstyle E_{-} - E\left(K\right) + \hbar \omega\left(k\right)\right] \, \hat{f}_{2}\left(\vec{k}, \lambda\right) \\
  \times \left(-\ii\right) \hbar \mathcal{E}\left(k\right) \frac{\hbar}{\sqrt{m_{\mathrm{e}} K}} \smashoperator{\sum_{m_{l}'}} Y_{1 m_{l}'}\bigl(\uvec{K}\bigr) \smashoperator{\sum_{j' m_{j}'}} \, \clebschgordan*{1 m_{l}'}{\tfrac{1}{2} m_{s}}{j' m_{j}'} \\
  \times A_{1}\left[1 j' m_{j}', 0 \tfrac{1}{2} m_{j}; \uvec{\varepsilon}\left(\vec{k}, \lambda\right)\right] \, D_{1}\left[E\left(K\right) 1 j', n 0 \tfrac{1}{2}\right],
\end{multline}
where
\begin{subequations}
  \begin{align}
    A_{1}\left(l' j' m_{j}', l j m_{j}; \uvec{\varepsilon}\right) & \; \parbox[b]{.6\textwidth}{$\displaystyle = \sqrt{\tfrac{2 l + 1}{2 l' + 1}} \, \clebschgordan*{l 0}{1 0}{l' 0} \smashoperator[l]{\sum_{m_{l}^{\vphantom{\prime}} m_{l}'}} \smashoperator[r]{\sum_{q=-1 \vphantom{m_{l}'}}^{1}} \varepsilon_{q} \, \clebschgordan*{l m_{l}^{\vphantom{\prime}}}{1 q}{l' m_{l}'}$} \notag \\
    & \; \parbox[b]{.6\textwidth}{\raggedleft $\displaystyle \times \sum_{m_{s}} \, \clebschgordan*{l' m_{l}'}{\tfrac{1}{2} m_{s}}{j' m_{j}'} \, \clebschgordan*{l m_{l}}{\tfrac{1}{2} m_{s}}{j m_{j}},$}
    \label{eq:A1}
    \intertext{is the first-order angular dipole matrix element and}
    D_{1} \left(E' l' j', n l j\right) & \; \parbox[b]{.6\textwidth}{$\displaystyle = e \smashoperator{\int_{0}^{\infty}} \mathrm{d}x \, x^{3} R_{E' l' j'}^{*} \left(x\right) R_{n l j}^{\vphantom{*}} \left(x\right)$}
    \label{eq:D1}
  \end{align}
\end{subequations}
is the first-order radial integral \cite{Fermi:1930qa,Seaton:1951mi}.
The $D_{1}$ integrals,
not explicitly calculated here,
serve as empirical parameters below.

\begin{figure}
  \centering%
  \begingroup%
    \iftikz%
      \tikzexternalenable%
      \input{tikz/selection_rules.tex}%
    \else%
      \includegraphics{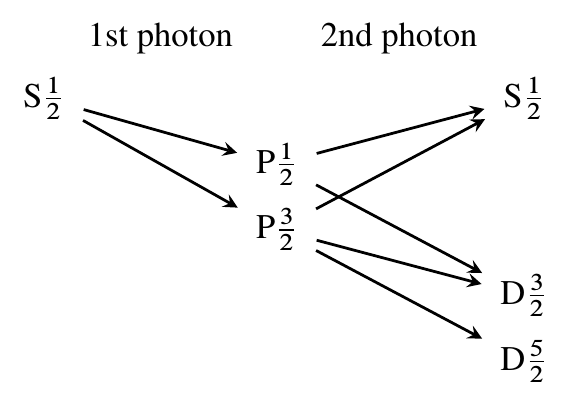}%
    \fi%
  \endgroup%
  \caption{%
    Electric dipole selection rules between
    $l j$-states in one- and two-photon absorption.
    These two processes are distinguishable by
    the final orbital angular momenta,
    in particular,
    the separation into states with even
    ($\mathrm{S}$, $\mathrm{D}$)
    and those with odd parity ($\mathrm{P}$).%
  }%
  \label{fig:selection_rules}%
\end{figure}

Figure \ref{fig:selection_rules}
illustrates the selection rules for the single-photon transition,
which are enforced by $A_{1}$,
\ie{},
$T_{1}$ only connects the ground state to
$\mathrm{P}$ continuum states with $l' = 1$.
Moreover, $A_{1}$ determines the contributions of
$\sigma$ ($\Delta m_{j} = \pm 1$)
and $\pi$ transitions ($\Delta m_{j} = 0$)
to the transition amplitude.
These primarily depend on the projections
$\varepsilon_{q} = \uvec{\varepsilon} \cdot \uvec{e}_{q}^{*}$
of the field polarization $\uvec{\varepsilon}$
onto the spherical basis vectors
$\uvec{e}_{\pm 1} = \mp \left(\uvec{x} \pm \uvec{y}\right) / \sqrt{2}$,
$\uvec{e}_{0} = \uvec{z}$.
Figure \ref{fig:selection_rules} also shows the selection
rules for the two-photon process (only partial waves with $l' = 0$ or $2$
contribute) arising from the second-order angular dipole matrix element
\begin{subequations}
  \begin{multline}
    A_{2}\left(l' j' m_{j}', l'' j'', l j m_{j}; \uvec{\varepsilon}, \uvec{\epsilon}'\right) = \sum_{m_{j}''} \, A_{1}\left(l' j' m_{j}', l'' j'' m_{j}''; \uvec{\varepsilon}\right) \\
    \times A_{1}\left(l'' j'' m_{j}'', l j m_{j}; \uvec{\varepsilon}'\right),
    \label{eq:A2}
  \end{multline}
  that, together with the second-order radial integral
  \begin{multline}
    D_{2}\left(E' l' j', l'' j'', n l j; k\right) = \sum_{n''} \frac{D_{1}\left(E' l' j', n'' l'' j''\right) D_{1}\left(n'' l'' j'', n l j\right)}{E_{n l j} - E_{n'' l'' j''} + \hbar c k + \ii 0} \\
    + \smashoperator{\int_{0}^{\infty}} \mathrm{d}E'' \, \frac{D_{1} \left(E' l' j', E'' l'' j''\right) D_{1}\left(E'' l'' j'', n l j\right)}{E_{n l j} - E'' + \hbar c k + \ii 0},
    \label{eq:D2}
  \end{multline}
\end{subequations}
enters the two-photon transition amplitude
\begin{multline}
  T_{2}\bigl[\hat{f}_{1}\bigr]\left(\vec{K} m_{s}, m_{j}\right) = - 2 \pi \ii \\
  \times \smashoperator{\int} \mathrm{d}\vec{k}_{1} \smashoperator{\sum_{\lambda_{1}}} \smashoperator{\int} \mathrm{d}\vec{k}_{2} \smashoperator{\sum_{\lambda_{2}}} \delta\bigl[\textstyle E_{-} - E\left(K\right) + \sum_{i = 1}^{2} \hbar \omega\left(k_{i}\right)\bigr] \\
  \times \hat{f}_{1}\left(\vec{k}_{1}, \lambda_{1}\right) \, \left(-\ii\right) \hbar \mathcal{E}\left(k_{1}\right) \, \hat{f}_{1}\left(\vec{k}_{2}, \lambda_{2}\right) \, \left(-\ii\right) \hbar \mathcal{E}\left(k_{2}\right) \, \frac{\hbar}{\sqrt{m_{\mathrm{e}} K}} \smashoperator{\sum_{l' m_{l}'}} Y_{l' m_{l}'}\bigl(\uvec{K}\bigr) \\
  \times \smashoperator{\sum_{j' m_{j}'}} \, \clebschgordan*{l' m_{l}'}{\tfrac{1}{2} m_{s}}{j' m_{j}'} \smashoperator{\sum_{j''}} A_{2}\left[l' j' m_{j}', 1 j'', 0 \tfrac{1}{2} m_{j}; \uvec{\varepsilon}\left(\vec{k}_{1}, \lambda_{1}\right), \uvec{\varepsilon}\left(\vec{k}_{2}, \lambda_{2}\right)\right] \\
  \times D_{2}\left[E\left(K\right) l' j', 1 j'', n 0 \tfrac{1}{2}; k_{2}\right].
\end{multline}
These expressions were derived from Eq.~\eqref{eq:TN},
and the arguments of $A_{2}$ and $D_{2}$
refer to the particular photoionization channels.
For example,
$D_{2}\left[E\left(K\right) 2 \tfrac{5}{2}, 1 \tfrac{3}{2}, n 0 \tfrac{1}{2}\right]$
is the radial matrix element
leading from the $\mathrm{S} \tfrac{1}{2}$ ground state
$\ket[\big]{n 0 \tfrac{1}{2} m_{j}^{\vphantom{\prime}}}$
via intermediate $\mathrm{P} \tfrac{3}{2}$ orbitals
to the $\mathrm{D} \tfrac{5}{2}$ continuum state
$\ket[\big]{E\left(K\right) \mathrm{D} \tfrac{5}{2} m_{j}'{}^{+}}$.

Since the states accessed by the two ionization pathways are 
of different parity and angular momentum, 
the outgoing matter wave is a welcher-weg marker
that, in principle, allows one to acquire absolute path knowledge.
As discussed in Sec.~\ref{sec:quantum_erasure:ww_marking},
this situation excludes phase control completely.
This can be rectified by rendering
the path knowledge physically inaccessible,
\ie{}, by quantum erasure.
This procedure is applied below
to obtain phase control.

The welcher-weg information is encoded in
the matter angular distribution.
Hence, the measurement of $M$ has to be combined 
with a measurement of an observable that is
sensitive to the electron's angular spread.
However, the observable has to be such that the ionization pathways are indistinguishable
in at least one of its eigenspaces.
The simplest (though not optimal) observable
fulfilling this requirement corresponds
to a single segment of the spherically shaped detector,
chosen so that,
if it detects an electron,
there is no way of telling whether its matter wave
had even or odd parity
(\ie{}, whether the electron was emitted in
a single- or a two-photon process).

Effectively, with this prescription,
the measurement is restricted to a narrow solid angle
around a single direction,
denoted $\uvec{K}_{+}$.
Consequently, $\mathcal{D}_{+}$ is confined in both energy
and direction $\uvec{K}$.
We define
$\uvec{K}_{+} = \vec{K}_{+} / K_{+}$
with
$\vec{K}_{+} = \abs*{\mathcal{D}_{+}}^{-1} \int_{\mathcal{D}_{+}} \mathrm{d}\vec{K} \vec{K}$,
the average of $\vec{K}$ over $\mathcal{D}_{+}$,
$K_{+} = \norm{\vec{K}_{+}}$ its length, and
$\abs*{\mathcal{D}_{+}} = \int_{\mathcal{D}_{+}} \mathrm{d}\vec{K}$, 
the volume of $\mathcal{D}_{+}$.
That is, we achieve quantum erasure
by measuring the differential cross section.

The states $\ket[\big]{\tilde{\psi}^{1}_{m_{s}, m_{j}}}$
and $\ket[\big]{\tilde{\psi}^{1}_{m_{s}, m_{j}}}$
depend on the size and shape of $\mathcal{D}_{+}$,
but given the narrow confinement of $\mathcal{D}_{+}$,
it is justified to assume that
$T_{1}$ and $T_{2}$
vary weakly over $\mathcal{D}_{+}$.
Hence,
$T_{1}$ and $T_{2}$ are evaluated
at $\vec{K}_{+}$ instead of $\vec{K}$, \ie{},
we approximate
\begin{subequations}%
  \begin{align}%
    \ket[\big]{\tilde{\psi}^{1}_{m_{s}, m_{j}}} & \simeq T_{1}\bigl[\hat{f}_{2}\bigr]\left(\vec{K}_{+} m_{s}, m_{j}\right) \smashoperator{\int_{\mathcal{D}_{+}}} \mathrm{d}\vec{K} \, \ket*{\vec{K} m_{s}{}^{+}}
    \shortintertext{and}
    \ket[\big]{\tilde{\psi}^{2}_{m_{s}, m_{j}}} & \simeq T_{2}\bigl[\hat{f}_{1}\bigr]\left(\vec{K}_{+} m_{s}, m_{j}\right) \smashoperator{\int_{\mathcal{D}_{+}}} \mathrm{d}\vec{K} \, \ket*{\vec{K} m_{s}{}^{+}}.
  \end{align}%
\end{subequations}%

\subsection{Implementation of The Open And The Closed CCI Configurations}
\label{sec:application:open_and_closed_configuration}

Below we show how to implement
the open and the closed configuration
with the photoelectron spin CCI.
These configurations are needed for
the quantum erasure and the quantum delayed choice
coherent control experiments described above.

\begin{figure}
  \centering%
  \begingroup%
    \iftikz%
    \tikzexternalenable%
     \input{tikz/cci_open_alkaliatom.tex}%
  \else%
    \includegraphics{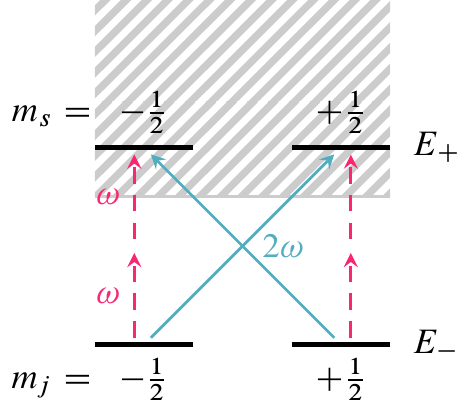}%
  \fi%
  \endgroup%
  \caption{%
    (color online)
    In the open configuration $o$,
    single-photon ionization (solid blue) only undergoes
    a $\sigma$-transition with $\abs[\big]{m_{j} - m_{s}'} = 1$,
    whereas two-photon ionization (dashed red)
    leaves the electronic spin unchanged.
    The processes are distinguishable from one another
    due to their different outcomes.%
  }%
  \label{fig:cci:open:alkali_atom}%
\end{figure}

Consider first the open CCI.
As explained in Sec.~\ref{sec:path_distinguishability:open_config},
the open configuration is maximally biased.
This is reflected in the following prescription:
if the atom's initial state is spin up (or down),
then a photoelectron emitted in single-photon ionization
has to arrive at the detector with spin down (up),
and a photoelectron emitted in double-photon ionization
must arrive with spin up (down).
That is:
\begin{subequations}%
  \begin{align}
    T_{1}\bigl[\hat{f}_{2 o}\bigr]\left(\vec{K}_{+} m_{s}, m_{j}\right) & \propto \delta_{m_{s}, -m_{j}} \, , \\
    T_{2}\bigl[\hat{f}_{1 o}\bigr]\left(\vec{K}_{+} m_{s}, m_{j}\right) & \propto \delta_{m_{s}, m_{j}} \, ,
  \end{align}%
  \label{eqs:T12:open:want:bias}%
\end{subequations}%
\ie{}, 
Eqs.~\eqref{eqs:transition_amplitudes_mutually_exclusive}
applied to the photoelectron spin control scenario.
In addition, we stipulate the symmetry condition
\begin{gather}%
  \textstyle \sqrt{\overlap{\tilde{\chi}^{1}_{o}}{\tilde{\chi}^{1}_{o}}} \, \abs[\big]{T_{1}\bigl[\hat{f}_{2 o}\bigr]\left(\vec{K}_{+} \,\mathopen{}-m_{j}, m_{j}\right)} = \sqrt{\overlap{\tilde{\chi}^{2}_{o}}{\tilde{\chi}^{2}_{o}}} \, \abs[\big]{T_{2}\bigl[\hat{f}_{1 o}\bigr]\left(\vec{K}_{+} m_{j}, m_{j}\right)}
  \label{eq:T12:open:want:symmetry}%
\end{gather}%
adapted from Eq.~\eqref{eqs:symmetric_conditions}.
With Eq.~\eqref{eq:T12:open:want:symmetry},
the single- and the double-photon process
have equal \emph{a priori} probabilities.

In the following, we specify suitable parameters
that allow $T_{1}$ and $T_{2}$
to fulfill the conditions \eqref{eqs:T12:open:want:bias}
and \eqref{eq:T12:open:want:symmetry},
the direction $\uvec{K}_{+}$ pointing to the detector,
the wavepackets' directions of incidence,
and the wavepackets' polarizations.
Wavepacket properties are described
by the amplitude functions $f_{1 o}$, $f_{2 o}$.

For the direction of detection, we choose
\begin{align}
  \uvec{K}_{+} = \frac{1}{\sqrt{3}} \bigl(\sqrt{2} \, \uvec{x} - \uvec{z}\bigr) \, ,
  \label{eq:Kplus}%
\end{align}
a choice that will work for both open
and closed configurations.

\begin{figure}
  \centering%
  \begingroup%
    \iftikz%
      \tikzexternalenable%
      \input{tikz/geometry_open_alkaliatom.tex}%
    \else%
      \includegraphics{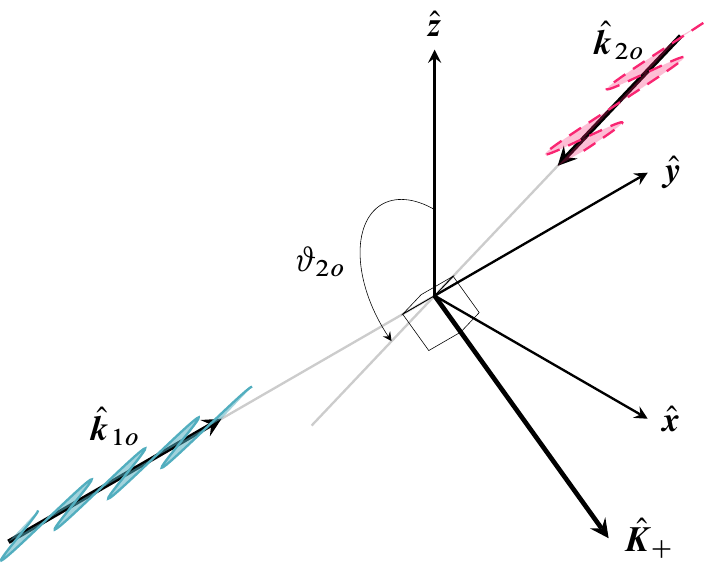}%
    \fi%
  \endgroup%
  \caption{%
    (color online)
    Scattering geometry of the open configuration $o$
    with incident light wavevectors
    $\uvec{k}_{1 o} = \uvec{y}$
    and $\uvec{k}_{2 o}$
    with projection
    $\uvec{k}_{2 o} \cdot \uvec{z} = \cos \vartheta_{2 o} = - \sqrt{2 / 3}$,
    The emitted photoelectron is detected
    in the direction $\uvec{K}_{+}$, which,
    together with $\uvec{k}_{2 o}$ and $\uvec{k}_{1 o}$,
    forms a right-handed orthogonal triad
    that is rotated from the laboratory's reference frame
    by $3 \pi / 2 - \vartheta_{2 o} = \pi - \arctan \sqrt{2}$
    around the $\uvec{y}$-axis.
    The alkali atom is located at the origin.
    The polarization of the incoming radiation
    is indicated in color (shades of gray).
    The open configuration employs linearly polarized light.
  }%
  \label{fig:geometry:open}%
\end{figure}

With regard to the wavepackets,
we concentrate on the simplifying case
where the amplitude functions
cluster around a single direction of incidence
$\bigl(\uvec{k}_{1 o}$ and $\uvec{k}_{2 o}$, respectively$\bigr)$
and a single polarization
($\uvec{\varepsilon}_{1 o}$, $\uvec{\varepsilon}_{2 o}$).
In such a case, we can approximate
\begin{align}
  f_{2 o}\left(\vec{k}, \lambda\right) & \simeq \alpha_{2 o} \, \sqrt{\ell_{2 o}^{-3} \, u_{2 o}^{\vphantom{-1}}\left[\ell_{2 o}^{-1} \left(\vec{k} - \vec{k}_{2 o}\right)\right]} \; \uvec{\varepsilon}_{2 o} \cdot \uvec{\varepsilon}\left(\vec{k}_{2 o}, \lambda\right)
\end{align}
(and equivalently $f_{1 o}$),
where $\alpha_{2 o}$ is an average amplitude
with phase $\varphi_{2 o}$ and absolute value
$\abs*{\alpha_{2 o}} = \norm{f_{2 o}}$.
Here, $u_{2 o}$ is a positive bump function
that minimizes the error of the approximation.
Such a function has compact support around the origin,
a unit volume integral,
and a unit average extent
that is scaled to $\ell_{2 o}$.
In essence, $\ell_{2 o}$ quantifies the average extent of $f_{2 o}$
in $\vec{k}$-space.
Since $f_{2 o}$ is concentrated around $\uvec{k}_{2 o}$,
$\ell_{2 o}$ is small.

The open configuration is then engineered by setting
\begin{subequations}%
  \begin{align}%
    \uvec{k}_{2 o} & = \uvec{y}, &
    \uvec{\varepsilon}_{2 o} & = \frac{1}{\sqrt{3}} \bigl(\uvec{x} + \sqrt{2} \, \uvec{z}\bigr), \\
    \uvec{k}_{1 o} & = - \frac{1}{\sqrt{3}} \bigl(\uvec{x} + \sqrt{2} \, \uvec{z}\bigr), &
    \uvec{\varepsilon}_{1 o} & = \uvec{y},
  \end{align}%
  \label{eqs:settings:open}%
\end{subequations}%
where the incoming light is chosen as linearly polarized.
The geometry of the experiment is visualized
in Fig.~\ref{fig:geometry:open}.
With the above parameter choices,
the transition amplitudes become
\begin{subequations}%
  \begin{gather}%
    T_{1}\bigl[\hat{f}_{2 o}\bigr]\left(\vec{K}_{+} \,\mathopen{}+\tfrac{1}{2}, +\tfrac{1}{2}\right) = T_{1}\bigl[\hat{f}_{2 o}\bigr]\left(\vec{K}_{+} \,\mathopen{}-\tfrac{1}{2}, -\tfrac{1}{2}\right) = 0,
  \end{gather}%
  \begin{multline}%
    T_{1}\bigl[\hat{f}_{2 o}\bigr]\left(\vec{K}_{+} \,\mathopen{}+\tfrac{1}{2}, -\tfrac{1}{2}\right) = - T_{1}\bigl[\hat{f}_{2 o}\bigr]\left(\vec{K}_{+} \,\mathopen{}-\tfrac{1}{2}, +\tfrac{1}{2}\right) = \ee^{\ii \varphi_{2 o}} \\
    \times \int \mathrm{d}\vec{k} \, \delta\bigl[\textstyle E_{-} - E_{+} + \hbar \omega\left(k\right)\bigr] \, \sqrt{u_{2 o}\left[\ell_{2 o}^{-1} \left(\vec{k} - \vec{k}_{2 o}\right)\right]} \\
    \times \, \frac{\sqrt{\pi} \hbar^{2} \mathcal{E}\left(k\right)}{3 \sqrt{\ell_{2 o}^{3} m_{\mathrm{e}} K_{+}}} \, \left[-D_{1}\left(E_{+} 1 \tfrac{1}{2}, n 0 \tfrac{1}{2}\right) + D_{1}\left(E_{+} 1 \tfrac{3}{2}, n 0 \tfrac{1}{2}\right)\right] \, ,
  \end{multline}%
  \begin{multline}%
    T_{2}\bigl[\hat{f}_{1 o}\bigr]\left(\vec{K}_{+} \,\mathopen{}+\tfrac{1}{2}, +\tfrac{1}{2}\right) = T_{2}\bigl[\hat{f}_{1 o}\bigr]\left(\vec{K}_{+} \,\mathopen{}-\tfrac{1}{2}, -\tfrac{1}{2}\right) = \ee^{2 \ii \varphi_{1 o}} \\ \times
    \smashoperator{\int} \mathrm{d}\vec{k}_{1} \smashoperator{\int} \mathrm{d}\vec{k}_{2} \, \delta\bigl[\textstyle E_{-} - E_{+} + \sum_{i = 1}^{2} \hbar \omega\left(k_{i}\right)\bigr] \\
    \times \sqrt{u_{1 o}\left[\ell_{1 o}^{-1} \left(\vec{k}_{1} - \vec{k}_{1 o}\right)\right]} \sqrt{u_{1 o}\left[\ell_{1 o}^{-1} \left(\vec{k}_{2} - \vec{k}_{1 o}\right)\right]} \, \frac{\sqrt{\pi} \ii \hbar^{3} \mathcal{E}\left(k_{1}\right) \mathcal{E}\left(k_{2}\right)}{\sqrt{45} \ell_{1 o}^{3}  \sqrt{m_{\mathrm{e}} K_{+}}} \\
    \begin{aligned}[b]
      \times \bigl[ 5 D_{2}\left(E_{+} 0 \tfrac{1}{2}, 1 \tfrac{1}{2}, n 0 \tfrac{1}{2}; k_{2}\right) & + 10 D_{2}\left(E_{+} 0 \tfrac{1}{2}, 1 \tfrac{3}{2}, n 0 \tfrac{1}{2}; k_{2}\right) \\
      - 5 D_{2}\left(E_{+} 2 \tfrac{3}{2}, 1 \tfrac{1}{2}, n 0 \tfrac{1}{2}; k_{2}\right) & - D_{2}\left(E_{+} 2 \tfrac{3}{2}, 1 \tfrac{3}{2}, n 0 \tfrac{1}{2}; k_{2}\right) \\
      & - 9 D_{2}\left(E_{+} 2 \tfrac{5}{2}, 1 \tfrac{3}{2}, n 0 \tfrac{1}{2}; k_{2}\right) \bigr] \, ,
    \end{aligned}
  \end{multline}%
  and
  \begin{gather}%
    T_{2}\bigl[\hat{f}_{1 o}\bigr]\left(\vec{K}_{+} \,\mathopen{}+\tfrac{1}{2}, -\tfrac{1}{2}\right) = T_{2}\bigl[\hat{f}_{1 o}\bigr]\left(\vec{K}_{+} \,\mathopen{}-\tfrac{1}{2}, +\tfrac{1}{2}\right) = 0,
  \end{gather}%
  \label{eqs:T12:open:is}%
\end{subequations}%
where $E_{+} = E\left(K_{+}\right)$.
Note that, if the spin-orbit coupling were absent,
then the radial integrals
$D_{1}$ and $D_{2}$
would add to zero
and $T_{1}$ and $T_{2}$ would vanish.
However, spin-orbit coupling is typically present,
and $T_{1}$ and $T_{2}$ are finite
with, in general, the second-order amplitude $T_{2}$
much smaller than its counterpart $T_{1}$.

In Eqs.~\eqref{eqs:T12:open:is},
the integrals over the bump functions $u_{2 o}$ and $u_{1 o}$
ensure the agreement between the photon energy
$2 \hbar \omega$ and the ionization energy $E_{+} - E_{-}$
in relation to the energy spreading,
$\hbar c \ell_{2 o}$ and $\hbar c \ell_{1 o}$,
respectively.
If $\ell_{2 o}$ (or $\ell_{1 o}$) is small enough,
all functions under the integral, except the
$\delta$-term and $u_{2 o}$ ($u_{1 o}$),
can be approximated by their value at $k_{2 o}$ ($k_{1 o}$).
The integral then reduces to a numerical factor
$\eta_{2 o}^{2} I_{2 o} / \hbar c$
$\bigl(\eta_{1 o}^{5} I_{1 o} / \hbar c\bigr)$,
where $I_{2 o}$ ($I_{1 o}$) is dimensionless.

To conclude the open configuration discussion,
note, from Eqs.~\eqref{eqs:T12:open:is}, that
$T_{1}$ and $T_{2}$ fulfill 
conditions \eqref{eqs:T12:open:want:bias}.
Condition \eqref{eq:T12:open:want:symmetry}
can be implemented by adjusting
$\sqrt{\overlap{\tilde{\chi}^{1}_{o}}{\tilde{\chi}^{1}_{o}}}$ and
$\sqrt{\overlap{\tilde{\chi}^{2}_{o}}{\tilde{\chi}^{2}_{o}}}$
that, for coherent states
[as in Eqs.~\eqref{eqs:gi:coherent_state}], become equal to
$\abs*{\alpha_{1 o}}^{2}$ and $\abs*{\alpha_{2 o}}$, respectively.

\begin{figure}
  \centering%
  \begingroup%
    \iftikz%
      \tikzexternalenable%
      \input{tikz/cci_closed_alkaliatom.tex}%
    \else%
      \includegraphics{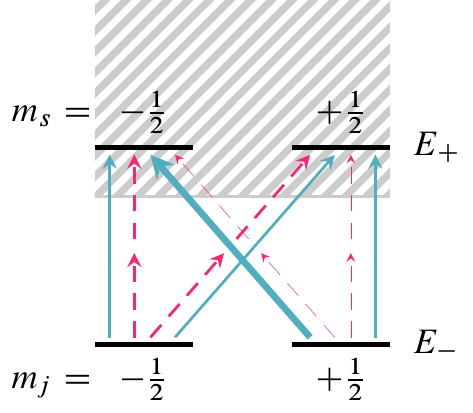}%
    \fi%
  \endgroup%
  \caption{%
    In the closed configuration $c$,
    the processes are not distinguishable by their outcomes,
    since the transition amplitudes between
    all spin combinations are finite and unbiased.
    Our proposed implementation deviates
    from the demanded behavior.
    Only for $m_j = - \frac{1}{2}$
    $\bigl[$and $D_{2}\left(E_{+} 2 \tfrac{5}{2}, 1 \tfrac{3}{2}, n 0 \tfrac{1}{2}; k_{2}\right) = 0\bigr]$,
    the amplitudes are unbiased.
    However, for $m_j = + \frac{1}{2}$, they are biased,
    illustrated here by the width of the arrows.
  }%
  \label{fig:cci:closed:alkali_atom}%
\end{figure}

Consider now the closed configuration which,
according to Sec.~\ref{sec:path_distinguishability:closed_config},
is characterized by
completely unbiased, symmetric transition amplitudes
[Eqs.~\eqref{eqs:transition_amplitudes_unbiased_eta}
and \eqref{eqs:symmetric_conditions}].
In the photoelectron spin control scenario,
these conditions are
\begin{subequations}%
  \begin{align}%
    T_{1}\bigl[\hat{f}_{2 c}\bigr]\left(\vec{K}_{+} \,\mathopen{}+\tfrac{1}{2}, m_{j}\right) & = T_{1}\bigl[\hat{f}_{2 c}\bigr]\left(\vec{K}_{+} \,\mathopen{}-\tfrac{1}{2}, m_{j}\right), \\
    T_{2}\bigl[\hat{f}_{1 c}\bigr]\left(\vec{K}_{+} \,\mathopen{}+\tfrac{1}{2}, m_{j}\right) & = - T_{2}\bigl[\hat{f}_{1 c}\bigr]\left(\vec{K}_{+} \,\mathopen{}-\tfrac{1}{2}, m_{j}\right)
  \end{align}%
  \label{eqs:T12:closed:want:bias}
\end{subequations}%
and
\begin{gather}%
  \textstyle \sqrt{\overlap{\tilde{\chi}^{1}_{c}}{\tilde{\chi}^{1}_{c}}} \, \abs[\big]{T_{1}\bigl[\hat{f}_{2 c}\bigr]\left(\vec{K}_{+} \,\mathopen{}+\tfrac{1}{2}, m_{j}\right)} = \sqrt{\overlap{\tilde{\chi}^{2}_{c}}{\tilde{\chi}^{2}_{c}}} \, \abs[\big]{T_{2}\bigl[\hat{f}_{1 c}\bigr]\left(\vec{K}_{+} \,\mathopen{}+\tfrac{1}{2}, m_{j}\right)}.
  \label{eq:T12:closed:want:symmetry}
\end{gather}%
Fig.~\ref{fig:cci:closed:alkali_atom} provides
a graphical representation of Eqs.~\eqref{eqs:T12:closed:want:bias}.
The challenge is to find
settings $\uvec{K}_{+}$, $\hat{f}_{1 c}$, and $\hat{f}_{2 c}$
so that $T_{1}$ and $T_{2}$ satisfy
Eqs. \eqref{eqs:T12:closed:want:bias} and
\eqref{eq:T12:closed:want:symmetry},
independent of the values of the radial integrals
$D_{1}$ and $D_{2}$
that depend on the energy $E\left(K\right)$
and on the particular alkali atom used.

\begin{figure}
  \centering%
  \begingroup%
    \iftikz%
      \tikzexternalenable%
      \input{tikz/geometry_closed_alkaliatom.tex}%
    \else%
      \includegraphics{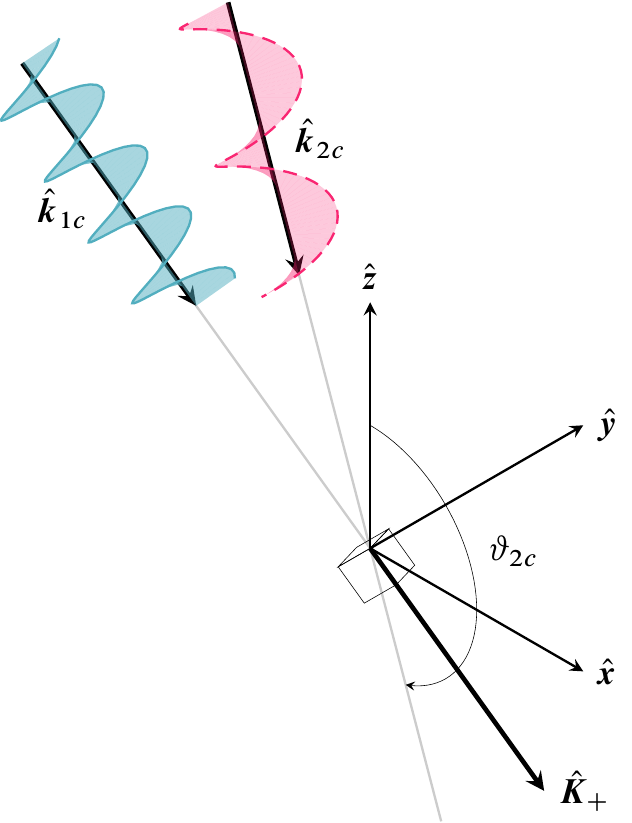}%
    \fi%
  \endgroup%
  \caption{%
    (color online)
    Scattering geometry of the closed configuration $c$
    with the alkali atom at the center,
    Wavepacket polarizations are indicated in color (shades of gray).
    The amplitude function $f_{1 c}$ of the first wavepacket
    is concentrated around
    the incident direction $\uvec{k}_{1 c} = \uvec{K}_{+}$.
    and the circular polarization $\uvec{\varepsilon}_{1 c}$.
    $\uvec{k}_{1 c}$ makes an angle of
    $\pi - \arccsc\left(3\right) \simeq 2.80 \mathrm{rad}$
    with the $\uvec{z}$-axis.
    The second wavepacket (described by $f_{2 c}$)
    is incident from the direction $\uvec{k}_{2 c}$, for which
    $\uvec{k}_{2 c} \cdot \uvec{z} = \cos \vartheta_{2 c} = - 2 \sqrt{2} / 3$.
    Its polarization $\uvec{\varepsilon}_{2 c}$ is elliptical.
  }%
  \label{fig:geometry:closed}%
\end{figure}

We adopt a similar approximation as
above for the open configuration.
That is, we approximate
\begin{align}
  f_{2 c}\left(\vec{k}, \lambda\right) & \simeq \alpha_{2 c} \, \sqrt{\ell_{2 c}^{-3} \, u_{2 c}^{\vphantom{-1}}\left[\ell_{2 c}^{-1} \left(\vec{k} - \vec{k}_{2 c}\right)\right]} \; \uvec{\varepsilon}_{2 c} \cdot \uvec{\varepsilon}\left(\vec{k}_{2 c}, \lambda\right)
\end{align}
and equivalently $f_{1 c}$.
We then set
\begin{subequations}%
  \begin{align}%
    \uvec{k}_{2 c} & = \frac{1}{\sqrt{3}} \bigl(\sqrt{2} \, \uvec{x} - \uvec{z}\bigr), \\
    \uvec{\varepsilon}_{2 c} & = \frac{1}{\sqrt{14}} \, \left[{\textstyle \sqrt{5 + 3 \sqrt{2}}} \; \frac{1}{\sqrt{3}} \bigl(- \uvec{x} - \sqrt{2} \, \uvec{z}\bigr) - \ii {\textstyle \sqrt{9 - 3 \sqrt{2}}} \, \uvec{y}\right], \\
    \uvec{k}_{1 c} & = \frac{1}{3} \bigl(\uvec{x} - 2 \sqrt{2} \, \uvec{z}\bigr), \\
    \uvec{\varepsilon}_{1 c} & = \frac{1}{\sqrt{2}} \left[\frac{1}{3} \bigl(2 \sqrt{2} \, \uvec{x} + \uvec{z}\bigr) + \ii \uvec{y}\right].
  \end{align}%
  \label{eqs:settings:closed}%
\end{subequations}%
$\uvec{K}_{+}$ stays as it is
[\cf{} Eq.~\eqref{eq:Kplus}]
as shown in Fig.~\ref{fig:geometry:closed}.
The transition amplitudes then become
\begin{subequations}%
  \begin{multline}%
    T_{1}\bigl[\hat{f}_{2 c}\bigr]\left(\vec{K}_{+} \,\mathopen{}+\tfrac{1}{2}, +\tfrac{1}{2}\right) = \frac{1}{1 + \sqrt{2}} \, T_{1}\bigl[\hat{f}_{2 c}\bigr]\left(\vec{K}_{+} \,\mathopen{}-\tfrac{1}{2}, +\tfrac{1}{2}\right) = \\
    - T_{1}\bigl[\hat{f}_{2 c}\bigr]\left(\vec{K}_{+} \,\mathopen{}+\tfrac{1}{2}, -\tfrac{1}{2}\right) = - T_{1}\bigl[\hat{f}_{2 c}\bigr]\left(\vec{K}_{+} \,\mathopen{}-\tfrac{1}{2}, -\tfrac{1}{2}\right) = \ee^{\ii \varphi_{2 c}} \\
    \times \int \mathrm{d}\vec{k} \, \delta\bigl[\textstyle E_{-} - E_{+} + \hbar \omega\left(k\right)\bigr] \, \sqrt{u_{2 c}\left[\ell_{2 c}^{-1} \left(\vec{k} - \vec{k}_{2 c}\right)\right]} \\
    \times \frac{\sqrt{\pi} \hbar^{2} \mathcal{E}\left(k\right)}{3 \sqrt{7 \ell_{2 c}^{3} m_{\mathrm{e}} K}} \, {\textstyle \sqrt{3 - \sqrt{2}}} \, \left[D_{1}\left(E_{+} 1 \tfrac{1}{2}, n 0 \tfrac{1}{2}\right) - D_{1}\left(E_{+} 1 \tfrac{3}{2}, n 0 \tfrac{1}{2}\right)\right]
    \label{eq:T1:closed:is}%
  \end{multline}%
  and%
  \begin{multline}%
    T_{2}\bigl[\hat{f}_{1 c}\bigr]\left(\vec{K}_{+} m_{s}, m_{j}\right) = \ee^{2 \ii \varphi_{1 c}} \smashoperator{\int} \mathrm{d}\vec{k}_{1} \smashoperator{\int} \mathrm{d}\vec{k}_{2} \, \delta\bigl[\textstyle E_{-} - E_{+} + \sum_{i = 1}^{2} \hbar \omega\left(k_{i}\right)\bigr] \\
    \times \sqrt{u_{1 c}\left[\ell_{1 c}^{-1} \left(\vec{k}_{1} - \vec{k}_{1 c}\right)\right]} \sqrt{u_{1 c}\left[\ell_{1 c}^{-1} \left(\vec{k}_{2} - \vec{k}_{1 c}\right)\right]} \, \frac{\sqrt{\pi} \ii \hbar^{3} \mathcal{E}\left(k_{1}\right) \mathcal{E}\left(k_{2}\right)}{90 \sqrt{m_{\mathrm{e}} K_{+}}} \\
    \times \begin{dcases}
      \!\begin{multlined}[b]
        \bigl(-1 + \sqrt{2}\bigr) \bigl[5 D_{2}\left(E_{+} 2 \tfrac{3}{2}, 1 \tfrac{1}{2}, n 0 \tfrac{1}{2}; k_{2}\right) \\[.35\baselineskip]
        + D_{2}\left(E_{+} 2 \tfrac{3}{2}, 1 \tfrac{3}{2}, n 0 \tfrac{1}{2}; k_{2}\right) \\
        - 3 \bigl(7 + 5 \sqrt{2}\bigr) D_{2}\left(E_{+} 2 \tfrac{5}{2}, 1 \tfrac{3}{2}, n 0 \tfrac{1}{2}; k_{2}\right)\bigr]
      \end{multlined} & \text{if $m_{s} = \tfrac{1}{2}$ and $m_{j} = \tfrac{1}{2}$,} \\[.35\baselineskip]
      \!\begin{multlined}[b]
        \bigl(-1 + \sqrt{2}\bigr) \bigl[-5 D_{2}\left(E_{+} 2 \tfrac{3}{2}, 1 \tfrac{1}{2}, n 0 \tfrac{1}{2}; k_{2}\right) \\[.35\baselineskip]
        - D_{2}\left(E_{+} 2 \tfrac{3}{2}, 1 \tfrac{3}{2}, n 0 \tfrac{1}{2}; k_{2}\right) \\
        + 6 D_{2}\left(E_{+} 2 \tfrac{5}{2}, 1 \tfrac{3}{2}, n 0 \tfrac{1}{2}; k_{2}\right)\bigr]
      \end{multlined} & \text{if $m_{s} = -\tfrac{1}{2}$ and $m_{j} = \tfrac{1}{2}$,} \\[.35\baselineskip]
      \!\begin{multlined}[b]
        \bigl(1 + \sqrt{2}\bigr) \bigl[5 D_{2}\left(E_{+} 2 \tfrac{3}{2}, 1 \tfrac{1}{2}, n 0 \tfrac{1}{2}; k_{2}\right) \\[.35\baselineskip]
        + D_{2}\left(E_{+} 2 \tfrac{3}{2}, 1 \tfrac{3}{2}, n 0 \tfrac{1}{2}; k_{2}\right) \\
        - 6 D_{2}\left(E_{+} 2 \tfrac{5}{2}, 1 \tfrac{3}{2}, n 0 \tfrac{1}{2}; k_{2}\right)\bigr]
      \end{multlined} & \text{if $m_{s} = \tfrac{1}{2}$ and $m_{j} = -\tfrac{1}{2}$,} \\[.35\baselineskip]
      \!\begin{multlined}[b]
        \bigl(1 + \sqrt{2}\bigr) \bigl[-5 D_{2}\left(E_{+} 2 \tfrac{3}{2}, 1 \tfrac{1}{2}, n 0 \tfrac{1}{2}; k_{2}\right) \\[.35\baselineskip]
        - D_{2}\left(E_{+} 2 \tfrac{3}{2}, 1 \tfrac{3}{2}, n 0 \tfrac{1}{2}; k_{2}\right) \\
        + 3 \bigl(7 - 5 \sqrt{2}\bigr) D_{2}\left(E_{+} 2 \tfrac{5}{2}, 1 \tfrac{3}{2}, n 0 \tfrac{1}{2}; k_{2}\right)\bigr]
      \end{multlined} & \text{if $m_{s} = -\tfrac{1}{2}$ and $m_{j} = -\tfrac{1}{2}$.}
    \end{dcases}
    \label{eq:T2:closed:is}%
  \end{multline}%
  \label{eqs:T12:closed:is}%
\end{subequations}%
Evidently, $T_{1}$ and $T_{2}$
do not yet fulfil the conditions \eqref{eqs:T12:closed:want:bias}.
First, in Eq.~\eqref{eq:T1:closed:is},
$T_{1}\bigl[\hat{f}_{2 c}\bigr]\left(\vec{K}_{+} \,\mathopen{}-\tfrac{1}{2}, +\tfrac{1}{2}\right)$
deviates from the required behavior, being
too large by a factor of $(1 + \sqrt{2})$.
By contrast,
the $T_1$ 
for the other values of $m_{s}$ and $m_{j}$
are as unbiased as required,
and they have the correct signs.
Therefore, the issue can be avoided
by using the input port $m_{j} = - \frac{1}{2}$, \ie{}, by 
setting $p_{\frac{1}{2}} = 0$ and $p_{-\frac{1}{2}} = 1$
in $\varrho_{\mathcal{M}}\left(-\infty\right)$
[Eq.~\eqref{eq:varrhoM:mj} on page \pageref{eq:varrhoM:mj}].

Second, as can be seen from Eq.~\eqref{eq:T2:closed:is},
$T_{2}$ is different for all four possible combinations
of $m_{s} = \pm \frac{1}{2}$ and $m_{j} = \pm \frac{1}{2}$,
because the second-order radial integral
$D_{2}\left(E_{+} 2 \tfrac{5}{2}, 1 \tfrac{3}{2}, n 0 \tfrac{1}{2}; k_{2}\right)$
enters in all four cases with a different prefactor. 
This problem cannot be solved by altering $\uvec{K}_{+}$,
$\uvec{k}_{2 c}$, $\uvec{\varepsilon}_{2 c}$,
$\uvec{k}_{1 c}$, or $\uvec{\varepsilon}_{1 c}$.
In order to fulfill the conditions
\eqref{eqs:T12:closed:want:bias},
we have to impose a requirement on
$D_{2}\left(E_{+} 2 \tfrac{5}{2}, 1 \tfrac{3}{2}, n 0 \tfrac{1}{2}; k_{2}\right)$, \eg{},
\begin{gather}
  D_{2}\left(E_{+} 2 \tfrac{5}{2}, 1 \tfrac{3}{2}, n 0 \tfrac{1}{2}; k_{2}\right) = 0.
\end{gather}
That is, one has to find an energy $E_{+}$
or system, where the transition to $\mathrm{D} \frac{5}{2}$
is negligible compared to the transition to $\mathrm{D} \frac{3}{2}$.%
\footnote{If only the input port $m_{j} = -\frac{1}{2}$ is used,
then the stipulation that
\begin{multline*}
  D_{2}\left(E_{+} 2 \tfrac{5}{2}, 1 \tfrac{3}{2}, n 0 \tfrac{1}{2}; k_{2}\right) = \frac{2}{93} \bigl(9 + 5 \sqrt{2}\bigr) \\
  \times \bigl[5 D_{2}\bigl(E_{+} 2 \tfrac{3}{2}, 1 \tfrac{1}{2}, n 0 \tfrac{1}{2}; k_{2}\bigr) + D_{2}\bigl(E_{+} 2 \tfrac{3}{2}, 1 \tfrac{3}{2}, n 0 \tfrac{1}{2}; k_{2}\bigr)\bigr]
\end{multline*}
is a possible solution to the bias issue as well.}

The final task is to
balance the magnitudes of
$T_{1}\bigl[\hat{f}_{2 c}\bigr]\left(\vec{K}_{+} \,\mathopen{}+\tfrac{1}{2}, m_{j}\right)$ and
$T_{2}\bigl[\hat{f}_{1 c}\bigr]\left(\vec{K}_{+} \,\mathopen{}+\tfrac{1}{2}, m_{j}\right)$
in accordance with Eq.~\eqref{eq:T12:closed:want:symmetry}.
This can be achieved by adjusting
$\sqrt{\overlap{\tilde{\chi}^{1}_{c}}{\tilde{\chi}^{1}_{c}}}$
and $\sqrt{\overlap{\tilde{\chi}^{2}_{c}}{\tilde{\chi}^{2}_{c}}}$
(or $\abs*{\alpha_{1 c}}^{2}$ and $\abs*{\alpha_{2 c}}$,
respectively, in the case of coherent states of incoming light).

In summary,
this section shows how to build an open and a closed CCI
for the specific example of photoelectron spin control.
With the settings as in Eqs.~\eqref{eq:Kplus}
and \eqref{eqs:settings:closed},
it is possible to do the
quantum erasure coherent control experiment
of Sec.~\ref{sec:quantum_erasure}.
The quantum delayed choice coherent control scenario
of Sec.~\ref{sec:delayed_choice}
can be implemented with a combination of Eqs.~\eqref{eq:Kplus},
\eqref{eqs:settings:open} and \eqref{eqs:settings:closed}.
The existence of a single direction of detection
$\uvec{K}_{+}$ [Eq.~\eqref{eq:Kplus}]
that works for both the open and the closed configuration
makes this possible.
See Ref.~\citealp{Scholak:2014aa}
for further details.

\section{Facing the Loopholes}
\label{sec:application:discussion}

In any test of a Bell inequality,
one has to deal with three loopholes:
locality,
fair-sampling,
and freedom-of-choice.
Only if the Bell violation occurs under
simultaneous closing of these loopholes,
does a Bell inequality provide
a stringent test for local realism.
As discussed below,
this presents a significant challenge
in the case of photoelectron spin control.
Indeed, closing these loopholes had been
one of the major problems in fundamental physics
for a long time \cite{Miller:2016tg}.
\begin{description}
  \item[Locality loophole.]
    In both the quantum erasure and the quantum delayed choice
    coherent control scenario,
    the locality loophole arises from the possibility that
    the measurement outcome of the spin polarization is
    influenced by the measurement of the light,
    or vice versa.
    The skeptic's suspicion is that the choice
    of the measurement observable, or of its measurement outcome,
    could have been heralded
    with subluminal or luminal speed
    from the location of the photodetection to the location
    of the spin detection.
  
    This locality loophole can be closed by
    separating both the choice of observable and the measurement events
    of either subsystem by space-like distances,
    excluding any causal influence.
    Additional confidence could be built
    by deliberately randomizing the observable choices
    using random numbers generated
    by quantum random number generators.
    In the case of photoelectron spin control,
    closing the locality loophole
    seems like an almost insurmountable challenge.
  \item[Fair-sampling or detection loophole.]
    This loophole recognizes that every detection is imperfect,
    and electron spin detectors are particularly flawed:
    only a small fraction of the photoelectrons can be collected,
    and an even smaller fraction has a discernible spin.
    With such a small detection efficiency,
    the detected electron spins may not be representative
    of the entire statistical ensemble, and they
    may suggest a violation of local realism
    despite the fact that the entirety of spins still obeys it.
    Equivalent issues, albeit less severe,
    exist for the detection of the light.
    
    The detection loophole can be dealt with in two ways.
    Either one relies on ``fair sampling''
    (\ie{}, the supplementary hypothesis that
    the sample of detected spins
    is representative of the whole sample),
    or one closes the loophole by increasing the efficiency
    above a certain threshold
    ($2 \sqrt{2} - 2 \simeq 83\%$ \cite{Garg:1987di}).
    However, with present technology,
    this threshold is unattainable for electron spin detection.
    Thus, in the photoelectron spin experiment,
    the assumption of fair sampling would have to be adopted
    to close this loophole.
  \item[Freedom-of-choice loophole.]
    This loophole arises from the assumption
    that the choice of the observables on the electron spin
    and the light is independent and free
    of properties of the physical system being measured,
    or of some local hidden variable.
    If this freedom were denied, and if all choices
    were completely deterministic, then
    the outcome of the Bell measurements would
    have been determined in advance.
    
    This is a loophole for local realism
    and, in a completely deterministic world,
    it cannot be closed \cite{Scheidl:2010vo}.
    If nature is nondeterministic, however,
    then the problem of freedom-of-choice
    can be solved by having a space-like separation
    between the control event and
    the choice of the measurement observables \cite{Hensen:2015lf}.
    But again, for photoelectron spin control,
    this is a rather difficult undertaking.
\end{description}
Closing all these loopholes simultaneously
(for any physical process)
had been one of the most significant and difficult problems
in foundational physics.
It has only been overcome recently
for entangled electron spins \cite{Hensen:2015lf}
and entangled photon pairs
\cite{Giustina:2015sl,Shalm:2015sl}.
With all the specific attributes of
photoelectron spin control, however,
repeating such a success would be expremely difficult.
Furthermore,
we could have discussed far simpler examples of
quantum erasure \cite{Chiao:1995qn,Walborn:2002gd,Aharonov:2005cs},
delayed choice \cite{Kim:2000th,Jacques:2007uq,Manning:2015aa,Ma:2016aa},
and quantum delayed choice \cite{Tang:2012kx,Kaiser:2012fk,Peruzzo:2012mw},
based on established technology like linear quantum optics
that is used routinely in applications of
quantum information theory.
Indeed, we do not propose photoelectron spin control
to show that quantum erasure
and quantum delayed choice are possible
or to provide further evidence that nature violates local realism.
Rather, we set out to find the most quantum of
the quantum phase control scenarios
and the limits of their verification.
It appears we have successfully examined these limits.

\section{Conclusions}
\label{sec:conclusions}

We began this article by noting
that two-color weak-field phase control
is analogous to a classical control phenomenon.
Hence, it cannot be advertised as a true quantum interference effect.
In a nutshell,
there are several reasons why this is the case.
\begin{enumerate}
  \item
    The control features of two-color weak-field phase control
    can be derived perturbatively from nonlinear response theory,
    an approach that applies to both classical and quantum matter.
    In both cases, nonlinear response theory predicts both the
    nonadditivity and the phase sensitivity of the response.
    (Nonadditivity implies that the collective response to
    the incident driving fields differs from
    the sum of the individual responses to each field component.
    Phase sensitivity refers to
    the dependence of the response on the
    phases of the field components.
    These are the features of two-color phase control.)
  \item
    Perturbative nonlinear response theory
    separates the response into
    material and nonmaterial contributions.
    The phase sensitivity originates exclusively
    from the nonmaterial contributions.
    Since these contributions are classical,
    differences between a classical and quantum response
    can only result from quantitative differences
    in the material contribution.
    These differences are not related to either
    phase sensitivity or to quantum interference.
  \item
    Formally, claiming that the observation of
    a phase-sensitive and nonadditive response
    in experiments of coherent phase control
    is a demonstration of quantum interference
    is a logical fallacy known as the
    ``affirmation of the consequent''.
    Agreement and consistency with a theory
    does not constitute a proof,
    even if no other explanation is known.
\end{enumerate}

These points raised the question of
how a conventional coherent control experiment in general,
and a phase control experiment in particular, can be studied
from a foundational perspective and developed into
a genuinely and verifiably quantum interference phenomenon.
This is the issue that was explored in the
main part of this article.
Specifically, we first introduced a general, unifying framework for
studying all-optical control of matter
using two-color light fields.
The theory was derived under minimal assumptions.
It is applicable to a wide variety of control targets
including atoms, molecules, and bulk solids.
The control field was treated in the continuous multimode formalism,
a useful mathematical tool to account for realistic pulse profiles
that are compact both in space and in frequency.
Furthermore, the approach allowed us to consider any choice of
the light's photon statistics.
We thus derived expressions that apply in equal measure to
semiclassical and to quantum states of light,
such as coherent state, Fock states, or squeezed states.

At the heart of this framework is a new kind of formal interferometer,
the coherent control interferometer (CCI),
which allowed us to reach a significant level of abstraction.
It is defined in terms of its input and output ports
as well as via a special set of scattering states
that play the role of a binary path degree of freedom.
By means of these prescriptions, knowing the path traversed in the CCI
is equivalent to knowing the frequency of the light with which
the matter has interacted.

Through the introduction of the CCI,
the problem of phase control was translated
into the language of common two-way interferometry,
including interferometers such as the Young double-slit,
the Mach-Zehnder interferometer,
the Stern-Gerlach apparatus,
and the Ramsey interferometer.
This allowed detailed clarification of
the concepts of waves, particles,
and their complementarity in two-color phase control scenarios.
These fundamental concepts are the keys
to demonstrating quantum interference
in coherent phase control.
They are also of practical relevance,
since, as we found,
the success of phase control is always limited by
the simultaneously acquirable amount of path knowledge.
The development of the CCI allowed
an approach to systematically exploring this relationship.

To this end,
we proposed the quantum erasure coherent phase control scenario,
in which the CCI is subject to which-path marking.
Unlike conventional phase control scenarios,
this one makes use of quantum light,
which both controls the dynamics
and becomes entangled with the matter in the process.
As a consequence, the path taken through the CCI is marked,
providing path knowledge that can be extracted,
if not in practice then at least in principle.
Phase control, therefore, has to be poor, unless
the CCI is joint with a quantum erasure measurement
of the outgoing quantum light.
In this case, we found that control can be improved at the
expense of the simultaneously acquired path knowledge,
with limits set by quantitative wave-particle complementarity.
We further found that, in the setting of the CCI,
the Englert-Greenberger-Yasin duality relation
$\mathfrak{C}_{\mathcal{R}}^{2} + \mathfrak{P}^{2} \le 1$
provides a quantitative relationship
between the maximum controllability $\mathfrak{C}_{\mathcal{R}}$
and the minimum path knowledge $\mathfrak{P}$
acquirable in any measurement of the outgoing light.
We derived the measurement observable that
saturates the above duality relation,
giving best possible phase control.
We also showed how quantum erasure can be achieved
in a homodyne measurement scheme
with a so-called click/no-click photodetector
that does not measure the number of photons.
With quantitative wave-particle complementarity and quantum erasure,
this proposed coherent control scenario
can demonstrate two hallmark features of quantum mechanics.

As a second application of the CCI formalism we proposed the
quantum delayed-choice coherent control scenario,
designed to demonstrate Wheeler's idea of delayed choice,
another hallmark of quantum mechanics.
This scenario, which employs quantum light as well,
realizes two complementary situations simultaneously.
In the first, phase control is absent,
characterized by particle-like CCI output port statistics.
In the second, phase control is present,
and the CCI displays wave-like statistics.
These two situations are coherently blended
and conditioned on marker states of the outgoing light.
We showed that this can be described in terms of entanglement between
the radiation field and the control target.
Our theory for quantum delayed-choice coherent control scenario predicts that,
prior to measurement,
it is impossible to know
whether there will be phase control or not.
Upon measurement, the case that applies will be random.
This is the essence of
quantum delayed-choice coherent phase control.

In the last section,
we showed how these two scenarios,
quantum erasure and the quantum delayed choice coherent control,
are not only consistent with quantum theory,
but are in fact inconsistent with
every conceivable---or inconceivable---classical description.
To this end, we constructed Bell inequality tests
for each of the two proposed scenarios.
For the quantum erasure scenario, we showed that
a violation of the Bell inequality certifies that
the erased path distinguishability
has been of nonclassical origin.
For the quantum delayed-choice scenario,
on the other hand,
we showed that a failed Bell test suggests that
the outcome of the experiment has not only been unpredictable,
but also nonclassically random.
Thus, in both scenarios,
a Bell inequality violation demonstrates that
phase control originates from nontrivial and
nonclassical interference effects.
It is quantum coherent control.

To demonstrate the utility of this framework
and the feasibility of the two proposed control scenarios,
we examined the specific case of
photoelectron spin control,
where a heavy alkali atom is ionized by weak coherent
$(\omega + 2 \omega)$ radiation.
With this example, we expanded on the
seminal two-color coherent phase control experiments of Elliott \etal{},
in which, according to our discussion,
control is analogous to a classical control phenomenon.
In doing so, we showed how
a well-known coherent phase control experiment
can be modified to display quantum coherent control.

In conclusion,
we proposed a general framework to examine various degrees of ``quantumness'' 
in phase control scenarios and paved the way for new coherent control experiments
in which phase control will be a genuinely and verifiably quantum interference phenomenon.
Applications of this approach to other molecular control scenarios
is expected to be highly enlightening.

\section*{Acknowledgments}
\addcontentsline{toc}{section}{Acknowledgments}

Financial support from
the Natural Sciences and Engineering Research Council of Canada
is gratefully acknowledged.
T.S.~thanks Christine Tewfik for her patience and support, and
P.B.~thanks Professor A.~Steinberg for numerous discussions
preliminary to this work.

\begingroup
\appendix%
\renewcommand*{\thesection}{\Alph{section}}%

\section{The Longtime Limit of The Total Density Operator}
\label{app:derivation_total_density_operator}

Below, perturbation theory is used to
solve the light-matter scattering problem
for weak field unitary $1$ \vs{} $N$ coherent phase control dynamics,
where $N = 2, 3, \ldots$~.
Although standard quantum structures are simpler \cite{Shapiro:1988cr},
we use the Liouville representation of quantum mechanics
to allow a future extension to nonunitary dynamics.
Here, Liouville space is a vector space
of operators on the regular Hilbert space
(such as Hamiltonians, observables, density operators, \etc{}),
and superoperators are the operators acting on Liouville space.
The time evolution superoperator
$\mathscr{U}(t, t') = \exp[- \ii \mathscr{L} (t-t') / \hbar]$
solves the Liouville equation
\begin{align}
  \ii \hbar \partial_{t} \mathscr{U}(t, t') & = \mathscr{L} \mathscr{U}(t, t'),
  \label{eq:liouville}
\end{align}
where
$\mathscr{L} = \mathscr{L}_{0} + \mathscr{V}$
the total Liouville superoperator of the system,
corresponds to the commutator
$\left[H, . \, \right] = \left[H_{0} + V, . \, \right]$.
Equation \eqref{eq:liouville} is equivalent to the von Neumann equation
for the density operator $\varrho$.
For convenience,
we define a free time evolution superoperator
$\mathscr{U}_{0}(t,t')$ that solves the free Liouville equation
$\ii \hbar \partial_{t} \mathscr{U}_{0}(t, t') = \mathscr{L}_{0} \mathscr{U}_{0}(t, t')$,
with
$\mathscr{L}_{0} = \mathscr{L}_{\mathcal{M}} + \mathscr{L}_{\mathcal{R}} = \left[H_{\mathcal{M}} + H_{\mathcal{R}}, . \, \right] = \left[H_{0}, . \, \right]$
in which the material part of the system, $\mathcal{M}$,
and the radiation field, $\mathcal{R}$,
evolve independently of each other.
The perturbation
$\mathscr{V} = \left[V, . \, \right]$
describes the light-matter interaction,
treated within the electric dipole approximation.
The initial state of the system
is given by
\begin{subequations}%
  \begin{align}
    \lket*{\varrho\left(-\infty\right)} & = \lket*{\varrho_{\mathcal{M}}\left(-\infty\right)} \otimes \lket*{\varrho_{\mathcal{R}}\left(-\infty\right)},
    \shortintertext{where}%
    \lket*{\varrho_{\mathcal{M}}\left(-\infty\right)} & = \smashoperator{\sum_{\nu \in \mathcal{D}_{-}}} p_{\nu} \lket*{\nu \nu}
    \shortintertext{and}%
    \lket*{\varrho_{\mathcal{R}}\left(-\infty\right)} & = \lket*{\chi \chi}.
  \end{align}%
  \label{eq:varrhominusinfty:liouville}%
\end{subequations}%
These vectors are the Liouville representations of
Eqs.~\eqref{eq:varrhominusinfty},
\eqref{eq:varrhoM}, and \eqref{eq:chi},
respectively, \ie{}, $\lket*{\nu \nu}$ corresponds to $\ket*{\nu}\bra*{\nu}$,
\etc.

Consider then the scattering problem.
Initially, as well as finally,
the electromagnetic fields are far away from the
material target and therefore do not interact with it.
Scattering theory provides a map, the scattering superoperator,
\begin{align}
  \mathscr{S} & = \lim_{t \to \infty} \mathscr{U}_{0}\left(0, t/2\right) \, \mathscr{U}\left(t/2, -t/2\right) \, \mathscr{U}_{0}\left(-t/2, 0\right),
\end{align}
that relates a past asymptotic in-state
$\lket*{\varrho\left(-\infty\right)}$
to its future asymptotic out-state
$\lket*{\varrho\left(\infty\right)}$:
\begin{align}
  \lket*{\varrho\left(\infty\right)} = \mathscr{S} \lket*{\varrho\left(-\infty\right)}.
\end{align}
The superoperator $\mathscr{S}$ has a perturbative expansion,
\begin{align}
  \mathscr{S} & = 1 - 2 \pi \ii \int \mathrm{d}E \, \delta\left(E - \mathscr{L}_{0}\right) \, \mathscr{T}\left(E + \ii 0\right) \, \delta\left(\mathscr{L}_{0} - E\right),
\end{align}
with
\begin{align}
  \mathscr{T}\left(z\right) & = \mathscr{V} + \mathscr{V} \, \mathscr{G}_{0} \left(z\right) \, \mathscr{T} \left(z\right) = \sum_{n = 0}^{\infty} \mathscr{V} \left(\mathscr{G}_{0} \left(z\right) \mathscr{V}\right)^{n} = \sum_{n = 0}^{\infty} \mathscr{T}_{n}\left(z\right)
\end{align}
the transition superoperator, and with
$\mathscr{G}_{0} \left(z\right) = \left(z - \mathscr{L}_{0}\right)^{-1}$,
the unperturbed resolvent superoperator.

Restricting attention to only interesting asymptotic out-states
(\ie{}, those characterized by the quantum number set $\mathcal{D}_{+}$),
is equivalent to postselection,
$\lket*{\varrho\left(\infty\right)} \mapsto \lket*{\tilde{\varrho}\left(\infty\right)}$,
of the form
\begin{subequations}%
  \begin{align}%
    \lket*{\tilde{\varrho}\left(\infty\right)} & = \tilde{p}^{-1} \mathscr{P}_{\mathcal{D}_{+}} \lket*{\varrho\left(\infty\right)} \\
    & = \tilde{p}^{-1} \, \smashoperator{\sumint_{\mathcal{D}_{+}}} \mathrm{d}\mu \smashoperator{\sumint_{\mathcal{D}_{+}}} \mathrm{d}\mu' \; \lket*{\mu \mu'} \loverlap*{\mu \mu'}{\varrho\left(\infty\right)},
  \end{align}%
  where
  \begin{align}%
    \tilde{p} & = \smashoperator{\sumint_{\mathcal{D}_{+}}} \mathrm{d}\mu \, \loverlap*{\mu \mu, \operatorone_{\mathcal{R}}}{\varrho\left(\infty\right)}.
  \end{align}%
  \label{eq:tilde_varrho_infty:initial_definition}%
\end{subequations}%
That is, the superoperator
$\mathscr{P}_{\mathcal{D}_{+}}$
projects into the subspace spanned by $\lket*{\mu \mu'}$, with
$\mu, \mu'$ in the domain $\mathcal{D}_{+}$.

Below we calculate the leading terms
in the perturbative expansion of
$\lket*{\tilde{\varrho}\left(\infty\right)}$.
In coherent control,
the material ground state $\lket*{\nu \nu}$
and the relevant asymptotic out-states $\lket*{\mu \mu}$
(and out-coherences $\lket*{\mu \mu'}$)
do not overlap. Hence
$\mathscr{P}_{\mathcal{D}_{+}} \lket*{\varrho_{\mathcal{M}}\left(-\infty\right)} = 0$,
\ie{}, the zeroth-order term vanishes.
The first-order term vanishes as well
since it is proportional to matrix elements of the form
$\lbracket{\mu \mu', \ldots}{\mathscr{V}}{\nu \nu, \ldots}$
that evaluate to zero.
The first nonzero term is, therefore, of at least second order.
The first step in the calculation of that term
is the evaluation of
the second-order $\mathscr{S}$-matrix element
\begin{multline}
  - 2 \pi \ii \int \mathrm{d}E \, \lbra*{\mu \mu', \{\vec{q} \kappa\} \{\vec{q}' \kappa'\}} \\
  \times \delta\left(E - \mathscr{L}_{0}\right) \, \mathscr{T}_{2}\left(E + \ii 0\right) \, \delta\left(\mathscr{L}_{0} - E\right) \lket*{\nu \nu, \{\vec{k} \lambda\} \{\vec{k}' \lambda'\}},
  \label{eq:2ndorderterm}
\end{multline}
where the Liouville notations
$\lket*{\nu \nu, \{\vec{k} \lambda\} \{\vec{k}' \lambda'\}}$
and $\lket*{\mu \mu', \{\vec{q} \kappa\} \{\vec{q}' \kappa'\}}$
correspond to
\begin{subequations}%
  \begin{gather}%
    \ket{\nu}\bra{\nu} \otimes \ket{\vec{k}_{1}\lambda_{1}, \ldots, \vec{k}_{n}\lambda_{n}}\bra{\vec{k}'_{1}\lambda'_{1}, \ldots, \vec{k}'_{n'}\lambda'_{n'}}
    \shortintertext{and}
    \ket{\mu}\bra{\mu'} \otimes \ket{\vec{q}_{1}\kappa^{\vphantom{\prime}}_{1}, \ldots, \vec{q}_{m}\kappa^{\vphantom{\prime}}_{m}}\bra{\vec{q}'_{1}\kappa'_{1}, \ldots, \vec{q}'_{m'}\kappa'_{m'}},
  \end{gather}%
\end{subequations}%
respectively.
The numbers $n$ and $n'$ ($m$ and $m'$)
are the initial (final) photon numbers in the field.
The above vectors are eigenvectors of
the free Liouville operator $\mathscr{L}_{0}$.
In particular,
\begin{subequations}%
  \begin{gather}%
    \mathscr{L}_{0} \lket*{\nu \nu, \{\vec{k} \lambda\} \{\vec{k}' \lambda'\}} = \hbar \omega\left(\{k\}, \{k'\}\right) = \sum_{i = 1}^{n} \hbar \omega\left(k_{i}\right) - \sum_{i' = 1}^{n'} \hbar \omega\left(k'_{i'}\right)
  \end{gather}%
  and
  \begin{multline}%
    \mathscr{L}_{0} \lket*{\mu \mu', \{\vec{q} \kappa\} \{\vec{q}' \kappa'\}} = \Delta_{\mu \mu'} + \hbar \omega\left(\{q\}, \{q'\}\right) = E_{\mu} - E_{\mu'} \\
    + \sum_{j = 1}^{m} \hbar \omega\left(q_{j}\right) - \sum_{j' = 1}^{m'} \hbar \omega\left(q'_{j'}\right),
  \end{multline}%
\end{subequations}%
where $E_{\mu}$ denotes the eigenenergy of the
eigenstate $\ket{\mu}$ of the
material Hamiltonian $H_{\mathcal{M}}$
and $\hbar \omega(k) = \hbar c k$ is the energy of
single-photon eigenstate $\ket{\vec{k} \lambda}$
of the field Hamiltonian $H_{\mathcal{R}}$.

With these,
the second order $\mathscr{S}$-matrix element
[Eq.~\eqref{eq:2ndorderterm}]
becomes
\begin{multline}
  - 2 \pi \ii \, \delta\left[\Delta_{\mu \mu'} + \hbar \omega\left(\{q\}, \{q'\}\right) - \hbar \omega\left(\{k\}, \{k'\}\right)\right] \\ \times \Bigl\{\bracket*{\mu, \{\vec{q} \kappa\}}{V}{\nu, \{\vec{k} \lambda\}} \, \frac{1}{\Delta_{\mu' \nu} + \hbar \omega\left(\{q'\}, \{k\}\right) + \ii 0} \, \bracket*{\nu, \{\vec{k}' \lambda'\}}{V}{\mu', \{\vec{q}' \kappa'\}} \\ + \bracket*{\nu, \{\vec{k}' \lambda'\}}{V}{\mu', \{\vec{q}' \kappa'\}} \, \frac{1}{- \Delta_{\mu' \nu} - \hbar \omega\left(\{q'\}, \{k\}\right) + \ii 0} \, \bracket*{\mu, \{\vec{q} \kappa\}}{V}{\nu, \{\vec{k} \lambda\}}\Bigr\}.
  \label{eq:2ndorderterm2}
\end{multline}
The terms in the braces,
$\{\cdots+\cdots\}$,
can be illustrated graphically:
\begingroup
\tikzset{baseline={($(current bounding box.center)-(0,2.5pt)$)},->,>=stealth',shorten >=.5pt,node distance=1.55ex,every state/.style={rectangle split,rectangle split ignore empty parts,rectangle split horizontal,rectangle split part align=base,inner sep=0ex,minimum size=2.35ex,text width=5.5ex,align=center,draw}}
\begin{align}
  \begin{tikzpicture}
    \node[state] (n_nunup) {{\scriptsize $\nu \{\vec{k} \lambda\}$} \nodepart{two} {\scriptsize $\nu \{\vec{k}^{\mathclap{\,\prime}} \lambda^{\mathclap{\prime}}\}$}};
    \node[state] (n_numup) [right=of n_nunup] {{\scriptsize $\nu \{\vec{k} \lambda\}$} \nodepart{two} {\scriptsize $\mu^{\mathclap{\,\prime}} \{\vec{q}^{\mathclap{\,\prime}} \kappa^{\mathclap{\prime}}\}$}};
    \node[state] (n_mumup) [below=of n_numup] {{\scriptsize $\mu \{\vec{q} \kappa\}$} \nodepart{two} {\scriptsize $\mu^{\mathclap{\,\prime}} \{\vec{q}^{\mathclap{\,\prime}} \kappa^{\mathclap{\prime}}\}$}};
    \path[->] (n_numup) edge (n_mumup);
    \path[<-] (n_nunup) edge (n_numup);
  \end{tikzpicture} + \begin{tikzpicture}
    \node[state] (n_nunup) {{\scriptsize $\nu \{\vec{k} \lambda\}$} \nodepart{two} {\scriptsize $\nu \{\vec{k}^{\mathclap{\,\prime}} \lambda^{\mathclap{\prime}}\}$}};
    \node[state] (n_munup) [below=of n_nunup] {{\scriptsize $\mu \{\vec{q} \kappa\}$} \nodepart{two} {\scriptsize $\nu \{\vec{k}^{\mathclap{\,\prime}} \lambda^{\mathclap{\prime}}\}$}};
    \node[state] (n_mumup) [right=of n_munup] {{\scriptsize $\mu \{\vec{q} \kappa\}$} \nodepart{two} {\scriptsize $\mu^{\mathclap{\,\prime}} \{\vec{q}^{\mathclap{\,\prime}} \kappa^{\mathclap{\prime}}\}$}};
    \path[->] (n_nunup) edge (n_munup);
    \path[<-] (n_munup) edge (n_mumup);
  \end{tikzpicture} \, .
\end{align}
\endgroup
The diagrams represent the two possible pathways
through Liouville space that connect the in-vector
$\lket*{\nu \nu, \{\vec{k} \lambda\} \{\vec{k}' \lambda'\}}$
(shown in each case in the upper left corner)
to the out-vector
$\lket*{\mu \mu', \{\vec{q} \kappa\} \{\vec{q}' \kappa'\}}$
(in the lower right corner).
Liouville pathways describe
the joint evolution of both bras and kets
that comprise the system's density operator $\varrho$ \cite{Mukamel:1995po}.
The arrows correspond to transitions
mediated by the light-matter coupling $V$.
A horizontal arrow indicates action of $V$ from the right,
whereas a vertical arrow represents action from the left.
The above diagrams thus differ by the order
these interactions take place.

Straightforward algebraic manipulation reduces
Eq.~\eqref{eq:2ndorderterm2} to
\begin{multline}
  \left(- 2 \pi \ii\right) \, \delta\left[\Delta_{\nu \mu} + \hbar \omega\left(\{k\}, \{q\}\right)\right] \bracket*{\mu, \{\vec{q} \kappa\}}{V}{\nu, \{\vec{k} \lambda\}} \\
  \times 2 \pi \ii \, \delta\left[\Delta_{\mu' \nu} + \hbar \omega\left(\{q'\}, \{k'\}\right)\right] \bracket*{\nu, \{\vec{k}' \lambda'\}}{V}{\mu', \{\vec{q}' \kappa'\}} \, ,
  \label{eq:2ndorderterm3}
\end{multline}
an expression which describes scattering events
involving the emission
or absorption of a photon.
In the following,
we concentrate on absorption only and set
$m = n - 1$ and $m' = n' - 1$.
Energy conservation is guaranteed by the $\delta$-functions:
Eq.~\eqref{eq:2ndorderterm3} is nonzero only if
the energy of the absorbed photon energy resonant with material states
involved in the transition.
Consequently, the second-order term can only appear
in $\lket*{\tilde{\varrho}\left(\infty\right)}$
if at least one incoming wavepacket carries photons
with the required frequency.
This is indeed the case
in the $1$ \vs{} $N$ coherent phase control scenarios
with $N = 2, 3, \ldots$~.

The second-order term
\begin{gather}
  I_2 \equiv - 2 \pi \ii \int \mathrm{d}E \, \mathscr{P}_{\mathcal{D}_{+}} \delta\left(E - \mathscr{L}_{0}\right) \, \mathscr{T}_{2}\left(E + \ii 0\right) \, \delta\left(\mathscr{L}_{0} - E\right) \lket*{\varrho\left(-\infty\right)}
\end{gather}
can be calculated from Eq.~\eqref{eq:2ndorderterm3}
by expanding the initial state $\lket*{\varrho\left(-\infty\right)}$
in the Liouville basis
$\lket*{\nu \nu', \{\vec{k} \lambda\} \{\vec{k}' \lambda'\}}$,
\ie{} by using
\begin{multline}
  \lket*{\varrho\left(-\infty\right)} = \smashoperator{\sum_{\vphantom{\{\lambda'\}} \nu \in \mathcal{D}_{-}}} \int \mathrm{d}\{\vec{k}\} \sum_{\{\lambda\}} \int \mathrm{d}\{\vec{k}'\} \sum_{\{\lambda'\}} \lket*{\nu \nu, \{\vec{k} \lambda\} \{\vec{k}' \lambda'\}} \\
  \times \loverlap*{\nu \nu, \{\vec{k} \lambda\} \{\vec{k}' \lambda'\}}{\varrho\left(-\infty\right)} .
\end{multline}
Here we are interested in the result for the specific initial state
\eqref{eq:varrhominusinfty:liouville}
with the radiative part
$\lket*{\varrho_{\mathcal{R}}\left(-\infty\right)}$
corresponding to $\ket*{\chi} \bra*{\chi}$, where
\begin{align}
  \ket*{\chi} = g_{1}\bigl(a^{\dagger}\bigl[\hat{f}_{1}\bigr]\bigr) \, g_{2}\bigl(a^{\dagger}\bigl[\hat{f}_{2}\bigr]\bigr) \, \ket*{\mathrm{vac}} 
\end{align}
[see Eq.~\eqref{eq:chi} on page \pageref{eq:chi}].

After a lengthy calculation,
one obtains
\begin{multline}
  I_2 = \smashoperator{\sumint_{\mathcal{D}_{+}}} \mathrm{d}\mu \smashoperator{\sumint_{\mathcal{D}_{+}}} \mathrm{d}\mu' \, \ket*{\mu^{\vphantom{\prime}}} \bra*{\mu'} \, \otimes \, \smashoperator{\sum_{\nu \in \mathcal{D}_{-}}} \, p_{\nu} \, T_{1}\bigl[\hat{f}_{2}\bigr]\left(\mu, \nu\right) \, T_{1}^{*}\bigl[\hat{f}_{2}\bigr]\left(\mu', \nu\right) \\ \times g_{1}^{\vphantom{\left(2\right)}}\bigl(a^{\dagger}\bigl[\hat{f}_{1}\bigr]\bigr) \, g_{2}^{\left(1\right)}\bigl(a^{\dagger}\bigl[\hat{f}_{2}\bigr]\bigr) \ket*{\mathrm{vac}}\bra*{\mathrm{vac}} \, g_{1}^{\vphantom{\left(2\right)}}\bigl(a^{\dagger}\bigl[\hat{f}_{1}\bigr]\bigr)^{\dagger} \, g_{2}^{\left(1\right)}\bigl(a^{\dagger}\bigl[\hat{f}_{2}\bigr]\bigr)^{\dagger},
  \label{eq:2ndorderterm4}
\end{multline}
where $T_{1}\bigl[\hat{f}_{2}\bigr]\left(\mu, \nu\right)$
is as in Eq.~\eqref{eq:T1} on page \pageref{eq:T1}.
Eq.~\eqref{eq:2ndorderterm4} illustrates that,
to second order in $\mathscr{V}$,
only the second wavepacket
interacts with the material system,
whereas the first wavepacket passes
the matter unimpeded.
Given that the quantum resonance condition is fulfilled,
the second wavepacket undergoes change, from
$g_{2}^{\vphantom{\left(1\right)}}\bigl(a^{\dagger}\bigl[\hat{f}_{2}\bigr]\bigr) \ket*{\mathrm{vac}}$
to
$g_{2}^{\left(1\right)}\bigl(a^{\dagger}\bigl[\hat{f}_{2}\bigr]\bigr) \ket*{\mathrm{vac}}$.

In $1$ \vs{} $N$ scenarios,
there are only two other relevant orders in
the perturbative expansion of
$\lket*{\tilde{\varrho}\left(\infty\right)}$:
the $\left(N + 1\right)^{\text{th}}$ and
the $\left(2 N\right)^{\text{th}}$ order.
We focus first on the latter with $N = 2$
for simplicity.
Whereas the second-order term accounts for the
absorption of a single photon from the second wavepacket only,
the fourth-order term describes two-photon absorption
exclusively from the first wavepacket.
The fourth-order matrix element of $\mathscr{S}$ with respect to
$\lket*{\nu \nu, \{\vec{k} \lambda\} \{\vec{k}' \lambda'\}}$
and $\lket*{\mu \mu', \{\vec{q} \kappa\} \{\vec{q}' \kappa'\}}$
with initial photon numbers $n$ and $n'$
and final photon numbers $m = n - 2$ and $m' = n' - 2$,
respectively, reads
\begingroup
\tikzset{baseline={($(current bounding box.center)-(0,2.5pt)$)},->,>=stealth',shorten >=.5pt,node distance=1.55ex,every state/.style={rectangle split,rectangle split ignore empty parts,rectangle split horizontal,rectangle split part align=base,inner sep=0ex,minimum size=2.35ex,text width=5.5ex,align=center,draw}}
\begin{multline}
  2 \pi \ii \, \delta\left[\Delta_{\mu \mu'} + \hbar \omega\left(\{q\}, \{q'\}\right) - \hbar \omega\left(\{k\}, \{k'\}\right)\right] \, \smashoperator{\sum_{\xi, \xi'}} \int \mathrm{d}\{\vec{u}\} \sum_{\{\tau\}} \int \mathrm{d}\{\vec{u'}\} \\ \begin{aligned}[b]
    \times \sum_{\{\tau'\}} \, \left[\vphantom{\begin{tikzpicture}
      \node[state] (n_1) {{\scriptsize $1$} \nodepart{two} {\scriptsize $1$}};
      \node[state] (n_2) [below=of n_1] {{\scriptsize $2$} \nodepart{two} {\scriptsize $2$}};
      \node[state] (n_3) [below=of n_2] {{\scriptsize $3$} \nodepart{two} {\scriptsize $3$}};
    \end{tikzpicture}}\right. \begin{tikzpicture}
      \node[state] (n_nunup) {{\scriptsize $\nu \{\vec{k} \lambda\}$} \nodepart{two} {\scriptsize $\nu \{\vec{k}^{\mathclap{\,\prime}} \lambda^{\mathclap{\prime}}\}$}};
      \node[state] (n_nuxip) [right=of n_nunup] {{\scriptsize $\nu \{\vec{k} \lambda\}$} \nodepart{two} {\scriptsize $\xi^{\mathclap{\,\prime}} \{\vec{u}^{\mathclap{\,\prime}} \tau^{\mathclap{\prime}}\}$}};
      \node[state] (n_numup) [right=of n_nuxip] {{\scriptsize $\nu \{\vec{k} \lambda\}$} \nodepart{two} {\scriptsize $\mu^{\mathclap{\,\prime}} \{\vec{q}^{\mathclap{\,\prime}} \kappa^{\mathclap{\prime}}\}$}};
      \node[state] (n_ximup) [below=of n_numup] {{\scriptsize $\xi \{\vec{u} \tau\}$} \nodepart{two} {\scriptsize $\mu^{\mathclap{\,\prime}} \{\vec{q}^{\mathclap{\,\prime}} \kappa^{\mathclap{\prime}}\}$}};
      \node[state] (n_mumup) [below=of n_ximup] {{\scriptsize $\mu \{\vec{q} \kappa\}$} \nodepart{two} {\scriptsize $\mu^{\mathclap{\,\prime}} \{\vec{q}^{\mathclap{\,\prime}} \kappa^{\mathclap{\prime}}\}$}};
      \path[->] (n_numup) edge (n_ximup)
        (n_ximup) edge (n_mumup);
      \path[<-] (n_nunup) edge (n_nuxip)
        (n_nuxip) edge (n_numup);
    \end{tikzpicture} & \\[.35\baselineskip]
    + \, \begin{tikzpicture}
      \node[state] (n_nunup) {{\scriptsize $\nu \{\vec{k} \lambda\}$} \nodepart{two} {\scriptsize $\nu \{\vec{k}^{\mathclap{\,\prime}} \lambda^{\mathclap{\prime}}\}$}};
      \node[state] (n_nuxip) [right=of n_nunup] {{\scriptsize $\nu \{\vec{k} \lambda\}$} \nodepart{two} {\scriptsize $\xi^{\mathclap{\,\prime}} \{\vec{u}^{\mathclap{\,\prime}} \tau^{\mathclap{\prime}}\}$}};
      \node[state] (n_xixip) [below=of n_nuxip] {{\scriptsize $\xi \{\vec{u} \tau\}$} \nodepart{two} {\scriptsize $\xi^{\mathclap{\,\prime}} \{\vec{u}^{\mathclap{\,\prime}} \tau^{\mathclap{\prime}}\}$}};
      \node[state] (n_ximup) [right=of n_xixip] {{\scriptsize $\xi \{\vec{u} \tau\}$} \nodepart{two} {\scriptsize $\mu^{\mathclap{\,\prime}} \{\vec{q}^{\mathclap{\,\prime}} \kappa^{\mathclap{\prime}}\}$}};
      \node[state] (n_mumup) [below=of n_ximup] {{\scriptsize $\mu \{\vec{q} \kappa\}$} \nodepart{two} {\scriptsize $\mu^{\mathclap{\,\prime}} \{\vec{q}^{\mathclap{\,\prime}} \kappa^{\mathclap{\prime}}\}$}};
      \path[->] (n_nuxip) edge (n_xixip)
        (n_ximup) edge (n_mumup);
      \path[<-] (n_nunup) edge (n_nuxip)
        (n_xixip) edge (n_ximup);
    \end{tikzpicture} & \\[.35\baselineskip]
    + \, \begin{tikzpicture}
      \node[state] (n_nunup) {{\scriptsize $\nu \{\vec{k} \lambda\}$} \nodepart{two} {\scriptsize $\nu \{\vec{k}^{\mathclap{\,\prime}} \lambda^{\mathclap{\prime}}\}$}};
      \node[state] (n_xinup) [below=of n_nunup] {{\scriptsize $\xi \{\vec{u} \tau\}$} \nodepart{two} {\scriptsize $\nu \{\vec{k}^{\mathclap{\,\prime}} \lambda^{\mathclap{\prime}}\}$}};
      \node[state] (n_xixip) [right=of n_xinup] {{\scriptsize $\xi \{\vec{u} \tau\}$} \nodepart{two} {\scriptsize $\xi^{\mathclap{\,\prime}} \{\vec{u}^{\mathclap{\,\prime}} \tau^{\mathclap{\prime}}\}$}};
      \node[state] (n_ximup) [right=of n_xixip] {{\scriptsize $\xi \{\vec{u} \tau\}$} \nodepart{two} {\scriptsize $\mu^{\mathclap{\,\prime}} \{\vec{q}^{\mathclap{\,\prime}} \kappa^{\mathclap{\prime}}\}$}};
      \node[state] (n_mumup) [below=of n_ximup] {{\scriptsize $\mu \{\vec{q} \kappa\}$} \nodepart{two} {\scriptsize $\mu^{\mathclap{\,\prime}} \{\vec{q}^{\mathclap{\,\prime}} \kappa^{\mathclap{\prime}}\}$}};
      \path[->] (n_nunup) edge (n_xinup)
        (n_ximup) edge (n_mumup);
      \path[<-] (n_xinup) edge (n_xixip)
        (n_xixip) edge (n_ximup);
    \end{tikzpicture} & \\[.35\baselineskip]
    + \, \begin{tikzpicture}
      \node[state] (n_nunup) {{\scriptsize $\nu \{\vec{k} \lambda\}$} \nodepart{two} {\scriptsize $\nu \{\vec{k}^{\mathclap{\,\prime}} \lambda^{\mathclap{\prime}}\}$}};
      \node[state] (n_nuxip) [right=of n_nunup] {{\scriptsize $\nu \{\vec{k} \lambda\}$} \nodepart{two} {\scriptsize $\xi^{\mathclap{\,\prime}} \{\vec{u}^{\mathclap{\,\prime}} \tau^{\mathclap{\prime}}\}$}};
      \node[state] (n_xixip) [below=of n_nuxip] {{\scriptsize $\xi \{\vec{u} \tau\}$} \nodepart{two} {\scriptsize $\xi^{\mathclap{\,\prime}} \{\vec{u}^{\mathclap{\,\prime}} \tau^{\mathclap{\prime}}\}$}};
      \node[state] (n_muxip) [below=of n_xixip] {{\scriptsize $\mu \{\vec{q} \kappa\}$} \nodepart{two} {\scriptsize $\xi^{\mathclap{\,\prime}} \{\vec{u}^{\mathclap{\,\prime}} \tau^{\mathclap{\prime}}\}$}};
      \node[state] (n_mumup) [right=of n_muxip] {{\scriptsize $\mu \{\vec{q} \kappa\}$} \nodepart{two} {\scriptsize $\mu^{\mathclap{\,\prime}} \{\vec{q}^{\mathclap{\,\prime}} \kappa^{\mathclap{\prime}}\}$}};
      \path[->] (n_nuxip) edge (n_xixip)
        (n_xixip) edge (n_muxip);
      \path[<-] (n_nunup) edge (n_nuxip)
        (n_muxip) edge (n_mumup);
    \end{tikzpicture} & \\[.35\baselineskip]
    + \, \begin{tikzpicture}
      \node[state] (n_nunup) {{\scriptsize $\nu \{\vec{k} \lambda\}$} \nodepart{two} {\scriptsize $\nu \{\vec{k}^{\mathclap{\,\prime}} \lambda^{\mathclap{\prime}}\}$}};
      \node[state] (n_xinup) [below=of n_nunup] {{\scriptsize $\xi \{\vec{u} \tau\}$} \nodepart{two} {\scriptsize $\nu \{\vec{k}^{\mathclap{\,\prime}} \lambda^{\mathclap{\prime}}\}$}};
      \node[state] (n_xixip) [right=of n_xinup] {{\scriptsize $\xi \{\vec{u} \tau\}$} \nodepart{two} {\scriptsize $\xi^{\mathclap{\,\prime}} \{\vec{u}^{\mathclap{\,\prime}} \tau^{\mathclap{\prime}}\}$}};
      \node[state] (n_muxip) [below=of n_xixip] {{\scriptsize $\mu \{\vec{q} \kappa\}$} \nodepart{two} {\scriptsize $\xi^{\mathclap{\,\prime}} \{\vec{u}^{\mathclap{\,\prime}} \tau^{\mathclap{\prime}}\}$}};
      \node[state] (n_mumup) [right=of n_muxip] {{\scriptsize $\mu \{\vec{q} \kappa\}$} \nodepart{two} {\scriptsize $\mu^{\mathclap{\,\prime}} \{\vec{q}^{\mathclap{\,\prime}} \kappa^{\mathclap{\prime}}\}$}};
      \path[->] (n_nunup) edge (n_xinup)
        (n_xixip) edge (n_muxip);
      \path[<-] (n_xinup) edge (n_xixip)
        (n_muxip) edge (n_mumup);
    \end{tikzpicture} & \\[.35\baselineskip]
    + \, \begin{tikzpicture}
      \node[state] (n_nunup) {{\scriptsize $\nu \{\vec{k} \lambda\}$} \nodepart{two} {\scriptsize $\nu \{\vec{k}^{\mathclap{\,\prime}} \lambda^{\mathclap{\prime}}\}$}};
      \node[state] (n_xinup) [below=of n_nunup] {{\scriptsize $\xi \{\vec{u} \tau\}$} \nodepart{two} {\scriptsize $\nu \{\vec{k}^{\mathclap{\,\prime}} \lambda^{\mathclap{\prime}}\}$}};
      \node[state] (n_munup) [below=of n_xinup] {{\scriptsize $\mu \{\vec{q} \kappa\}$} \nodepart{two} {\scriptsize $\nu \{\vec{k}^{\mathclap{\,\prime}} \lambda^{\mathclap{\prime}}\}$}};
      \node[state] (n_muxip) [right=of n_munup] {{\scriptsize $\mu \{\vec{q} \kappa\}$} \nodepart{two} {\scriptsize $\xi^{\mathclap{\,\prime}} \{\vec{u}^{\mathclap{\,\prime}} \tau^{\mathclap{\prime}}\}$}};
      \node[state] (n_mumup) [right=of n_muxip] {{\scriptsize $\mu \{\vec{q} \kappa\}$} \nodepart{two} {\scriptsize $\mu^{\mathclap{\,\prime}} \{\vec{q}^{\mathclap{\,\prime}} \kappa^{\mathclap{\prime}}\}$}};
      \path[->] (n_nunup) edge (n_xinup)
        (n_xinup) edge (n_munup);
      \path[<-] (n_munup) edge (n_muxip)
        (n_muxip) edge (n_mumup);
    \end{tikzpicture} & \left.\vphantom{\begin{tikzpicture}
      \node[state] (n_1) {{\scriptsize $1$} \nodepart{two} {\scriptsize $1$}};
      \node[state] (n_2) [below=of n_1] {{\scriptsize $2$} \nodepart{two} {\scriptsize $2$}};
      \node[state] (n_3) [below=of n_2] {{\scriptsize $3$} \nodepart{two} {\scriptsize $3$}};
    \end{tikzpicture}}\right]
  \end{aligned}.
  \label{eq:4thorderterm}
\end{multline}
\endgroup
The algebraic version of the first diagram in the brackets is
\begin{multline}
  \bracket*{\mu, \{\vec{q} \kappa\}}{V}{\xi, \{\vec{u} \tau\}} \, \frac{1}{\Delta_{\mu' \xi} + \hbar \omega\left(\{q'\}, \{u\}\right) + \ii 0} \, \bracket*{\xi, \{\vec{u} \tau\}}{V}{\nu, \{\vec{k} \lambda\}} \\
  \times \frac{1}{\Delta_{\mu' \nu} + \hbar \omega\left(\{q'\}, \{k\}\right) + \ii 0} \, \bracket*{\xi', \{\vec{u}' \tau'\}}{V}{\mu', \{\vec{q}' \kappa'\}} \\
  \times \frac{1}{\Delta_{\xi' \nu} + \hbar \omega\left(\{u'\}, \{k\}\right) + \ii 0} \, \bracket*{\nu, \{\vec{k}' \lambda'\}}{V}{\xi', \{\vec{u}' \tau'\}} \, .
\end{multline}
According to Eq.~\eqref{eq:4thorderterm},
there are six different time-orderings in which
the light-matter interaction $V$ can act on the density matrix,
each represented by a unique diagram.
The Liouville pathways described by these diagrams
can combine in various ways,
and certain individual contributions can cancel
when all paths are summed up.

The fourth-order term
\begin{align}
  I_4 \equiv - 2 \pi \ii \int \mathrm{d}E \, \mathscr{P}_{\mathcal{D}_{+}} \delta\left(E - \mathscr{L}_{0}\right) \, \mathscr{T}_{4}\left(E + \ii 0\right) \, \delta\left(\mathscr{L}_{0} - E\right) \lket*{\varrho\left(-\infty\right)}
\end{align}
can be obtained from Eq.~\eqref{eq:4thorderterm}
in a manner corresponding to the calculation of the
second-order term \eqref{eq:2ndorderterm3},
again using $\lket*{\varrho\left(-\infty\right)}$
from Eq.~\eqref{eq:varrhominusinfty:liouville}.
We arrive at
\begin{multline}
  I_4 = \smashoperator{\sumint_{\mathcal{D}_{+}}} \mathrm{d}\mu \smashoperator{\sumint_{\mathcal{D}_{+}}} \mathrm{d}\mu' \, \ket*{\mu^{\vphantom{\prime}}} \bra*{\mu'} \, \otimes \, \smashoperator{\sum_{\nu \in \mathcal{D}_{-}}} \, p_{\nu} \, T_{2}^{\vphantom{*}}\bigl[\hat{f}_{1}\bigr]\left(\mu, \nu\right) \, T_{2}^{*}\bigl[\hat{f}_{1}\bigr]\left(\mu', \nu\right) \\ \times g_{1}^{\left(2\right)}\bigl(a^{\dagger}\bigl[\hat{f}_{1}\bigr]\bigr) \, g_{2}^{\vphantom{\left(1\right)}}\bigl(a^{\dagger}\bigl[\hat{f}_{2}\bigr]\bigr) \ket*{\mathrm{vac}}\bra*{\mathrm{vac}} \, g_{1}^{\left(2\right)}\bigl(a^{\dagger}\bigl[\hat{f}_{1}\bigr]\bigr)^{\dagger} \, g_{2}^{\vphantom{\left(1\right)}}\bigl(a^{\dagger}\bigl[\hat{f}_{2}\bigr]\bigr)^{\dagger}.
  \label{eq:4thorderterm2}
\end{multline}
It is instructive to compare Eq.~\eqref{eq:4thorderterm2}
to the second-order result
given in Eq.~\eqref{eq:2ndorderterm4}.
Whereas, to second order in $\mathscr{V}$,
only the second wavepacket
scatters with the material,
here, to fourth order,
only the first wavepacket does so.
Since two photons are absorbed,
Eq.~\eqref{eq:4thorderterm2} features
second-order ($T_{2}$) instead of first-order ($T_{1}$)
transition amplitudes.
Analogously, the function $g_{1}$
appears in the form of its second derivative.
It is therefore a simple matter to correctly guess the
$\left(2 N\right)^{\text{th}}$-order result. It reads as
\begin{multline}
  \smashoperator{\sumint_{\mathcal{D}_{+}}} \mathrm{d}\mu \smashoperator{\sumint_{\mathcal{D}_{+}}} \mathrm{d}\mu' \, \ket*{\mu^{\vphantom{\prime}}}\bra*{\mu'} \, \otimes \, \smashoperator{\sum_{\nu \in \mathcal{D}_{-}}} \, p_{\nu} \, T_{N}^{\vphantom{*}}\bigl[\hat{f}_{1}\bigr]\left(\mu, \nu\right) \, T_{N}^{*}\bigl[\hat{f}_{1}\bigr]\left(\mu', \nu\right) \\ \times g_{1}^{\left(N\right)}\bigl(a^{\dagger}\bigl[\hat{f}_{1}\bigr]\bigr) \, g_{2}^{\vphantom{\left(1\right)}}\bigl(a^{\dagger}\bigl[\hat{f}_{2}\bigr]\bigr) \ket*{\mathrm{vac}}\bra*{\mathrm{vac}} \, g_{1}^{\left(N\right)}\bigl(a^{\dagger}\bigl[\hat{f}_{1}\bigr]\bigr)^{\dagger} \, g_{2}^{\vphantom{\left(1\right)}}\bigl(a^{\dagger}\bigl[\hat{f}_{2}\bigr]\bigr)^{\dagger}
  \label{eq:2Nthorderterm2}
\end{multline}
with $T_{N}\bigl[\hat{f}_{1}\bigr]\left(\mu, \nu\right)$
as in Eq.~\eqref{eq:TN} on page \pageref{eq:TN}.
This term is a leading contribution
to $1$ \vs{} $N$ coherent phase control
with $N > 1$.

The third and last contribution
comes from the cross term
that emerges in $(N + 1)^{\text{th}}$ order,
\eg{}, order $3$ for $N = 2$.
In calculating the
third-order $\mathscr{S}$-matrix elements,
two cases must be distinguished.
In the first,
the ket portion of the density matrix
is subject to two-photon absorption,
whereas the bra portion undergoes
single-photon absorption, \ie{},
 $m = n - 2$ and $m' = n' - 1$.
The result is
\begingroup
\tikzset{baseline={($(current bounding box.center)-(0,2.5pt)$)},->,>=stealth',shorten >=.5pt,node distance=1.55ex,every state/.style={rectangle split,rectangle split ignore empty parts,rectangle split horizontal,rectangle split part align=base,inner sep=0ex,minimum size=2.35ex,text width=5.5ex,align=center,draw}}
\begin{multline}
  - 2 \pi \ii \, \delta\left[\Delta_{\mu \mu'} + \hbar \omega\left(\{q\}, \{q'\}\right) - \hbar \omega\left(\{k\}, \{k'\}\right)\right] \, \smashoperator{\sum_{\xi}} \int \mathrm{d}\{\vec{u}\} \\ \begin{aligned}[b]
    \times \sum_{\{\tau\}} \, \left[\vphantom{\begin{tikzpicture}
      \node[state] (n_1) {{\scriptsize $1$} \nodepart{two} {\scriptsize $1$}};
      \node[state] (n_2) [below=of n_1] {{\scriptsize $2$} \nodepart{two} {\scriptsize $2$}};
      \node[state] (n_3) [below=of n_2] {{\scriptsize $3$} \nodepart{two} {\scriptsize $3$}};
    \end{tikzpicture}}\right. \begin{tikzpicture}
      \node[state] (n_nunup) {{\scriptsize $\nu \{\vec{k} \lambda\}$} \nodepart{two} {\scriptsize $\nu \{\vec{k}^{\mathclap{\,\prime}} \lambda^{\mathclap{\prime}}\}$}};
      \node[state] (n_numup) [right=of n_nunup] {{\scriptsize $\nu \{\vec{k} \lambda\}$} \nodepart{two} {\scriptsize $\mu^{\mathclap{\,\prime}} \{\vec{q}^{\mathclap{\,\prime}} \kappa^{\mathclap{\prime}}\}$}};
      \node[state] (n_ximup) [below=of n_numup] {{\scriptsize $\xi \{\vec{u} \tau\}$} \nodepart{two} {\scriptsize $\mu^{\mathclap{\,\prime}} \{\vec{q}^{\mathclap{\,\prime}} \kappa^{\mathclap{\prime}}\}$}};
      \node[state] (n_mumup) [below=of n_ximup] {{\scriptsize $\mu \{\vec{q} \kappa\}$} \nodepart{two} {\scriptsize $\mu^{\mathclap{\,\prime}} \{\vec{q}^{\mathclap{\,\prime}} \kappa^{\mathclap{\prime}}\}$}};
      \path[->] (n_numup) edge (n_ximup)
        (n_ximup) edge (n_mumup);
      \path[<-] (n_nunup) edge (n_numup);
    \end{tikzpicture} & + \begin{tikzpicture}
      \node[state] (n_nunup) {{\scriptsize $\nu \{\vec{k} \lambda\}$} \nodepart{two} {\scriptsize $\nu \{\vec{k}^{\mathclap{\,\prime}} \lambda^{\mathclap{\prime}}\}$}};
      \node[state] (n_xinup) [below=of n_nunup] {{\scriptsize $\xi \{\vec{u} \tau\}$} \nodepart{two} {\scriptsize $\nu \{\vec{k}^{\mathclap{\,\prime}} \lambda^{\mathclap{\prime}}\}$}};
      \node[state] (n_ximup) [right=of n_xinup] {{\scriptsize $\xi \{\vec{u} \tau\}$} \nodepart{two} {\scriptsize $\mu^{\mathclap{\,\prime}} \{\vec{q}^{\mathclap{\,\prime}} \kappa^{\mathclap{\prime}}\}$}};
      \node[state] (n_mumup) [below=of n_ximup] {{\scriptsize $\mu \{\vec{q} \kappa\}$} \nodepart{two} {\scriptsize $\mu^{\mathclap{\,\prime}} \{\vec{q}^{\mathclap{\,\prime}} \kappa^{\mathclap{\prime}}\}$}};
      \path[->] (n_nunup) edge (n_xinup)
        (n_ximup) edge (n_mumup);
      \path[<-] (n_xinup) edge (n_ximup);
    \end{tikzpicture} \\[.35\baselineskip]
    & + \begin{tikzpicture}
      \node[state] (n_nunup) {{\scriptsize $\nu \{\vec{k} \lambda\}$} \nodepart{two} {\scriptsize $\nu \{\vec{k}^{\mathclap{\,\prime}} \lambda^{\mathclap{\prime}}\}$}};
      \node[state] (n_xinup) [below=of n_nunup] {{\scriptsize $\xi \{\vec{u} \tau\}$} \nodepart{two} {\scriptsize $\nu \{\vec{k}^{\mathclap{\,\prime}} \lambda^{\mathclap{\prime}}\}$}};
      \node[state] (n_ximup) [right=of n_xinup] {{\scriptsize $\xi \{\vec{u} \tau\}$} \nodepart{two} {\scriptsize $\mu^{\mathclap{\,\prime}} \{\vec{q}^{\mathclap{\,\prime}} \kappa^{\mathclap{\prime}}\}$}};
      \node[state] (n_mumup) [below=of n_ximup] {{\scriptsize $\mu \{\vec{q} \kappa\}$} \nodepart{two} {\scriptsize $\mu^{\mathclap{\,\prime}} \{\vec{q}^{\mathclap{\,\prime}} \kappa^{\mathclap{\prime}}\}$}};
      \path[->] (n_nunup) edge (n_xinup)
        (n_ximup) edge (n_mumup);
      \path[<-] (n_xinup) edge (n_ximup);
    \end{tikzpicture} \left.\vphantom{\begin{tikzpicture}
      \node[state] (n_1) {{\scriptsize $1$} \nodepart{two} {\scriptsize $1$}};
      \node[state] (n_2) [below=of n_1] {{\scriptsize $2$} \nodepart{two} {\scriptsize $2$}};
      \node[state] (n_3) [below=of n_2] {{\scriptsize $3$} \nodepart{two} {\scriptsize $3$}};
    \end{tikzpicture}}\right]
  \end{aligned},
\end{multline}
\endgroup
where the first diagram in the brackets can also be written as
\begin{multline}
  \bracket*{\mu, \{\vec{q} \kappa\}}{V}{\xi, \{\vec{u} \tau\}} \, \frac{1}{\Delta_{\mu' \xi} + \hbar \omega\left(\{q'\}, \{u\}\right) + \ii 0} \, \bracket*{\xi, \{\vec{u} \tau\}}{V}{\nu, \{\vec{k} \lambda\}} \\
  \times \frac{1}{\Delta_{\mu' \nu} + \hbar \omega\left(\{q'\}, \{k\}\right) + \ii 0} \, \bracket*{\nu, \{\vec{k}' \lambda'\}}{V}{\mu', \{\vec{q}' \kappa'\}} \, .
\end{multline}
The other case is the conjugate process,
one might say,
in which the ket loses a single photon
and the bra loses two of them, \ie{},
$m = n - 1$, $m' = n' - 2$.
As a result, the contribution is
\begingroup
\tikzset{baseline={($(current bounding box.center)-(0,2.5pt)$)},->,>=stealth',shorten >=.5pt,node distance=1.55ex,every state/.style={rectangle split,rectangle split ignore empty parts,rectangle split horizontal,rectangle split part align=base,inner sep=0ex,minimum size=2.35ex,text width=5.5ex,align=center,draw}}
\begin{multline}
  2 \pi \ii \, \delta\left[\Delta_{\mu \mu'} + \hbar \omega\left(\{q\}, \{q'\}\right) - \hbar \omega\left(\{k\}, \{k'\}\right)\right] \, \smashoperator{\sum_{\xi'}} \int \mathrm{d}\{\vec{u'}\} \\ \begin{aligned}[b]
    \times \sum_{\{\tau'\}} \, \left[\vphantom{\begin{tikzpicture}
      \node[state] (n_1) {{\scriptsize $1$} \nodepart{two} {\scriptsize $1$}};
      \node[state] (n_2) [below=of n_1] {{\scriptsize $2$} \nodepart{two} {\scriptsize $2$}};
    \end{tikzpicture}}\right. \begin{tikzpicture}
      \node[state] (n_nunup) {{\scriptsize $\nu \{\vec{k} \lambda\}$} \nodepart{two} {\scriptsize $\nu \{\vec{k}^{\mathclap{\,\prime}} \lambda^{\mathclap{\prime}}\}$}};
      \node[state] (n_nuxip) [right=of n_nunup] {{\scriptsize $\nu \{\vec{k} \lambda\}$} \nodepart{two} {\scriptsize $\xi^{\mathclap{\,\prime}} \{\vec{u}^{\mathclap{\,\prime}} \tau^{\mathclap{\prime}}\}$}};
      \node[state] (n_numup) [right=of n_nuxip] {{\scriptsize $\nu \{\vec{k} \lambda\}$} \nodepart{two} {\scriptsize $\mu^{\mathclap{\,\prime}} \{\vec{q}^{\mathclap{\,\prime}} \kappa^{\mathclap{\prime}}\}$}};
      \node[state] (n_mumup) [below=of n_numup] {{\scriptsize $\mu \{\vec{q} \kappa\}$} \nodepart{two} {\scriptsize $\mu^{\mathclap{\,\prime}} \{\vec{q}^{\mathclap{\,\prime}} \kappa^{\mathclap{\prime}}\}$}};
      \path[->] (n_numup) edge (n_mumup);
      \path[<-] (n_nunup) edge (n_nuxip)
        (n_nuxip) edge (n_numup);
    \end{tikzpicture} & \\[.35\baselineskip]
    + \, \begin{tikzpicture}
      \node[state] (n_nunup) {{\scriptsize $\nu \{\vec{k} \lambda\}$} \nodepart{two} {\scriptsize $\nu \{\vec{k}^{\mathclap{\,\prime}} \lambda^{\mathclap{\prime}}\}$}};
      \node[state] (n_nuxip) [right=of n_nunup] {{\scriptsize $\nu \{\vec{k} \lambda\}$} \nodepart{two} {\scriptsize $\xi^{\mathclap{\,\prime}} \{\vec{u}^{\mathclap{\,\prime}} \tau^{\mathclap{\prime}}\}$}};
      \node[state] (n_muxip) [below=of n_nuxip] {{\scriptsize $\mu \{\vec{q} \kappa\}$} \nodepart{two} {\scriptsize $\xi^{\mathclap{\,\prime}} \{\vec{u}^{\mathclap{\,\prime}} \tau^{\mathclap{\prime}}\}$}};
      \node[state] (n_mumup) [right=of n_muxip] {{\scriptsize $\mu \{\vec{q} \kappa\}$} \nodepart{two} {\scriptsize $\mu^{\mathclap{\,\prime}} \{\vec{q}^{\mathclap{\,\prime}} \kappa^{\mathclap{\prime}}\}$}};
      \path[->] (n_nuxip) edge (n_muxip);
      \path[<-] (n_nunup) edge (n_nuxip)
        (n_muxip) edge (n_mumup);
    \end{tikzpicture} & \\[.35\baselineskip]
    + \, \begin{tikzpicture}
      \node[state] (n_nunup) {{\scriptsize $\nu \{\vec{k} \lambda\}$} \nodepart{two} {\scriptsize $\nu \{\vec{k}^{\mathclap{\,\prime}} \lambda^{\mathclap{\prime}}\}$}};
      \node[state] (n_munup) [below=of n_nunup] {{\scriptsize $\mu \{\vec{q} \kappa\}$} \nodepart{two} {\scriptsize $\nu \{\vec{k}^{\mathclap{\,\prime}} \lambda^{\mathclap{\prime}}\}$}};
      \node[state] (n_muxip) [right=of n_munup] {{\scriptsize $\mu \{\vec{q} \kappa\}$} \nodepart{two} {\scriptsize $\xi^{\mathclap{\,\prime}} \{\vec{u}^{\mathclap{\,\prime}} \tau^{\mathclap{\prime}}\}$}};
      \node[state] (n_mumup) [right=of n_muxip] {{\scriptsize $\mu \{\vec{q} \kappa\}$} \nodepart{two} {\scriptsize $\mu^{\mathclap{\,\prime}} \{\vec{q}^{\mathclap{\,\prime}} \kappa^{\mathclap{\prime}}\}$}};
      \path[->] (n_nunup) edge (n_munup);
      \path[<-] (n_munup) edge (n_muxip)
        (n_muxip) edge (n_mumup);
    \end{tikzpicture} & \left.\vphantom{\begin{tikzpicture}
      \node[state] (n_1) {{\scriptsize $1$} \nodepart{two} {\scriptsize $1$}};
      \node[state] (n_2) [below=of n_1] {{\scriptsize $2$} \nodepart{two} {\scriptsize $2$}};
    \end{tikzpicture}}\right],
  \end{aligned}
\end{multline}
\endgroup
where we can also write
\begin{multline}
  \bracket*{\mu, \{\vec{q} \kappa\}}{V}{\nu, \{\vec{k} \lambda\}} \, \frac{1}{\Delta_{\mu' \nu} + \hbar \omega\left(\{q'\}, \{k\}\right) + \ii 0} \, \bracket*{\xi', \{\vec{u}' \tau'\}}{V}{\mu', \{\vec{q}' \kappa'\}} \\
  \times \frac{1}{\Delta_{\xi' \nu} + \hbar \omega\left(\{u'\}, \{k\}\right) + \ii 0} \, \bracket*{\nu, \{\vec{k}' \lambda'\}}{V}{\xi', \{\vec{u}' \tau'\}}
\end{multline}
for the first term in the brackets.
These expressions can be used
to derive the third-order contribution to
$\mathscr{P}_{\mathcal{D}_{+}} \lket*{\varrho\left(\infty\right)}$
with $\lket*{\varrho\left(-\infty\right)}$ as before.
Afterwards, it is straightforward
to obtain the terms that are
relevant in the $1$ \vs{} $2$ control scenario.
We skip these steps and directly generalize to the
$\left(N + 1\right)^{\text{th}}$-order result
for $1$ \vs{} $N$ coherent control.
We find
\begin{multline}
  - 2 \pi \ii \int \mathrm{d}E \, \mathscr{P}_{\mathcal{D}_{+}} \delta\left(E - \mathscr{L}_{0}\right) \, \mathscr{T}_{N + 1}\left(E + \ii 0\right) \, \delta\left(\mathscr{L}_{0} - E\right) \lket*{\varrho\left(-\infty\right)} \\
  = \smashoperator{\sumint_{\mathcal{D}_{+}}} \mathrm{d}\mu \smashoperator{\sumint_{\mathcal{D}_{+}}} \mathrm{d}\mu' \, \ket*{\mu^{\vphantom{\prime}}}\bra*{\mu'} \, \otimes \, \smashoperator{\sum_{\nu \in \mathcal{D}_{-}}} \, p_{\nu} \, T_{1}^{\vphantom{*}}\bigl[\hat{f}_{2}\bigr]\left(\mu, \nu\right) \, T_{N}^{*}\bigl[\hat{f}_{1}\bigr]\left(\mu', \nu\right) \\
  \times g_{1}^{\vphantom{\left(N\right)}}\bigl(a^{\dagger}\bigl[\hat{f}_{1}\bigr]\bigr) \, g_{2}^{\left(1\right)}\bigl(a^{\dagger}\bigl[\hat{f}_{2}\bigr]\bigr) \ket*{\mathrm{vac}}\bra*{\mathrm{vac}} \, g_{1}^{\left(N\right)}\bigl(a^{\dagger}\bigl[\hat{f}_{1}\bigr]\bigr)^{\dagger} \, g_{2}^{\vphantom{\left(1\right)}}\bigl(a^{\dagger}\bigl[\hat{f}_{2}\bigr]\bigr)^{\dagger} \\
  + \; \vphantom{\left[\sumint_{\mathcal{D}_{+}}\right]} \hc \, ,
  \label{eq:Np1thorderterm2}
\end{multline}
giving Eq.~\eqref{eq:tilde_varrho_infty}.

\endgroup

\bibliography{document}

\end{document}